\documentclass[12pt]{article}
\usepackage{graphicx}
\usepackage{epsf}
\usepackage{amsmath}
\usepackage{here}
\usepackage{setspace}
\usepackage{natbib}
\usepackage{xcolor}
\usepackage{varioref}
 \usepackage{wrapfig}
 \usepackage{threeparttable}
 \usepackage{dcolumn}
  \newcolumntype{d}{D{.}{.}{-1}}
 \usepackage{nomencl}
  \makeglossary
  \usepackage{subfigure}
 \usepackage{subfigmat}
 \usepackage{fancyvrb}
  \fvset{fontsize=\footnotesize,xleftmargin=2em}
 \usepackage{lettrine}

\usepackage{caption}
\usepackage{epstopdf}
\usepackage{lineno}

\begin{document}
\title{ Three-dimensional, Rotational Flamelet Closure Model with Two-way Coupling}
 \author{William A. Sirignano \\
  {\normalsize\itshape  Department of Mechanical and Aerospace Engineering }\\
   {\normalsize\itshape  University of California, Irvine, CA 92697}  \\
 }

\newcommand{\hs}{\mbox{\hspace{0.10in}}}
\newcommand{\hsm}{\mbox{\hspace{-0.06in}}}
\newcommand{\be}{\begin{equation}}
\newcommand{\ee}{\end{equation}}
\newcommand{\bea}{\begin{eqnarray}}
\newcommand{\eea}{\end{eqnarray}}

\maketitle

\begin{abstract}
A new flamelet model is developed for sub-grid modeling and coupled with the resolved flow for turbulent combustion. The model differs from current models in critical ways. (i) Non-premixed flames, premixed flames, or multi-branched flame structures are determined rather than prescribed. (ii) The effects of shear strain and vorticity  are determined. (iii) The strain rates and vorticity applied at the sub-grid level are directly determined from the resolved-scale strain rates and vorticity without a contrived progress variable. (iv) The flamelet model is three-dimensional . (v) The effect of variable density is addressed. Solutions to the multicomponent Navier-Stokes equations governing the flamelet model are obtained. By coordinate transformation, a similar solution is found for the model, through a system of ordinary differential equations. Vorticity  creates a centrifugal force on the sub-grid counterflow that modifies the molecular transport rates and burning rate.  Sample computations of the rotational flamelet model without coupling to the resolved flow are presented first to demonstrate the importance of the new features. Scaling laws are presented for relating strain rates and vorticity at the sub-grid level to quantities at the resolved-flow level for coupling with large-eddy simulations or  Reynolds-averaged flows. The time-averaged behavior of a simple turbulent flow is resolved with coupling to the rotational flamelet model. Specifically, a two-dimensional, multicomponent, time-averaged  planar shear layer with variable density and energy release is employed using a mixing-length description for the eddy viscosity. Needs for future study are identified.

\end{abstract}



\linespread{1.0}

\section{Introduction}

Combustion in high mass-flux chambers is the practical and major method for energy conversion for mechanical power and heating. Inherently, the high mass-flow rate leads to turbulent flow. Thereby, many length and time scales appear in the physics making serious challenges for both computational and experimental analyses. For computations where the smallest scales typically cannot be resolved, the method of large-eddy simulations (LES) is employed wherein the smaller scales are filtered via integration over a window size commensurate with the computational mesh size that allows affordable computations. Consequently, the essential, rate-controlling, physical and chemical processes that occur on shorter scales than the filter size must be modelled.  Those sub-grid models must be properly coupled to the resolved LES flow field.

Current flamelet models that are used for LES or Reynolds-averaged Navier-Stokes (RANS) methods have some advantages. Typically, the flamelet equations are a  system of ordinary differential equations (ODEs) that can be solved offline with solutions available in tabular form or through neural networks (NN). The flamelet models can handle multi-species, multi-step oxidation kinetics without requiring small time steps during the solution of the resolved-scale fluid dynamics. Thus, for several reasons, savings of computational resources can be huge compared to direct numerical simulation. We aim here to retain these very attractive features while removing some less desirable features. Already, some progress has been made in extending the fundamental flamelet theory beyond its long-term limitation of a single-flame structure, two-dimensional (or axisymmetric) configuration, and use of the uniform-density assumption. However, those advances still must be applied to LES or RANS. In addition, the flamelet theory must be advanced to consider shear strain and vorticity at the small scale of the flamelet; these are the vital forgotten physics in current flamelet modelling. Furthermore, the strain rates in the flamelet model are far from properly connected to the strain rates at the resolved scale. Attempts at corrections of these weaknesses are made here.

The goals in this paper are to improve the flamelet model by including  several important physical effects that are commonly neglected in present models and to identify other issues, related to the coupling between the sub-grid-scale physics and the resolved-scale (or time-averaged) physics, that require further study.

\subsection{Existing Flamelet Theory}

 There is need to understand the laminar mixing and combustion that commonly occur within the smallest turbulent eddies. These laminar flamelet sub-domains experience significant strain of all types, shear, tensile, and compressive. Some important works exist here but typically for either counterflows with only normal strain or simple vortex structures in two-dimensions or axisymmetry and often with a constant-density approximation. See \cite{Linan}, \cite{Marble}, \cite{Karagozian}), \cite{Cetegen1}, \cite{Cetegen2}, \cite{Peters}, and \cite{Pierce}. Linan and Peters focused on the counterflow configuration.  \cite{Williams1975} first established the concept of laminar flamelets in the turbulent diffusion flame structure.   Karagozian and Marble examined a three-dimensional flow with radial inward velocity, axial jetting, and a vortex centered on the axis. The flame sheet wrapped around the axis due to the vorticity. An interesting review of the early flamelet theory is given by \cite{Williams2000}. Generally, flamelet studies have focused on either premixed or nonpremixed flames; a unifying approach to premixed, nonpremixed, and multi-branched flames has not been developed. A unifying approach is taken here.

 Most flamelet studies have not directly considered vorticity interaction with the flamelet. See,  for example, \cite{Linan, Peters, Williams2000, Pierce}. \cite{Williams1975} first recognized the advantage of separating rotation (due to vorticity) and stretching by transformation to a rotating, non-Newtonian reference frame. He did not however examine the momentum consequences in the new reference frame which will be examined later here.  The other works that have examined vortex-flame interaction have not separated the effects of stretching and rotation. See \cite{Marble, Karagozian, Cetegen1, Cetegen2, Meneveau-Poinsot}.

 The two-dimensional planar or axisymmetric counterflow configuration has become a foundation for flamelet model. Local conversion to a coordinate system based on the principal strain-rate  directions can provide the counterflow configuration in a general flow.  Furthermore, the quasi-steady counterflow can be analyzed by ordinary differential equations because the dependence on the transverse coordinate is either constant or linear, depending on the variable.   Pierce and Moin modified the nonpremixed-flamelet counterflow configuration by fixing domain size and forcing flux to zero at the boundaries. Flamelet theory as a closure model for turbulent combustion is typically based on the tracking of two variables: a normalized conserved scalar and the strain rate; the latter is generally given indirectly through a progress variable. Mixture fraction is traditionally used for the conserved scalar.

The flamelet model has become a popular sub-grid model for gaseous combustors. Some development is also underway for the use of flamelets in spray combustion but here we will focus on the former type. The flamelet model for LES developed by \cite{Pierce} was a substantial advancement through the introduction of the flamelet progress variable (FPV). Their approach has also been used by \cite{Ihme2009}, \cite{Tuan1, Tuan2, Tuan3}, and others. Other works are based on the use of the original form developed by \cite{Peters}. Pierce and Moin extended that work in two ways. Firstly, the inclusion of both the upper and middle branches of the curve of flame temperature vs. scalar dissipation rate allowed better representation of the unsteady details in turbulent combustion such as extinction and re-ignition.  Secondly, the creation of the progress variable as a function (of the scalar dissipation rate) which is governed by a PDE (added to the LES equations) with a chemical-rate source term determined through the flamelet model. \cite{Tuan2} found that the inclusion of both branches for flame stability resulted in better agreement with experiment. Note that \cite{Tuan2} addressed rocket combustion instability which places a greater demand on the flamelet model than most other applications.  In addition to the velocity fluctuations due to turbulence, it becomes necessary to address very large fluctuations in both velocity, pressure, and temperature due to the nonlinear acoustics; thus, a large demand for data storage is created so that the flamelet model is able to cover the needed range of input variables coming from the LES.        \cite{Ihme2009} introduced the use of neural networks in place of the look-up table.          Recently, \cite{Shadram2021a, Shadram2021b} replaced the look-up table approach for flamelet models in rocket combustion instability studies with a neural-network approach; this reduces substantially the demand for data storage capacity. \cite{Mueller2020} presented the flamelet model in a somewhat different mathematical framework but without the addition of new physical description.

There are concerns with the existing model. While there is clear utility for the FPV approach, there are concurrently clear signs of incompleteness in the flamelet model and contradiction between the tenets of the model and the LES results produced with the model. A few examples of problems that require resolution are provided here. Firstly, the design of the flamelet model uses a counterflow configuration where only normal strain is imposed on the flame region by the ambient flow. Yet, the LES results show that the flame is embedded in a flow field with substantial shear strain rate, i.e., a flow with vorticity. Secondly, the current models assume that the inflowing streams in the counterflow are irrotational; yet, we know that the smaller scales in turbulent combustion are highly rotational. Thirdly, the classical flamelet model is two-dimensional (or axisymmetric) whereas it has been used in flows that are clearly three-dimensional. The model considers only a single-branched diffusion flame while its coupled use in LES commonly predicts multi-branched flames in qualitative agreement with experimental evidence. A fourth issue involves the quasi-steady assumption for the flamelet model that is used in a highly unsteady LES flow with a broadband noise.

These models are built around the postulate that the flamelets are always nonpremixed (i.e., diffusion) flames and subject to flow strain. However, evidence of both nonpremixed and premixed flames has been found in LES and experimental results. In fact, they can co-exist in a multi-branched structure. \cite{Tuan1} and \cite{Tuan2, Tuan3} employed the \cite{Pierce}  flamelet approach in the simulation of a single-injector rocket engine. They showed the importance of flamelets subject to high strain rates. However, contradictions occurred in that both premixed flames and nonpremixed flames appeared in the predictions.  In fact, they report multi-branched flames; in particular, the combination is often seen of a fuel-lean premixed-flame branch with a branch consisting of a merged diffusion flame and fuel-rich premixed flame. Note that the mixture fraction has been used widely as an independent variable to display non-premixed flamelet scalar variations; this cannot be useful for premixed flames. \cite{Sirignano2019b} has shown that any conserved scalar can serve well as an independent variable to present scalar results for nonpremixed and multi-branched flamelets.

Experiments and asymptotic analysis by \cite{Seshadri} showed that a partially premixed fuel-lean flame and a diffusion flame can co-exist in a counterflow with opposing streams of heptane vapor and methane-oxygen-nitrogen mixture.   Thus, a need exists for flamelet theory to address both premixed and non-premixed flames. Recently, \cite{Rajamanickam} provided an interesting three-dimensional triple-flame analysis.

The classical counterflow treatment by  \cite{Linan, Peters} has two opposing streams, fuel or fuel plus a chemically inert gas and oxidizer or oxidizer plus an inert gas. They considered uniform density. That critical assumption was relaxed by
\cite{Sirignano2019a}   for reacting flows and heated flows.
\cite{Sirignano2019b, Sirignano2020}   with one-step kinetics and  \cite{Lopez2019, Lopez2020} with detailed kinetics address that single diffusion-flame case. In addition, situations are addressed where the inflowing streams from $y_{\infty}$ and $y_{-\infty}$  may consist of a combustible mixture of fuel and oxidizer, thereby allowing another flame or two besides the simple diffusion flame to co-exist.    \cite{Sirignano2019b} provides  a counterflow analysis with three-dimensional strain and shows the possibility for a variety of flame configurations to exist depending on the compositions of the inflowing streams: (i) three flames including fuel-lean partially premixed, nonpremixed (i.e., diffusion-controlled), and fuel-rich partially premixed; (ii) nonpremixed and fuel-rich partially premixed; (iii) fuel-lean partially premixed and nonpremixed; (iv) nonpremixed; and (v) premixed. \cite{Lopez2019, Lopez2020} extended the counterflow analysis to consider detailed kinetics for methane-oxygen detailed chemical kinetics and confirmed that combinations of premixed and non-premixed flames could exist in a multi-flame counterflow.  \cite{Sirignano2019a,Sirignano2019b,  Sirignano2020}  has shown that any conserved scalar can serve the purpose and can  replace the mixture fraction as a convenient independent variable.

\subsection{Relative Orientations of Principal Strain Axes, Vorticity, and Scalar Gradients}\label{orientation}

Both normal strain rate and shear strain rate are important. There is a strong need to study mixing and combustion in three-dimensional flows with both imposed normal strain and shear strain and therein imposed vorticity with global circulation.   Shear strain can, in general, be decomposed into a normal strain and a rotation (whose rate is half of the vorticity magnitude).  For example, a rectangular shape that is changed  by shear strain can be viewed as a combination of  deformation to a parallelogram caused by normal strain perpendicular to the diagonal and rotation of the diagonal caused by vorticity. The behavior due to  the strain and rotation becomes especially important on the smallest scales of turbulence where mixing and chemical reaction occur. The magnitudes of strain rate and vorticity will increase as the eddy size (or wavelength) decreases in the turbulence energy cascade process.   The Kolmogorov scale  size is determined by the dissipation rate of turbulence kinetic energy and dynamic viscosity and is the smallest turbulence length scale. The final molecular mixing and chemical reaction in the combustion process occur on a still smaller scale, where  there will be an axis (or direction) of principal compressive normal strain and an orthogonal axis for principal tensile strain, the third orthogonal axis could be either tensile or compressive.  These axes would rotate under shear strain (or equivalently vorticity). Similarly, the direction of the scalar gradient rotates under shear.  A useful flamelet model must have a statistically accurate representation of the relative orientations on this smallest scale of the vorticity vector, scalar gradients, and the directions of the three principal axes for strain rate.  Several studies exist that are helpful in understanding this important alignment issue.

Generally and always for incompressible flow, one principal strain rate $\gamma$ locally will be compressive (corresponding to inflow in a counterflow configuration), another principal strain rate $\alpha$ will be tensile (also named extensional and corresponding to outflow), and the third can be either extensional or compressive and will have an intermediate  strain rate $\beta$ of lower magnitude than the other like strain rate. Specifically, $\alpha > \beta > \gamma , \;\alpha > 0 , \; \gamma <0,$ and, for incompressible flow, $\alpha + \beta + \gamma = 0$.  If the intermediate strain rate $\beta < 0$, there is inflow from two directions with outflow in one direction; a contracting jet flow occurs locally. Conversely, with $\beta > 0$, there is outflow in two directions and inflow in one direction; a counterflow  or, in other words, the head-on collision of two opposed jets occurs. \cite{Betchov}   has shown that, for homogeneous, isotropic turbulence in an incompressible flow, the situation with $\beta > 0$ and resulting counterflow is the most important for production of vorticity and the turbulence energy cascade to smaller scales.

Several interesting findings result from direct numerical simulations (DNS) for incompressible flows. Both \cite{Ashurstetal} and \cite{Nomura1992} compared a case of homogeneous sheared turbulence with a case of isotropic turbulence. They report that the vorticity alignment with the intermediate strain direction is most probable in both cases but especially in the case with shear. Furthermore, the intermediate strain rate is most likely to be extensive (positive) implying a counterflow configuration. \cite{Dresselhaus} uses a kinematic approach to study the stretching of material and vorticity in a fluid flow and predict the tendency towards alignment of the intermediate strain direction with the vorticity. If the vorticity had strong alignment with the major compressive or major tensile strain direction, the magnitude of helicity, the dot product of  velocity $\vec{u}$ and vorticity $\vec{\omega}$, would be large. \cite{Kerr} reports that large values of helicity are not found in the turbulence cascade process. \cite{Ashurstetal} note further that the positive intermediate strain rate has a significantly smaller magnitude than either of the other two principal strain rates. Furthermore, the time for alignment of the vorticity with that intermediate direction is short compared to the eddy-turnover time.

\cite{Nomura1993} studied reacting flow and show that in regions of exothermic reaction and variable density, alignment of the vorticity with the most tensile strain direction can occur. Still though as the strain rates increase, the intermediate direction becomes more favored for alignment with vorticity; that direction is also preferred in regions where mixing occurs without substantial divergence of the velocity due to chemical  reaction.

We may also expect that a material interface most probably aligns to be normal to the direction of the compressive normal strain. That is, the scalar gradient and the direction of compressive strain are aligned. See \cite{Ashurstetal, Nomura1992, Nomura1993} and \cite{Boratav1996, Boratav1998}. Authors agree that the most common intermittent vortex structures in regions of high strain rate are sheets or ribbons rather than tubes.

An important issue for flamelet modelling is the relative magnitudes of the vorticity and the rates of principal normal strain. For homogeneous, incompressible turbulence, \cite{Betchov} showed that, for the average across all length scales, these quantities are of the same order of magnitude. Of course, the smallest scales contribute more to the average since the velocity derivatives are larger on those scales. Also, for shear flows, we expect that the turbulence at the smaller scales will be isotropic  and behave more like the homogeneous flow. In our analysis for variable-density, reacting  shear flows, we will assume the same order-of-magnitude similarity between vorticity and  the rates of principal normal strain  applies for the smallest scales.

Based on those understandings concerning vector orientations, Sirignano (2020), extended flamelet theory in a second significant aspect beyond the inclusion of both premixed and non-premixed flame structures; namely, a model was created of a three-dimensional field with both shear and normal strains.  The three-dimensional problem is reduced to a two-dimensional form and then, for the counterflow or mixing-layer flow, to a one-dimensional similar form.  The system of ordinary differential equations (ODEs) is presented for the thermo-chemical variables and the velocity components. Conserved scalars are determined and can become the independent variable if they behave in a monotonic fashion.  \cite{Sirignano2020} also was able to use a velocity component as the independent variable for the flamelet model with shear strain and vorticity. The validity of the similar solution form for mixing layers with certain thin reaction zones was discussed using concepts from singular perturbation theory. The chemical kinetic model appears as a source term for an ODE. These new findings are very helpful in improving the foundations for flamelet theory and its use in sub-grid modeling for turbulent combustion.

Still, however, there has not been a flamelet model that connects well the scales on the resolved LES level to the sub-grid scales at the level of the flame structure.

Based on the observations of the needed improvements, the aim here is to develop a flamelet model that
(i) determines rather than prescribes the existence of non-premixed flames, premixed flames, or multi-branched flame structures; (ii) determines directly the the effect of shear strain and vorticity on the  flames; (iii) applies directly the resolved-scale strain rates and vorticity to the sub-grid level without the use of a contrived progress variable; (iv) employs a three-dimensional flamelet model; (v) considers the effect of variable density.  Furthermore, some discussion and primitive analysis of the in situ function of the flamelet model in a turbulent shear flow will be presented. The analysis will use one-step kinetics to avoid complications in this initial study; however, a clear template will exist for the employment of multi-step kinetics. The choice for the in situ study will be a mixing layer with the use of mixing-length theory. Here, the goal is not to advance the portion of the analysis of the resolved scale; rather, the method for coupling the sub-grid closure model will be made clear.

In Section \ref{scaling}, the  scaling and connections between the resolved shear flow and the flamelet behavior on the sub-grid scale  discussed.   Section \ref{flamelet} has the description of  a new sub-grid flame model that better handles connection with strain and vorticity on the resolved scale, three-dimensional character, and multibranched flame structure. The application of the sub-grid flamelet with a resolved shear flow is addressed in Section \ref{mixingshearflow}. Concluding comments are made in Section \ref{conclusions}.

\section{Scaling Between Flamelet Scale and Resolved Scale}\label{scaling}

Current flamelet theory for use in LES or RANS makes no substantial attempt to scale properly between the small scale of the flamelet and the larger flow  scales which pertain to the computational fluid dynamics analysis. A theoretical basis is needed to prescribe how to determine vorticity, strain rates, and scalar gradients on the flamelet scale given those properties on the resolved scale. Certainly, as mentioned in Subsection \ref{orientation} of the Introduction, a body of helpful literature exists on this subject. That shall be considered here. Specifically, we shall avoid the creation of arbitrary variables such as the flame progress variable (FPV) used in many publications. The FPV is construed as a measure of progress. However, in order to relate increasing temperature during  combustion  to the flamelet theory, scalar dissipation rate is obliged to increase as the FPV increases. Furthermore, scalar dissipation rate on the flamelet scale should be related to scalar and velocity gradients and not to temperature magnitude. The existing theory will not allow a hot gas to experience a reduction in the values of velocity and scalar gradients which surely is not consistent with general flow patterns in turbulent combustors.
Here, a new method will be developed for determining burning rate  from the flamelet theory and applying it to the resolved scale for LES or to the averaged flow field for RANS. During the burning process, temperature will be able to increase even if strain rate and the associated scalar dissipation rate might be decreasing.

\subsection{Scaling of Velocity, Strain Rate, and Vorticity}

Here, at first, we use approximate concepts which perhaps are reasonably well suited for a use of mixing-length theory to describe the time-averaged turbulent flow in a shear layer. The aim is to provide a simple framework for the first application and test of a new flamelet theory. More sophisticated and modern statistical approaches can be found, e.g., \cite{Pope2000}, and can be used in the future for examination of turbulent flows using RANS or LES.

The  shear-driven flow on the larger scale can be characterized by a length $\delta$ and a  time-averaged velocity difference $\Delta U$ across that particular length that ultimately relate to the magnitudes of the largest eddy size and the turbulence kinetic energy. Then, the velocity, length, and time scales for the resolved scales are $\Delta U, \delta,$ and $\delta/\Delta U$, respectively.  The rms velocity fluctuation $u'$, the  turbulence kinetic energy $k$, and the rate of dissipation of turbulence kinetic energy $\epsilon$ will have magnitudes of the order of $\Delta U, \Delta U^2,$ and $\Delta U^3/\delta$, respectively.  The Kolmogorov scale is the smallest scale in the turbulence energy cascade where the inertia and viscosity effects balance each other (\cite{Pope2000, White}). On that smallest scale, the characteristic velocity, length, and time scales  become $u_{\kappa} = (\nu \epsilon)^{1/4} ,     \kappa = (\nu^3/\epsilon)^{1/4} $, and $t_{\kappa} = (\nu / \epsilon)^{1/2},$ respectively, where $\nu$ is the kinematic viscosity of the fluid.
One can use these Kolmogorov scales to be the scales for flamelet analysis; a more refined approach might be to use the rate of dissipation for scalar quantities $\epsilon_c$ discussed by \cite{ElghobashiLaunder} instead of $\epsilon$.

Estimate a Reynolds number for the resolved flow using $Re \equiv \Delta U \delta/\nu$. Vorticity, rate of normal strain, and rate of shear strain will be $O(\Delta U/\delta)$.   If we estimate the magnitudes of rate of strain and vorticity on the flamelet scale to be given as
\begin{eqnarray}
S^* \equiv  \frac{\partial u_{\kappa}}{\partial \kappa} = O\bigg(\frac{u_{\kappa}}{\kappa}   \bigg)\;\; ; \;\; \frac{u_{\kappa}}{\kappa}
=\bigg(\frac{\epsilon}{\nu}\bigg)^{1/2} = \frac{\Delta U}{\delta} Re^{1/2}
\label{strain-epsilon}
\end{eqnarray}
Clearly, for high $Re$ values, we may expect vorticity and rate of strain on the flamelet scale to be orders higher  than found on the resolved scale. This scaling and the connection of the strain rates and vorticity on different scales has not been addressed in prior flamelet modeling.

Note that different estimates of a relevant resolved-scale or averaged-flow Reynolds number will be used at later points in this discussion. The intention, however, is to maintain the same order of magnitude.

If we examine a time-averaged shear flow, the quantity $\Delta U/\delta$ can be replaced by the magnitude of a velocity gradient for the  averaged flow treated through RANS simulations following a mixing-length concept.  For LES, that quantity can be related to a velocity gradient on the smallest resolved scale, following an approach similar to the Smagorinsky model for Reynolds stress. Specifically,
\begin{eqnarray}
\frac{\Delta U}{\delta} \equiv \bigg|\frac{\partial u}{\partial x}\bigg|  \;\;  ;  \;\;
Re \equiv   \bigg|\frac{\partial u}{\partial x}\bigg| \frac{\delta^2}{\nu} \;\;  ;   \;\;\nonumber \\
\bigg|\frac{\partial u_{\kappa}}{\partial \kappa}\bigg| = \bigg|\frac{\partial u}{\partial x}\bigg|Re^{1/2}  =  \bigg|\frac{\partial u}{\partial x}\bigg|^{3/2}\frac{\delta}{\nu^{1/2}} = \frac{S_{rs}^{*3/2} \delta}{\nu^{1/2}}
\label{strain}
\end{eqnarray}
Here, one can make a choice about the interpretation
of the resolved scale strain rate $ S^{*}_{rs} \equiv |\partial u / \partial x|$. The largest component of shear strain rate on the resolved scale or for the time-averaged flow is recommended for use.  $S^{*}_{rs}$ can vary with location in the flow and, for unsteady RANS and LES, can vary with time as well. The quantity $\partial u_{\kappa}/\partial \kappa $ will be imposed as the compressive normal strain rate in the flamelet model; it will be a negative number $-(S^*_1 + S^*_2)$. Thus, Equation (\ref{strain}) can relate the normal strain rate on the flamelet scale to the strain rate on the resolved scale or for the time-averaged flow. In the case with the two-equation RANS model using $k, \epsilon$ theory, Equation (\ref{strain-epsilon}) will yield $S^* = (\epsilon/ \delta^2)^{1/3}$.

A relation must be created between the dimensional vorticity $\omega^*$ on the resolved scale and the dimensional vorticity $\omega^*_{\kappa}$ on the flamelet sub-grid scale.
(Note that, in the development of the flamelet model, the sub-grid $\omega_{\kappa}$ is dimensionless.) A reasonable relationship, mimicking the strain rate relation, is given as $\omega^*_{\kappa} = \omega^{*3/2}\delta/ \nu^{1/2} = O(\omega^* Re^{1/2}) >> \omega^*.$

\subsection{Scaling of the Scalar Properties}

The inflow boundary conditions  for the scalar quantities in the flamelet calculation must be determined from the resolved scale behavior. The scalar gradients will be much larger on the flamelet scale due to the dynamics of the turbulent flow. As a first approximation, consider that the scalar  gradients scale in proportion to the strain rates.
For example, using the mass fraction of species $m$, we state
\begin{eqnarray}
\frac{ \frac{ \partial Y_{m, \kappa}} {\partial \kappa} }{\frac{ \partial Y_m} {\partial x} }=
\frac{ \frac{ \partial u_{\kappa}} {\partial \kappa} }{\frac{ \partial u} {\partial x} }= Re^{1/2}
\label{strainscale}
\end{eqnarray}
where the subscript $\kappa $ designates the flamelet scale.
The domain sizes between the Kolmogorov scale and the resolved scale as $\kappa = \delta Re^{-3/4} $. Setting the change in scalar value across the given domain as the product of domain size and its gradient, the result is $\Delta Y_{m, \kappa} = \Delta Y_m Re^{-1/4}$. So, the variation in scalar properties across the smallest eddy is smaller than the variation across the larger eddies. Although the gradient of the scalar property is much greater on the flamelet scale due to turbulent mixing, the variation across the domain is smaller due to the more greatly reduced domain size. Changes in enthalpy $h$ and density $\rho$ will be determined following the same pattern. The important implication is that, in general, the partial premixing on the smaller scales should be greater than experienced on the largest scales.

If on the resolved scale at a particular point $\vec{x}, t$, the scalar property is $Y_m$, the bounding values for the inflow of the flamelet counterflow will be taken as
$Y_{m, \kappa, \infty} = Y_{m}(\vec{x},t) + \Delta Y_{m, \kappa}$ and
 $Y_{m, \kappa, - \infty} = Y_{m}(\vec{x},t) - \Delta Y_{m, \kappa}$.  Again, the boundary values for other scalars will be handled identically. A consequence here will be that the incoming streams of the flamelet counterflow will more likely be fuel rich or fuel lean than pure fuel or pure oxidizer. Multi-branched-flame structures of the type found by \cite{Seshadri, Rajamanickam, Sirignano2019a, Sirignano2020, Lopez2019, Lopez2020} can be expected.

\subsection{Scaling of Energy Release Rate and Species Consumption and Production Rates}

The resolved scale will require input from the flamelet model for the quantities giving consumption  (or production) rates per unit volume for the chemical species and energy release rate per unit volume due to chemical reaction and perhaps also viscous dissipation rate.  The production and consumption rates within the flame are substantially higher than the average values over the counterflow volume. It is the average over the counterflow volume that should be used in the resolved-scale calculations. The same approach should be used for the viscous dissipation rate in high-speed flows where that is considered to be important.  These quantities to be used on the resolved scale are given as integrals  over the flamelet scale by Equation (\ref{rateintegral}).

The dimensional form of the rates will indicate that the integrated chemical rates are proportional to the magnitude of the compressive strain rate $S^*\equiv S^*_1 + S^*_2$ for the counterflow. The portion of the dimensional energy release rate due to viscous dissipation is proportional to  $S^{*2}$. That sub-grid strain rate will be much larger than the strain rate on the resolved scale as indicated by Equation (\ref{strainscale}). The scalar gradients have similar scaling relation between the sub-grid and the resolved scale. They impact diffusion  and therefore, in diffusion-controlled combustion, they determine burning rates.

\section{Sub-grid Flamelet Analysis}\label{flamelet}

We must formulate the problem in a quasi-steady three-dimensional form. The following alignments will be assumed. The direction of major compressive principal strain is orthogonal to the vorticity vector direction.  Specifically, the intermediate principal strain direction is aligned with the vorticity while the scalar gradient aligns with the principal compressive strain direction. These assumptions are consistent with the statistical findings of \cite{Nomura1993}.

\subsection{Coordinate Transformation} \label{transformation}

\begin{figure}
  \centering
  \includegraphics[height = 7.0cm, width=0.8\linewidth]{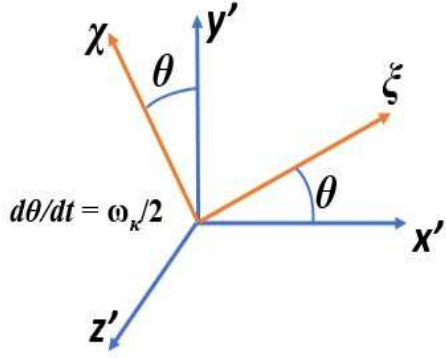}
  \caption{Transformation to $\xi, \chi, z'$ rotating coordinate system from $x', y', z'$ Newtonian system. $\theta$ increases in the counterclockwise direction.}
  \label{Coordinate}
\end{figure}

 A transformation displayed in Figure \ref{Coordinate} will be made from the Newtonian frame with rotating material (due to vorticity) to a rotating, non-Newtonian frame.  Let the vorticity direction be the $z'$ direction in an orthogonal framework. Any $x', y'$ plane contains the directions of scalar gradients, major principal axis for compressive strain, and major principal axis for tensile strain.  Note that the $x',y', z'$ directions are not correlated with coordinates on the resolved scale. $\omega_{\kappa}$ is the vorticity magnitude on this sub-grid (Kolmogorov) scale. $x', y', z'$ will be transformed to $\xi, \chi, z'$ wherein the material rotation is removed from the $\xi, \chi$ plane by having it rotate at angular velocity $d\theta/dt = \omega_{\kappa}/2$ relative to $x', y'$. Here, $\theta$ is the angle between the $x' $ and $\xi$ axes and simultaneously the angle between the $y'$ and $\chi$ axes. Clearly, we are taking the sub-grid domain sufficiently small to consider a uniform value of $\omega$ across it.

 The following relations apply.
\begin{eqnarray}
\xi &=& x' cos \theta + y' sin\theta \;\; ; \;\;
\chi = y' cos \theta - x' sin\theta \nonumber \\
\frac{\partial \xi}{\partial x'} &=&  cos \theta \;\; ; \;\; \frac{\partial \xi}{\partial y'}= sin\theta \;\; ; \;\;
\frac{\partial \chi}{\partial x'} = - sin \theta \;\; ; \;\; \frac{\partial \chi}{\partial y'} = cos \theta \nonumber \\
u_{\xi} &=& u \;cos \theta + v \; sin \theta  + \chi\frac{\omega_{\kappa}}{2}   \;\; ; \;\;
u_{\chi} = v\; cos \theta -u \;sin \theta  - \xi\frac{\omega_{\kappa}}{2}  \nonumber \\
\frac{\partial u}{\partial x'} &=& \frac{\partial u}{\partial \xi}cos \theta
- \frac{\partial u}{\partial \chi}sin \theta  \;\; ; \;\;
\frac{\partial u}{\partial y'} = \frac{\partial u}{\partial \xi}sin \theta
+ \frac{\partial u}{\partial \chi}cos \theta   \nonumber \\
\frac{\partial v}{\partial x'} &=& \frac{\partial v}{\partial \xi}cos \theta
- \frac{\partial v}{\partial \chi}sin \theta  \;\; ; \;\;
\frac{\partial v}{\partial y'} = \frac{\partial v}{\partial \xi}sin \theta
+ \frac{\partial v}{\partial \chi}cos \theta
\end{eqnarray}
Since
\begin{eqnarray}
\frac{\partial v}{\partial x'}  - \frac{\partial u}{\partial y'} = \omega_{\kappa}
\end{eqnarray}
it follows that
\begin{eqnarray}
\frac{\partial u_{\chi}}{\partial \xi}  - \frac{\partial u_{\xi}}{\partial \chi} = 0
\end{eqnarray}
Namely, the flow in the rotating frame of reference does not have the vorticity imposed on it.  However,  two points must be understood. Firstly, the frame is not Newtonian and a reversed (centrifugal) force is imposed. Secondly, the expansions due to combustion and energy release  can produce new vorticity but it will integrate to zero globally. That is, it will possess an antisymmetry.

\subsection{Governing Equations} \label{equations}

The governing equations for unsteady 3D flow in the non-Newtonian frame can be written with $u_i = u_{\xi}, u_{\chi}, w \; ; \; x_i = \xi, \chi, z$. The centrifugal acceleration $a_i = \xi \omega_{\kappa}^2/4 , \chi \omega_{\kappa}^2 /4, 0$.  The quantities $p, \rho, h, h_m, Y_m, \dot{\omega},  \mu, \lambda, D,$ and $ c_p  $ are pressure, density, specific enthalpy, heat of formation of species $m$, mass fraction of species $m$, chemical reaction rate of species $m$, dynamic viscosity, thermal conductivity, mass diffusivity, and specific heat, respectively. Furthermore, $\tau_{ij} $ is the viscous stress tensor and the Lewis number
$Le \equiv \lambda/(\rho D c_p )$.
\begin{eqnarray}
\frac{\partial \rho}{\partial t} + \frac{\partial (\rho u_j)}{\partial x_j} =0
\label{cont}
\end{eqnarray}
\begin{eqnarray}
\rho \frac{\partial u_i}{\partial t} + \rho u_j\frac{\partial u_i}{\partial x_j} +\frac{\partial p}{\partial x_i} = \frac{\partial \tau_{ij}}{\partial x_j} + \rho a_i
			\label{momentum}
		\end{eqnarray}
		where, following the Stokes hypothesis for a Newtonian fluid,
		\begin{eqnarray}
			\tau_{ij} = \mu \Big[ \frac{\partial u_i}{\partial x_j} + \frac{\partial u_j}{\partial x_i}
			-\frac{2}{3}\delta_{ij} \frac{\partial u_k}{\partial x_k}\Big]
			\label{tau}
		\end{eqnarray}
		\begin{eqnarray}
			\rho \frac{\partial h}{\partial t} + \rho u_j\frac{\partial h}{\partial x_j} - \frac{\partial p}{\partial t} - u_j\frac{\partial p}{\partial x_j}
			=\frac{\partial}{\partial x_j} \Big( \frac{\lambda}{c_p} \frac{\partial h}{\partial x_j}  \Big)
			+ \frac{\partial}{\partial x_j} \Big( \rho D (1 - Le)  \Sigma^N_{m=1}h_m \frac{\partial Y_m}{\partial x_j}  \Big) \nonumber \\
			-\rho \Sigma^N_{m=1}h_{f,m} \dot{\omega}_m
			+ \tau_{ij}\frac{\partial u_i}{\partial x_j}
			\label{energy}
		\end{eqnarray}
		\begin{eqnarray}
			\rho \frac{\partial Y_m}{\partial t} + \rho u_j\frac{\partial Y_m}{\partial x_j}  = \frac{\partial }{\partial x_j}\Big(\rho D \frac{\partial Y_m}{\partial x_j}\Big) + \rho \dot{\omega}_m   \;\;;\;\; m=1, 2, ...., N
			\label{species}
		\end{eqnarray}
		
		An alternative form of the energy equation can be developed to govern the total $H$ of the specific enthalpy, specific chemical energy, and kinetic energy per unit mass. That is, $H \equiv h + \Sigma_{m=1}^NY_m h_{f,m} +u_k u_k/2$. Specifically, the vector dot product of $u_i$ with Equation (\ref{momentum}) is used to substitute for $u_j \partial p/ \partial x_j $ in Equation (\ref{energy}) and Equation (\ref{species}) is used to substitute for $\dot{\omega}_m$ there. The Lewis number $Le =1$ is considered. It follows that
		\begin{eqnarray}
			\rho \frac{\partial H}{\partial t} + \rho u_j\frac{\partial H}{\partial x_j} - \frac{\partial p}{\partial t}
			=\frac{\partial }{\partial x_j} \Big( \frac{\lambda}{c_p} \frac{\partial (h+ \Sigma_{m=1}^NY_m h_{f,m}) }{\partial x_j} \Big)
			+\frac{\partial (u_i \tau_{ij})}{\partial x_j}+\rho u_j a_j
			\label{energy2}
		\end{eqnarray}
		
The energy source term $\rho u_j a_j = \rho (\omega_{\kappa}/2)^2(\xi u_{\xi} + \chi u_{\chi})$. If we neglect terms of the  order of the kinetic energy per mass, this effect disappears. We might still retain the viscous dissipation rate $\tau_{ij} \partial u_i/\partial x_j$ for cases where large strain rates are expected on the sub-grid scale, as suggested by \cite{Drozdachapter}. The viscous dissipation rate will have exactly the same value whether it is calculated in the Newtonian frame or the rotating frame; this result is expected because it relates to the thermodynamics where the laws are independent of the reference frame. The resulting equation becomes
\begin{eqnarray}
			\rho \frac{\partial h}{\partial t} + \rho u_j\frac{\partial h}{\partial x_j} - \frac{\partial p}{\partial t}
			\approx \frac{\partial}{\partial x_j} \Big( \frac{\lambda}{c_p} \frac{\partial h}{\partial x_j}  \Big)
			+ \frac{\partial}{\partial x_j} \Big( \rho D (1 - Le)  \Sigma^N_{m=1}h_m \frac{\partial Y_m}{\partial x_j}  \Big) \nonumber \\
			-\rho \Sigma^N_{m=1}h_{f,m} \dot{\omega}_m
			+ \tau_{ij}\frac{\partial u_i}{\partial x_j}
			\label{energy3}
		\end{eqnarray}
Here, we define the non-dimensional  Prandtl, Schmidt, and Lewis numbers:
$ Pr \equiv c_p \mu / \lambda$ ; $Sc \equiv \mu / (\rho D)$  ; and $ Le \equiv Sc/ Pr$.

		The non-dimensional forms of the above equations remain identical to the above forms if we choose certain reference values for normalization. In the remainder of this article, the non-dimensional forms of the above equations will be considered. The superscript $^*$ will be used to designate a dimensional property. The variables $u_i^*, t^*, x_i^*, \rho^*, h^*, p^*,$ and $\dot{\omega}_m^* ,$ and properties $\mu^*, \lambda^*/ c_p^*,$ and $ D^*$ are normalized respectively by
		$ [(S_1^* +S_2^*) \mu_{\infty}^*/ \rho_{\infty}^*]^{1/2}, (S_1^* + S_2^*)^{-1}, [ \mu_{\infty}^*/ (\rho_{\infty}^*(S_1^* +S_2^*))]^{1/2},
		\rho_{\infty}^*, (S_1^* +S_2^*) \mu_{\infty}^*/ \rho_{\infty}^*, (S_1^* + S_2^*)\mu_{\infty}^*, (S_1^* + S_2^*), \mu_{\infty}^*, \mu_{\infty}^*,$ and $\mu_{\infty}^*/ \rho_{\infty}^*$. The dimensional strain rates $S^*_1$ and $S^*_2$ and  vorticity $\omega_{\kappa}^*$ are normalized by $S^*_1 + S^*_2$.  It is understood that, for unsteady flow, the reference values for strain rates and far-stream variables and properties used for normalization  will be constants; for example, averages might be taken for fluctuating conditions. Note that the reference length $[ \mu_{\infty}^*/ (\rho_{\infty}^*(S_1^* +S_2^*))]^{1/2}$ is the estimate for the magnitude of the viscous-layer thickness.  In the following flamelet analysis, the vorticity $\omega_{\kappa}$ and the velocity derivatives $\partial u_i/ \partial x_j$ are non-dimensional quantities; their dimensional values can be obtained through multiplication by $S^*_1 + S^*_2$. In Section  \ref{scaling}, the algorithms are given  that relate dimensional vorticity and velocity derivatives on the resolved scale to dimensional vorticity and velocity derivatives on the sub-grid scale.
		
\subsection{Similar Form for the Velocity and Pressure} \label{similar}

		The stagnation point in either the steady counterflow or in the steady flow against a wall will be taken as the origin $\xi = \chi =z=0$. Along the line $\xi=z=0$ normal to the interface or wall, we can expect the first derivatives of $u_{\chi}, \rho, h, T,$ and $Y_m$ with respect to either $\xi$ or $z$ to be zero-valued. For unsteady cases, only symmetric situations will be considered so that the stagnation point remains at the origin and the wall or interface remains at $\chi=0$.  The velocity components $u_{\xi}$ and $w$ will be odd functions of $\xi$ and $z$, respectively, going through zero and changing sign at that line. Consequently, upon neglect of terms of $O(\xi^2)$ and $O(z^2)$, the variables $u_{\chi}, \rho, h, T,$ and $Y_m$ can be considered to be functions only of $t$ and $\chi$. For steady flow, the density-weighted \cite{Illingworth} transformation of $\chi$ can be used to replace $\chi$ with $\eta \equiv \int^y_0 \rho (\chi') d\chi'$. Neglect of the same order of terms implies that $u_{\xi}= S_1 \xi(df_1/d\eta)$ and $w= S_2 z(df_2/d\eta)$. Note $u_{\xi}$ is independent of $z$ and $w$ is independent of $\xi$ in this case where no shear strain is imposed on the incoming stream(s). At the edge of the viscous layer at large positive $\eta$,
		$df_1/d \eta \rightarrow 1, df_2/d \eta \rightarrow 1, f_1 \rightarrow \eta$, and $f_2 \rightarrow \eta$  .
		Ordinary differential equations are created here through the variable $\eta$ and the convenient definition is made that  $( )' \equiv d( )/d\eta$.
Note that other transformations of the $\chi$ coordinate can be made, e.g., weighting by transport properties \cite {Linanetal, Weissetal} rather than density.

		In the non-dimensional form given by Equations (\ref{cont}) through (\ref{energy3}), the dimensional strain rates $S_1^*$ and $S_2^*$ are each normalized by the dimensional sum $S_1^* + S_2^*$. Thus, the non-dimensional relation is $S_2 = 1 - S_1$ and only one independent non-dimensional strain-rate parameter is needed. Nevertheless, two strain rates are presented above and in the following analysis with the understanding that one depends on the other such that $S_1 + S_2 = 1$.  $S_1 + S_2$ will be explicitly stated in our analysis without substitution of the unity value.   This choice clarifies whether a particular  term when converted to a dimensional form depends on $S_1^*, S_2^*$, or the sum of the two strain rates.
		
		For steady state, the continuity equation (\ref{cont}) is readily integrated to give
		\begin{eqnarray}
			\rho u_{\chi} = -S_1 f_1(\eta) - S_2 f_2(\eta)
			\label{cont2}
		\end{eqnarray}
		and then
		\begin{eqnarray}
			u'_{\chi} =  \frac{S_1 f_1(\eta) + S_2 f_2(\eta)}{\rho^2 }
			\rho' -\frac{S_1 f_1'(\eta)+ S_2 f_2'(\eta)}{\rho}
			\label{dv}
		\end{eqnarray}
		Thus, the incoming inviscid flow outside the boundary layer is described by $u_{\chi}=-(S_1 + S_2)\eta$ for positive $\eta$ and $u_{\chi}=-(S_1 + S_2)\eta/ \rho_{-\infty}$ for negative $\eta$. Note that the same result is found for the unsteady or steady incompressible state where there is no need to use $\eta$ in place of $\chi$ since $\rho =1$ everywhere. Then, $u_{\chi} = -(S_1  + S_2 ) \chi$ for the external incoming flow.
		
		Equations (\ref{cont2}) and (\ref{dv}) may be substituted into Equation (\ref{momentum}) to determine the pressure gradient.
		\begin{eqnarray}
			\frac{\partial p}{\partial \xi} &=&  \rho [ \rho \mu S_1 f_1''' + S_1 f_1'' (\rho \mu)' +(S_1 f_1 + S_2f_2)S_1 f_1''  +  \bigg(\frac{\omega_{\kappa}}{2} \bigg)^2 - (S_1f_1')^2]\xi \nonumber  \\
			\frac{\partial p}{\partial \eta} &=& \frac{4}{3} \frac{d}{d \eta}( \rho \mu\frac{d u_{\chi}}{d \eta})
			-\frac{2}{3}(S_1f_1' + S_2 f_2')\frac{d\mu}{d\eta} +\frac{ \mu}{3}(S_1 f_1'' + S_2 f_2'') +  \bigg(\frac{\omega_{\kappa}}{2} \bigg)^2\chi
			+(S_1 f_1 + S_2 f_2)\frac{\partial u_{\chi}}{\partial \eta} \nonumber \\
			\frac{\partial p}{\partial z} &=&      \rho [ \rho \mu S_2 f_2''' + S_2 f_2'' (\rho \mu)' +(S_1 f_1 + S_2f_2)S_2 f_2''- (S_2f_2')^2]z  \nonumber \\
			\label{pressure}
		\end{eqnarray}
		It follows from the $\eta$ pressure-gradient in Equation (\ref{pressure}) that $\partial p/ \partial \eta$ is a function only of $\eta$. Therefore, $\partial^2 p/ \partial \xi \partial \eta = 0$ and $\partial^2 p/ \partial z \partial \eta = 0$. Now, the coefficient of $\xi$ on the right side of the $\xi$ pressure-gradient in Equation (\ref{pressure}) must be constant. The same conclusion is made for the coefficient of $z$ on the right side of the $z$ pressure-gradient in Equation (\ref{pressure}).  At $\eta =\infty$, $f_1' =f_2'=1$ and $f_1'' = f_2'' = f_1''' = f_2''' =0$ which allows the two constants to be determined.  Specifically, we obtain
		\begin{eqnarray}
			\rho \mu  f_1''' & + & f_1'' (\rho \mu)'+(S_1 f_1 + S_2f_2) f_1'' + S_1\Big(\frac{1}{\rho} -(f_1')^2\Big) +\frac{\omega_{\kappa}^2}{4S_1}\bigg(1 - \frac{1}{\rho}\bigg)= 0 \nonumber \\
			\rho \mu  f_2'''  &+&  f_2'' (\rho \mu)'+(S_1 f_1 + S_2f_2) f_2'' + S_2\Big(\frac{1}{\rho}-(f_2')^2\Big) = 0
			\label{ODEs}
		\end{eqnarray}
The boundary conditions will use the assumption that two velocity components asymptote to
the constant values
$ u_{\xi}(\infty), u_{\xi}(-\infty), w(\infty),$ and $w(-\infty)$ at large magnitudes of $\eta$.   The stream function bounding the two incoming streams is arbitrarily given a zero value and placed at $\eta =0$.	
\begin{eqnarray}
f'_1(\infty) &=& 1 \;\; ; \;\;
f'_1(-\infty)= \sqrt{\frac{1}{\rho(-\infty)}+  \bigg(\frac{\omega_{\kappa}}{2 S_1}\bigg)^2
\bigg(  1 -   \frac{1}{\rho(-\infty)} \bigg)  }         \;\; ; \;\;
 f_1(0) = 0  \;\; ; \;\;    \nonumber \\
f_2(\infty) &=& 1 \;\; ; \;\; f_2(-\infty)= \frac{1}{\sqrt{\rho(-\infty)}}\;\; ; \;\; f_2(0) =0
\label{BCs}
\end{eqnarray}

In the incompressible case where density is uniform throughout the flow, i.e. $\rho =1$, the solutions become simply that $f_1(\eta)=1$ and $f_2(\eta) =1$ everywhere.	When density varies through the flow because of heating or variation of composition, $u_{\xi}$ and $w$ vary with $\chi$, thereby creating a shear stress and vorticity albeit that the frame  transformation removed vorticity and shear  from the incoming flow. The vorticity will be created in an antisymmetric manner since the two velocity components are odd functions of $\xi$ and $z$, respectively. Thereby, the circulation remains zero for the transformed counterflow.

		For steady flows, $S_1 + S_2 =1$. The dependence of $u_{\chi}$ on $f\equiv S_1f_1 + S_2 f_2$ is shown by Equation (\ref{cont2}). Thus, the function $f$ will be important in determining both the field for $u_{\chi}$ and the scalar fields.  From Equation (\ref{ODEs}), an equation for $f$ can be formed.
		\begin{eqnarray}
			\rho \mu  f'''  +  f'' (\rho \mu)'+f f'' + \frac{1}{\rho} -(f')^2 +\frac{\omega_{\kappa}^2}{4}\bigg(1 - \frac{1}{\rho}\bigg)= 2S_1S_2\Big(\frac{1}{\rho} -f_1'f_2'\Big)
			\label{ODEf}
		\end{eqnarray}
		Consequently,  $f$ as well as $f_1$ and $f_2$ will depend on both $S_1$ and $S_2$, not merely on $S_1 + S_2$. That is, the particular distribution of the normal strain rate between the two transverse direction will matter. $f$ and $f_1$ will also depend directly on $\omega_{\kappa}$ (unless $ S_1 = 0$). $f_2$ will depend on $\omega_{\kappa}$ through its coupling with $f_1$.  In our calculations, we will consider a planar case ($S_1 = 1.0 , S_2 = 0$) where the product $S_1S_2$ is minimized and the vorticity vector is normal to the plane with strain. The case where $S_1 = S_2 = 0.5$ has the maximum value for the product $S_1 S_2$ would be axisymmetric if $\omega = 0$. However, symmetry is lost if the rotation exists.

		In the outer  flow where variability of viscosity and density may be neglected
and $u_{\chi} \rightarrow \eta/ \rho_{\infty}$ as $\eta \rightarrow \infty$, the $\chi$-momentum equation from (\ref{pressure}) becomes
		\begin{eqnarray}
			\frac{\partial }{\partial \eta} \Big[p - \frac{4\rho \mu}{3}\frac{\partial u_{\chi}}{\partial \eta}   \Big]  &\rightarrow&
			(S_1 f_1 + S_2 f_2)\frac{\partial u_{\chi}}{\partial \eta} + \bigg( \frac{\omega_{\kappa}}{2}\bigg)^2 \chi \rightarrow (S_1 + S_2 )\eta \frac{\partial u_{\chi}}{\partial \eta}
		+ \bigg(\frac{\omega_{\kappa}}{2}\bigg)^2 \frac{\eta}{\rho_{\infty}}   \nonumber \\
\frac{\partial p}{\partial \eta} 	&\rightarrow& \bigg[\bigg(\frac{\omega_{\kappa}}{2}\bigg)^2 -(S_1 + S_2)^2 \bigg] \frac{\eta}{\rho_{\infty}}
			\label{pressure2}
		\end{eqnarray}
The negative pressure gradient will be reduced by the centrifugal effect. Essentially, the pressure gradient serves to decelerate the incoming stream in a counterflow; here, it is helped by the imposed acceleration.

The variable-density-and-viscosity case requires some   couplings with Equations (\ref{energy}) and (\ref{species}) and with an equation of state and fluid-property laws which affect $\rho$ and $\mu$.
	
	The pressure derivative can be determined by substituting from Equation (\ref{cont2}) and (\ref{dv}) into (\ref{pressure}).
	\begin{eqnarray}
		\frac{\partial p}{\partial \eta} &=&\frac{4}{3} \frac{d}{d \eta}\Big( \rho \mu\Big[ \frac{S_1 f_1(\eta) + S_2 f_2(\eta)}{\rho^2 }
		\frac{d \rho }{d \eta }-\frac{S_1 f_1'(\eta)+ S_2 f_2'(\eta)}{\rho}
		\Big]\Big)+  \bigg(\frac{\omega_{\kappa}}{2} \bigg)^2\chi
		\nonumber \\
		&& +\frac{ \mu}{3}(S_1 f_1'' + S_2 f_2'')-\frac{2}{3}(S_1f_1' + S_2 f_2')\frac{d\mu}{d\eta}\nonumber  \\
		&& +(S_1 f_1 + S_2 f_2)\Big[\frac{S_1 f_1(\eta) + S_2 f_2(\eta)}{\rho^2 }
		\frac{d \rho }{d \eta }-\frac{S_1 f_1'(\eta)+ S_2 f_2'(\eta)}{\rho}\Big]
		\nonumber \\
		p(x, \eta, z) &=& p_{ref} +  \frac{4\mu(\eta)}{3\rho(\eta)} \Big[ (S_1 f_1(\eta) + S_2 f_2(\eta))
		\frac{d \rho }{d \eta }
		-\rho(\eta)(S_1 f_1'(\eta)+ S_2 f_2'(\eta))\Big] \nonumber \\
		&& + \frac{4 \mu(0)}{3}(S_1f_1'(0) + S_2f_2'(0))
		\nonumber \\
		&&+  \int_{0}^{\eta}\Big[\frac{\mu(\zeta)}{3}(S_1 f_1''(\zeta) + S_2 f_2''(\zeta))
		-\frac{2}{3}(S_1f_1'(\zeta) + S_2 f_2'(\zeta))\frac{d\mu}{d\zeta} \Big] d\zeta   \nonumber  \\
		&& +\int_{0}^{\eta}(S_1 f_1(\zeta) + S_2 f_2(\zeta))\Big[\frac{S_1 f_1(\zeta) + S_2 f_2(\zeta)}{\rho(\zeta)^2 }
		\frac{d \rho }{d \zeta }-\frac{S_1 f_1'(\zeta)+ S_2 f_2'(\zeta)}{\rho(\zeta)}\Big] d\zeta  \nonumber \\
		&&- \frac{S_1^2 x^2}{2} -  \frac{S_2^2 z^2}{2}  +  \bigg(\frac{\omega_{\kappa}}{2} \bigg)^2 \int_0^{\eta} \chi(\zeta) d \zeta
\nonumber \\
		\label{pressure1}
	\end{eqnarray}
	
The viscous dissipation $\Phi$ can be determined as follows:
\begin{eqnarray}
\Phi & \equiv & \tau_{ij}\frac{\partial u_i}{\partial x_j} = \mu \bigg[\frac{4}{3} \bigg(  \big(\frac{\partial u_{\xi}}{\partial \xi}\big)^2  +  \big(\frac{\partial u_{\chi}}{\partial \chi}\big)^2   +  \big(\frac{\partial w}{\partial z}\big)^2
- \frac{\partial u_{\xi}}{\partial \xi}\frac{\partial u_{\chi}}{\partial \chi}
- \frac{\partial w}{\partial z}\frac{\partial u_{\chi}}{\partial \chi}
- \frac{\partial w}{\partial z}\frac{\partial u_{\xi}}{\partial \xi}
\bigg) \nonumber \\
&+&  \big(\frac{\partial u_{\xi}}{\partial \chi}\big)^2  +  \big(\frac{\partial w}{\partial \chi}\big)^2        \bigg]  \nonumber \\
&=& \mu\bigg( 4(S_1f'_1)^2 + 4(S_2f'_2)^2 +  \frac{8}{3}S_1 S_2 f'_1 f'_2 - 4\frac{(S_1f_1 + S_2f_2)(S_1f'_1 + S_2f'_2)}{\rho}\frac{d\rho}{d \eta} \nonumber \\
  &+& \frac{4}{3}\big(\frac{S_1f_1 + S_2f_2}{\rho} \frac{d\rho}{d \eta}  \big)^2        \bigg)
\label{dissipation}
\end{eqnarray}
Here, neglect is made of the quantities $\mu(\partial u_{\xi}/\partial  \chi)^2$ and $\mu(\partial w/\partial  \chi)^2$ because they are of $O(\xi^2)$ and $O(z^2)$, respectively.

	An exact solution of the variable-density Navier-Stokes equation has been obtained subject to determination of $\rho$ and $\mu$ through solutions of the energy and species equations as discussed below. There has been no need for use of a boundary-layer approximation. Thus, the solution here is the natural solution, subject to neglect of terms of $O(\xi^2)$ and $O(z^2)$. Unlike the incompressible counterflow, a viscous layer exists with the  three normal strains and normal viscous stresses varying through the layer due to varying density and viscosity. Shear strain does not appear explicitly in the incoming flow because it has been removed through the coordinate transformation to the rotating frame. Nevertheless,its effect appears through the rotational term.
	
\subsection{Similar Form for the Scalar Fields} \label{similar2}

	Consider  the reacting, steady case keeping viscous dissipation. Assume Prandtl number $Pr$ is constant.
	Substitution from Equations (\ref{energy}) and (\ref{species}) yields the following ordinary differential equations:
	\begin{eqnarray}
	( \rho \mu h' )' + Pr f h'
			+ ( \rho^2 D (Pr - Sc)  \Sigma^N_{m=1}h_m Y'_m )'
			= Pr  \Sigma^N_{m=1}h_{f,m} \dot{\omega}_m
			- Pr \frac{\Phi}{\rho}
\nonumber \\
(\rho^2 D Y'_m )' + f Y'_m = - \dot{\omega}_m   \;\;;\;\; m=1, 2, ...., N
		\label{energy4}
	\end{eqnarray}

If $Le =1$, i.e., $Pr = Sc$, the  new scalar
$\tilde{h} \equiv h + \Sigma_{m=1}^N h_{f,m} Y_m$ is governed by
\begin{eqnarray}
(\rho \mu \tilde{h}')' + Pr f \tilde{h}'= - Pr\frac{\Phi}{\rho}
\label{htilde}
\end{eqnarray}
When, the viscous dissipation is negligible, $\tilde{h}$ is a conserved scalar indicating that the sum   of thermal energy plus chemical energy is in an advective-diffusive balance.

 It remains to use thermodynamic relations to substitute for  $\rho$ and $ \mu  $ in terms of $h $ and $p$.
	
	Now, we will address the special case where $\rho \mu =1$. The perfect gas law and the assumption of constant specific heat $c_p$ will give the relation that $1/ \rho = h$. Then, Equations (\ref{ODEs}),(\ref{BCs}),(\ref{energy4}), and \ref{htilde} yield
\begin{eqnarray}
		f_1'''  +f f_1'' + S_1[ h -(f_1')^2]    &+& \frac{\omega_{\kappa}^2}{4S_1}(1 - h)
=0 \nonumber \\
		f_2'''  +f f_2'' + S_2[h-(f_2')^2] &=& 0 \nonumber \\
 Y''_m  + Pr f Y'_m &=& - Pr\dot{\omega}_m   \;\;;\;\; m=1, 2, ...., N  \nonumber  \\
 h'' + Pr f h' + (Pr - Sc)  \Sigma^N_{m=1}h_m Y''_m
&=& Pr  \Sigma^N_{m=1}h_{f,m} \dot{\omega}_m - Pr \frac{\Phi}{\rho}  \nonumber  \\
		\tilde{h}''  +Pr f \tilde{h}'  &=& -\frac{\Phi}{\rho}
\label{ODEs3}
\end{eqnarray}

The boundary conditions are
\begin{eqnarray}
		f_1'(\infty)&=&  1 \;\; ;\;\; f_1'(-\infty)  = \sqrt{h_{-\infty}+  \bigg(\frac{\omega_{\kappa}}{2 S_1}\bigg)^2
(  1 -   h_{-\infty})  }       \;\;   ; \; \;  	f_1(0)  =0 \;\;  ; \;\;  \nonumber \\	
f_2' (\infty)&=& 1 \;\; ;\;\;  f_2'(-\infty) = \sqrt{h_{-\infty}}  \;\;\; ; f_2(0)  =0 \;\;  ; \;\; \nonumber \\
h(\infty) &=& 1  \;\; ; \; \; h(-\infty)= \frac{1}{\rho_{-\infty}}\;\; ;\;\; \nonumber \\
Y_m (\infty) &=& Y_{m,\infty} \;\; ; \;\; Y_m (-\infty) = Y_{m,-\infty} \;\; ;\;\;\nonumber \\
\tilde{h}(\infty) &=& 1 + \Sigma_{m=1}^N h_{f,m}Y_{m, \infty} \;\;  ; \;\;
\tilde{h}(-\infty) = h_{-\infty}+ \Sigma_{m=1}^N h_{f,m}Y_{m, -\infty}
		\label{BCs3}
	\end{eqnarray}

	Equations (\ref{ODEs})  indicates a dependence of the heat and mass transport on $f\equiv S_1f_1 + S_2f_2$. Manipulation of the first two equations of (\ref{ODEs3}) leads to an  ODE for $f$ with $S_1
S_2$ and $S_1S_2f_1'f_2'$ as parameters, clearly indicating that generally $f$ will have a dependence on $S_1S_2$. Thus, the behavior for the counterflow can vary from the planar value of $S_1 =1, S_2 =0$ (or vice versa) or from the case $S_1 = S_2 =1/2$.  This clearly shows that distinctions must be made amongst the various possibilities for three-dimensional strain fields as $S_1S_2$ varies between large negative numbers and $1/4$. An exception will be the incompressible case with constant properties where the $S_1S_2$ terms cancel in the equation for $f$.

	The vorticity $\omega_{\kappa}$ will impact directly $f_1$ and $f$; thereby, it is affecting the velocity field. Then, through the advection of the scalar properties, there is impact on mass fractions and enthalpy.   If the vorticity $\omega_{\kappa} =0$, a simple inspection of the governing ODEs leads to the conclusion that the values for $f_1, f_1', f_2, f_2',u/x,$ and $w/z$ can be interchanged with the values for $f_2, f_2', f_1, f_1',w/z,$ and $u/x$, respectively, when $S_1$ and $S_2$ are replaced by $1-S_1$ and $1-S_2$, respectively.

	Note that, for $S_1 >1$ or $S_2 >1$ (which imply $S_2 < 0$ or $S_1 <0$, respectively), there would be incoming streams from two directions. One incoming stream would have a prescribed velocity profile in the viscous layer determined as a local exact solution to the Navier-Stokes conditional on its matching the profile determined by upstream conditions for the flow in that direction; this situation is too highly contrived and is not considered here. Thus, $S_1$ and $S_2$ are always each non-negative and bounded above by unity value in our considerations here. The figures show results for three strain rates; $S_1 = 0$ (planar case); $S_1 =0.25$ (3D strain); and $S_1 = 0.5$ (axi-symmetric case).

If the viscous dissipation is negligible, $\tilde{h}$ becomes a conserved scalar. Other conserved scalars can be created. Mixture fraction is a popular choice. For case of one-step kinetics with $Le =1$ and negligible viscous dissipation, the Shvab-Zel'dovich variables, $\alpha \equiv Y_F - \nu_S Y_O$ and $\beta \equiv h - \nu_S Y_O Q$ become conserved scalars. $Y_F, Y_O, \nu_S, $ and $Q$ are fuel mass fraction, oxygen mass fraction, stoichiometric mass ratio, and chemical energy per unit mass of fuel, respectively. For steady-state and time-averaged flows, these conserved scalars vary monotonically across a flow field and can be used to replace one of the spatial coordinates. This coordinate transformation in the counterflow configuration results in a new form of the scalar equations where the advective term is not present; a reactive-diffusive balance results. This result has classically been used \cite{Peters, Pierce} together with an incompressible-flow assumption which gives an overly simplistic solution for the velocity field. \cite{Sirignano2019a, Sirignano2019b, Sirignano2020} has shown the transformation gains little when the variable density is considered; furthermore, the variable density significantly affects the reacting counterflow.

Consider the production or consumption rate of a particular species over the counterflow volume. We can either integrate over a volume using the original form in Equation (\ref{species}) or, more conveniently, using Equation (\ref{ODEs3}) to get exactly the same result. Consider the volume $-a < \xi < a, -b < y < b, -c <z < c$. The choices of lengths $a$ and $c$ will not matter on a per-unit-volume basis  since mass fraction $Y_m$ and reaction rate $\dot{\omega}_m$ do not vary with $x$ or $z$. $c$ is chosen to be of the order of the Kolmogorov scale. Volume $V = 8abc$, $\widetilde{\Phi}$  is the volume averaged viscous dissipation rate; and $ \widetilde{\rho\dot{\omega}_m} $  is the average mass production rate over the volume. It follows from integration of Equation (\ref{energy4}) after multiplication by density $\rho$ and division by $Pr V$  that
\begin{eqnarray}
\int^a_{-a}\int^b_{-b}\int^c_{-c}  \frac{\rho}{Pr V} [Y''_m  &+&  Pr f Y'_m + Pr\dot{\omega}_m]dx dy dz = 0   \;\;;\;\; m=1, 2, ...., N  \nonumber  \\
\widetilde{\rho\dot{\omega}_m} &\equiv& \frac{1}{V}\int_V \rho \dot{\omega}_m dV= - \frac{1}{2b} \int^{\eta(b)}_{\eta(-b)}fY'_m d\eta \;\;;\;\; m=1, 2, ...., N  \nonumber  \\
\int^a_{-a}\int^b_{-b}\int^c_{-c}  \frac{\rho}{Pr V}  [ h'' &+& Pr f h'
- Pr  \Sigma^N_{m=1}h_{f,m} \dot{\omega}_m + Pr\frac{\Phi}{\rho}] dx dy dz =0
\nonumber \\
\Sigma^N_{m=1}h_{f,m} \widetilde{\rho \dot{\omega}_m }- \tilde{\Phi} &=& \frac{1}{2b}\int^{\eta(b)}_{\eta(-b)} fh' d\eta
\label{rateintegral}
\end{eqnarray}
Here, $b$ has been considered large enough so that $Y'_m =0$ and $h'=0$ at those boundaries are good approximations.  However, the value for $ \widetilde{\rho \dot{\omega}_m }$ will depend strongly on the chosen domain size $2b$, which will have  a value of $O(10)$ typically in our analysis.    $Le =1$ has also been considered. The volume average viscous dissipation rate $\widetilde{\Phi}$ may be obtained by integration of Equation (\ref{dissipation}).

Consider a species $m$ that is flowing inward away from $\eta = \infty$ towards $\eta = 0$. If it is being produced (consumed), the derivative $Y'_m$ in Equation (\ref{rateintegral}) will be negative (positive) for $\eta > 0$ where velocity $v<0$ and $f>0$. The signs are opposite for a species flowing inward away from $\eta = -\infty$ and towards $\eta = 0$.  The equation provides two ways to evaluate the average production (consumption) rate for species $m$. The volume integral of the reaction rate will have highly nonuniform integrand values over the space while the outflow integral over $\eta$ will have a smoother variation of the integrand.

The flamelet model requires inputs that are scaled from the resolved flow. Specifically, rate of strain and vorticity, pressure, and the inflowing scalar values for the counterflow are needed. The magnitude of the resolved scale velocity is not relevant because the sub-grid velocities are measured relative to a frame moving with a Galilean transformation. The flamelet will give back to the resolved flow the instantaneous value for energy release rate.

\subsection{Chemical Kinetics Model}

The above equations can be readily applied for diffusion-flame counterflows and partially-premixed-flame counterflows as will be explained in the following sections. They can also describe situations where multi-branched flames exist. For case of zero vorticity, i.e., $\omega =0$, the generality has been shown by \cite{Sirignano2019b, Sirignano2019a}.  Although the analysis allows for the use of detailed chemical kinetics in the future, we will focus here on propane-oxygen flows with one-step kinetics. However, results are expected to be qualitatively more general, applying to situations with more detailed kinetics and to other  hydrocarbon /oxygen-or-air combination.
 \cite{Westbrook_Dryer:1984}
kinetics are used; they were developed for premixed flames but any error for nonpremixed flames is viewed as tolerable often here because diffusion would generally be rate-controlling.  Using astericks to denote dimensional quantities, one may deduce that
\begin{eqnarray}
\omega^*_F = - A^* {\rho^*}^{0.75} Y_F^{0.1} Y_O^{1.65} e^{-50.237/\tilde{h}}
\label{onestep}
\end{eqnarray}
where the ambient temperature is set at 300 K and density $\rho^*$ is to be given in units of kilograms per cubic meter. Here, $A^* =4.79\times10^8 (kg/m^3)^{-0.75}/s$. The dimensional strain rate $S^*_1 + S^*_2$ (at the sub-grid scale) is used to normalize time and reaction rate.
In non-dimensional terms,
\begin{eqnarray}
\omega_F &=& - \frac{A^* {\rho_{\infty}^*}^{0.75}}{S_1^* + S^*_2}\tilde{h}^{-0.75} Y_F^{0.1} Y_O^{1.65} e^{-50.237/\tilde{h}}   \nonumber  \\
\omega_F &=& - \frac{Da}{\tilde{h}^{0.75}} Y_F^{0.1} Y_O^{1.65} e^{-50.237/\tilde{h}}
\label{Damkohler}
\end{eqnarray}
The above equation defines the Damk\"{o}hler number $Da$. Furthermore, we set $Da \equiv K Da_{ref}$ where
\begin{eqnarray}
Da_{ref} \equiv \frac{ \tilde{A}(10 kg/m^3)^{0.75}}{(10^4/s)} =2.693 \times 10^5 \;\;  ;  \;\;
 K \equiv \Big[\frac{\rho_{\infty}^*}{10 kg/m^3}\Big]^{0.75}\frac{10^4/s}{S_1^* + S_2^*}
\end{eqnarray}
10 $kg/m^3$ and 10,000$/s$ are conveniently chosen as reference values for density and strain rate, respectively.    The reference value for density implies an elevated pressure. The strain-rate reference value is in the middle of an interesting range for this chemical reaction. Suppose the resolved scale has a velocity, a mixing length, and kinematic viscosity with the following respective orders of magnitude: $10 m/s$, $0.1 m,$ and $10^{-4} m^2/s$. Then, we may estimate $Re =10^4$ and resolved scale strain rate $\partial u / \partial x = 100/s$. Thereby, Equation (\ref{strainscale}) yields that $\partial u_{\kappa}/\partial \kappa = Re^{1/2}\partial u / \partial x = S^*_1 +S^*_2 = 10^4/s$.

 Clearly, there is no need to set pressure (or its proxy, density) and the strain rate separately for a one-step reaction. For propane and oxygen, the mass stoichiometric ratio $\nu =0.275$.

The non-dimensional parameter $K$ will increase (decrease) as the strain rate decreases (increases) and/ or the pressure increases (decreases). $K =1$ is our reference case and the range covered will include $O(10^{-1}) \leq K\leq  O(10^2)$, allowing for the needed variation in strain rate and pressure to sustain premixed flamelets, diffusion flamelets, and multi-branched flamelets.

The system of ordinary differential equations is solved numerically  using a relaxation method and central differences. Solution over the range $-5 \leq \eta \leq 5$ provides adequate fittings to the asymptotic behaviors. The parameters that are varied are $ K, Pr,$ and $S_1$ (and thereby $S_2 = 1 -S_1$). Most calculations have $Pr = 1$ with emphasis on the effect of variation in $K$, i.e., pressure and strain rate.

\subsection{Uncoupled Diffusion Flamelet Calculations}\label{DiffFlame}
		
\begin{figure}
  \centering
 \subfigure[enthalpy, $h/h_{\infty}$ ]{
  \includegraphics[height = 4.6cm, width=0.45\linewidth]{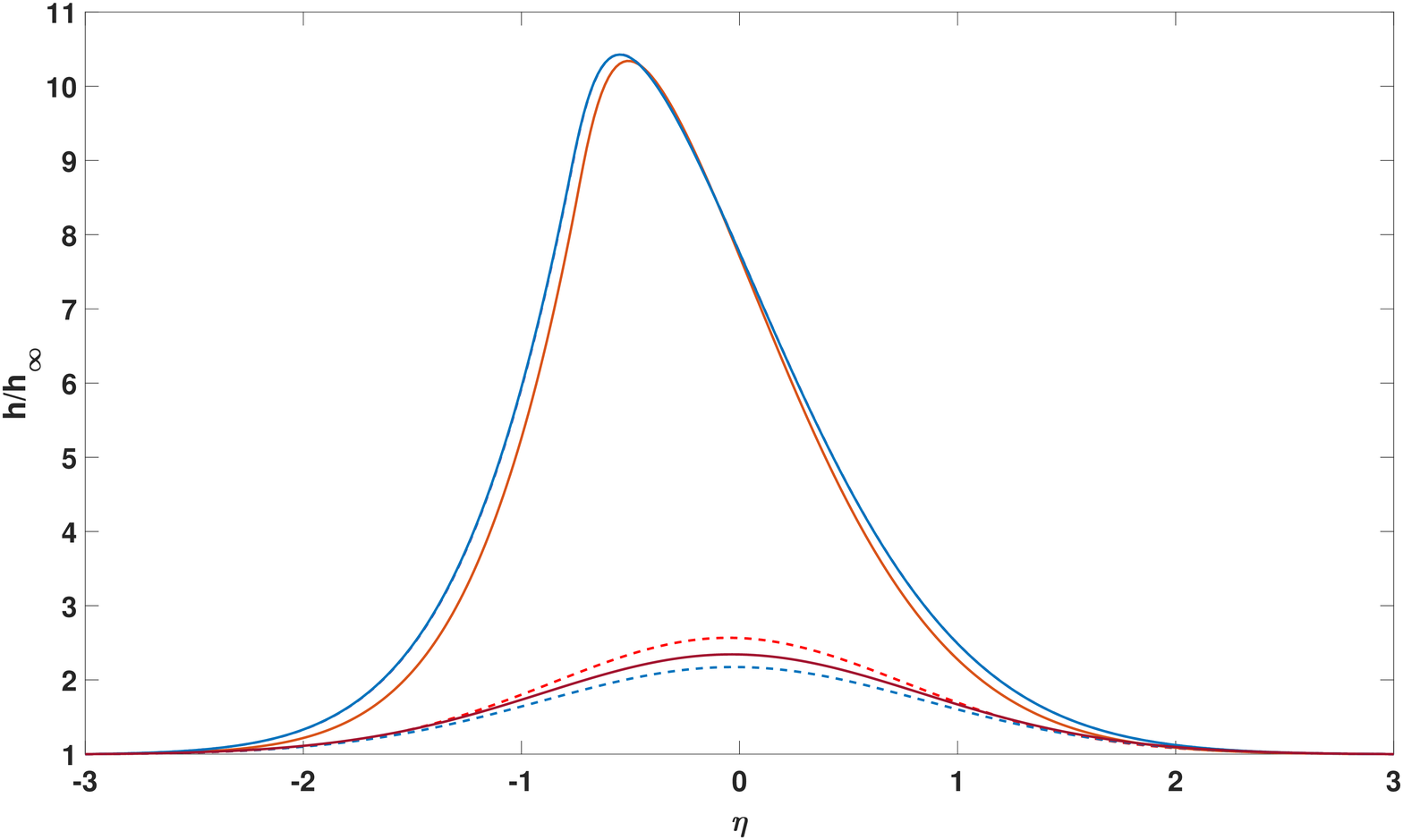}}
  \subfigure[fuel mass fraction, $Y_F$]{
  \includegraphics[height = 4.6cm, width=0.45\linewidth]{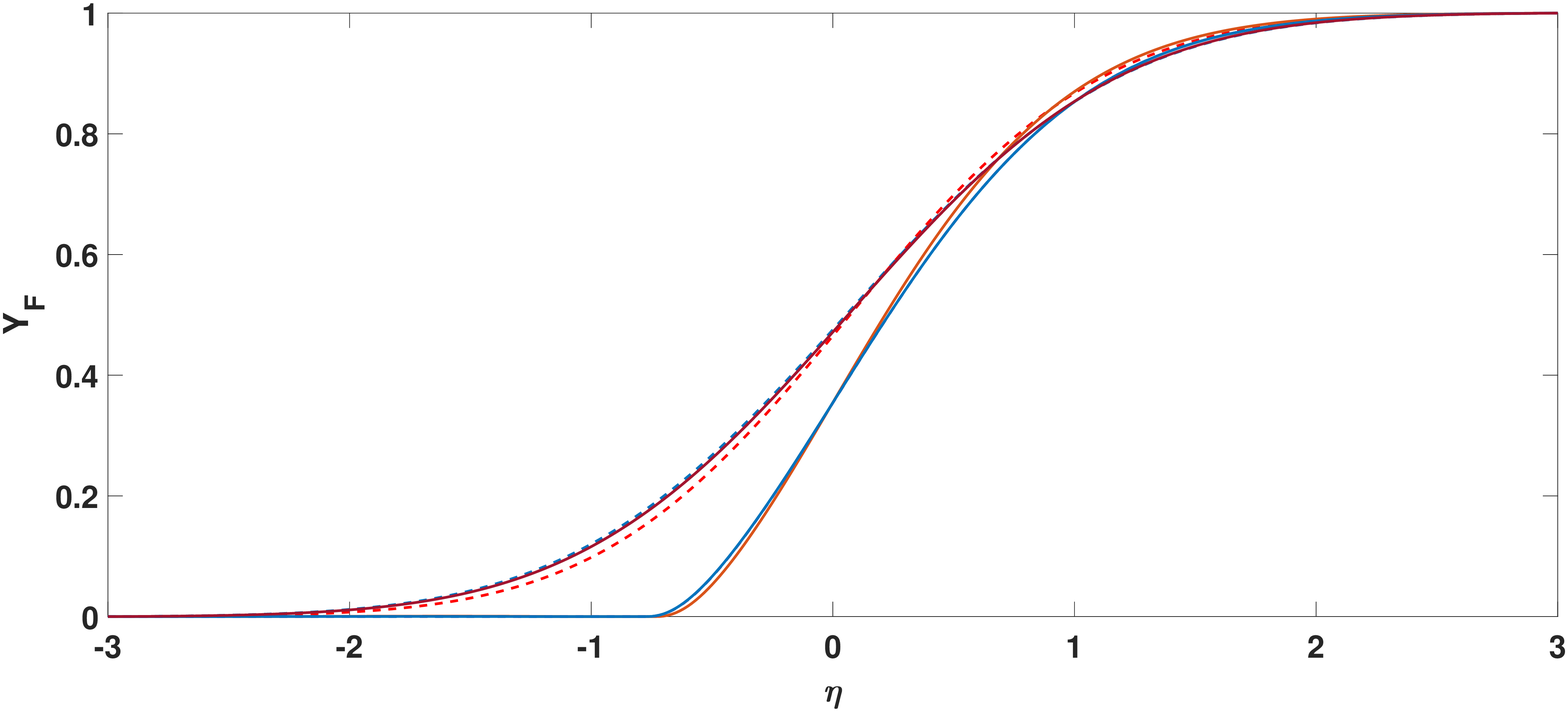}}     \\
  \vspace{0.2cm}
  \subfigure[ mass ratio x oxygen mass fraction, $\nu Y_O$]{
  \includegraphics[height = 4.6cm, width=0.45\linewidth]{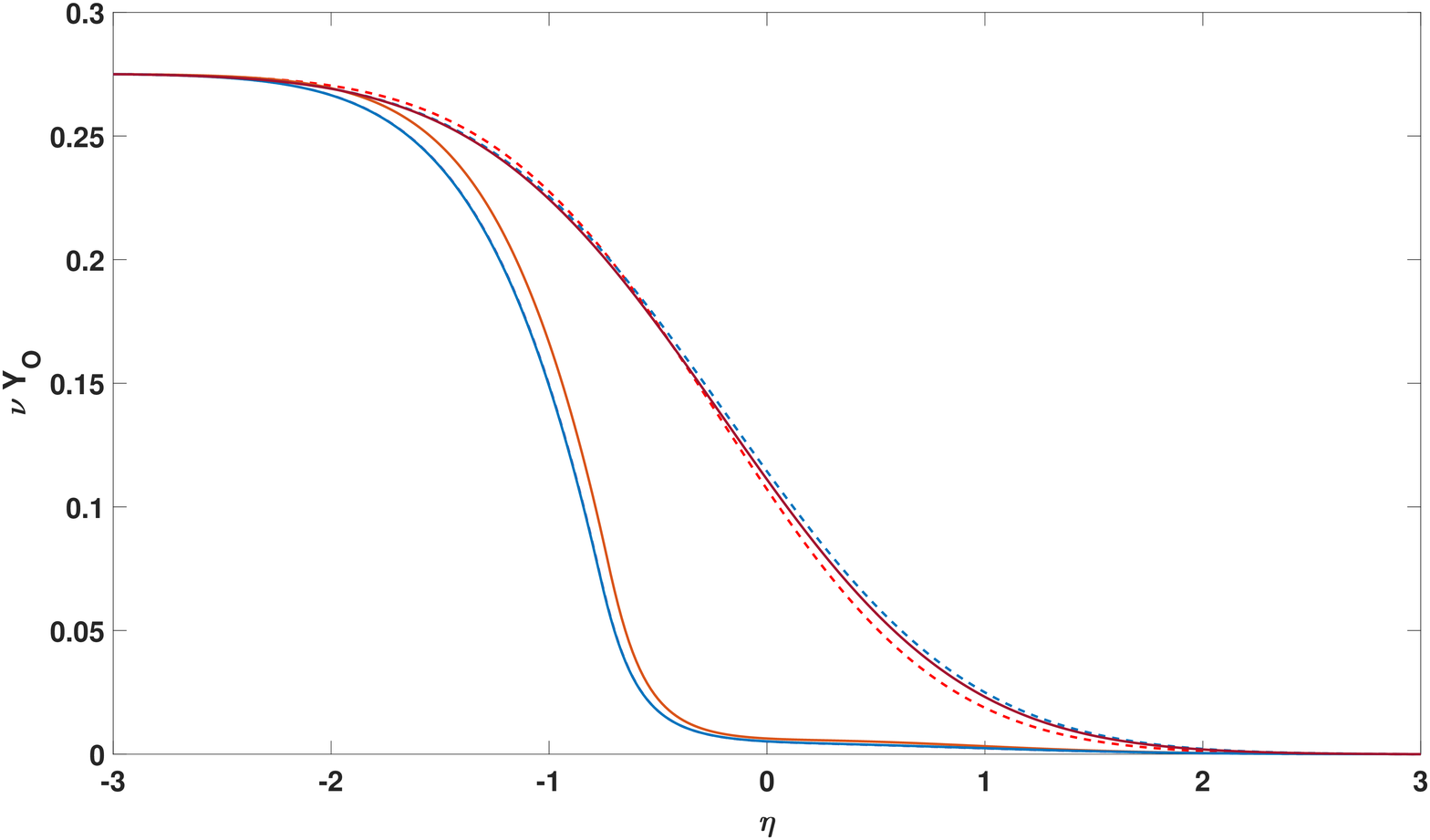}}
  \subfigure[integral of reaction rate, $\int \dot{\omega}_F d \eta$]{
  \includegraphics[height = 4.6cm, width=0.45\linewidth]{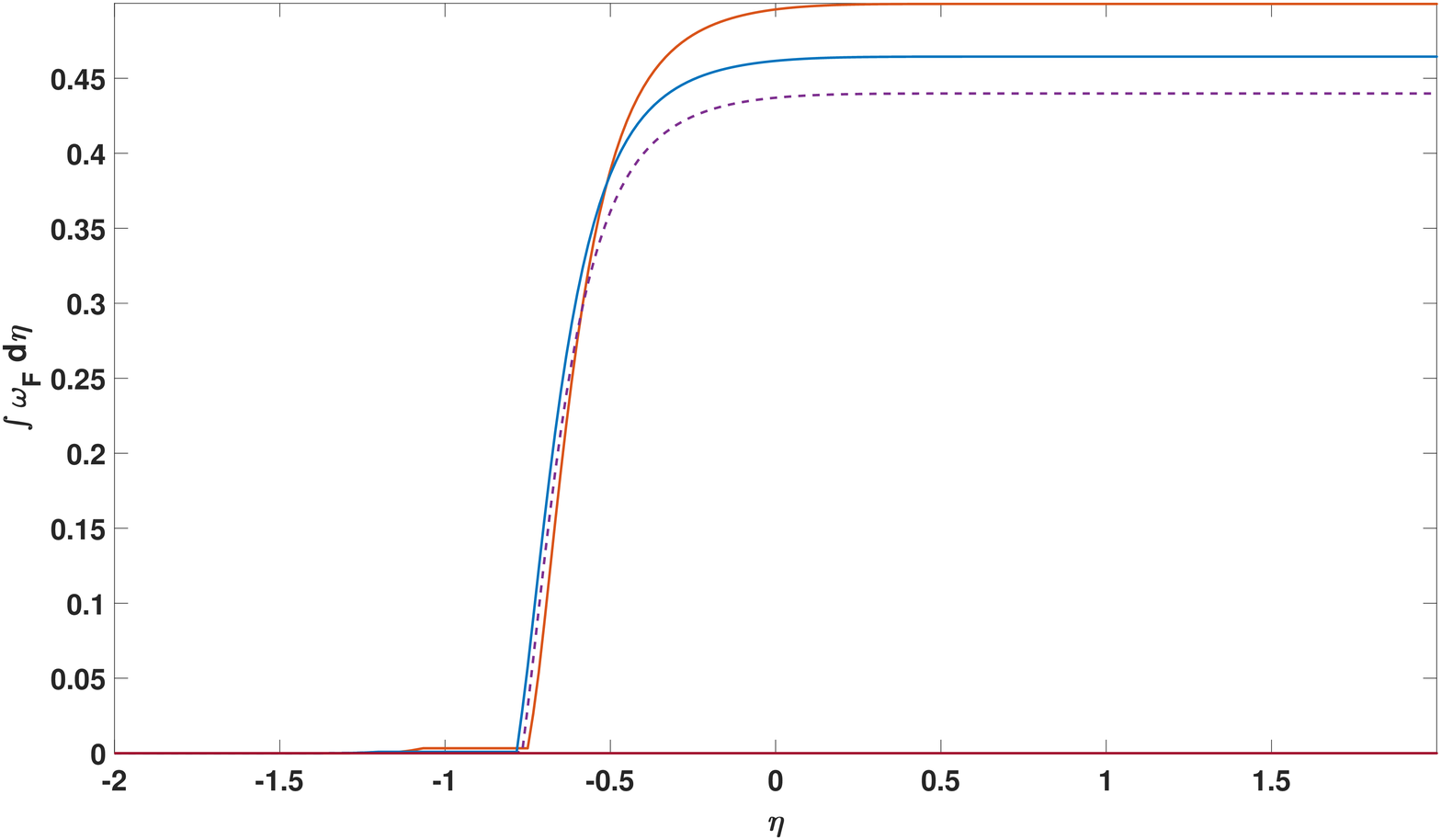}}     \\
  \vspace{0.2cm}
  \subfigure[reaction rate, $\dot{\omega}_F $]{
  \includegraphics[height = 4.6cm, width=0.45\linewidth]{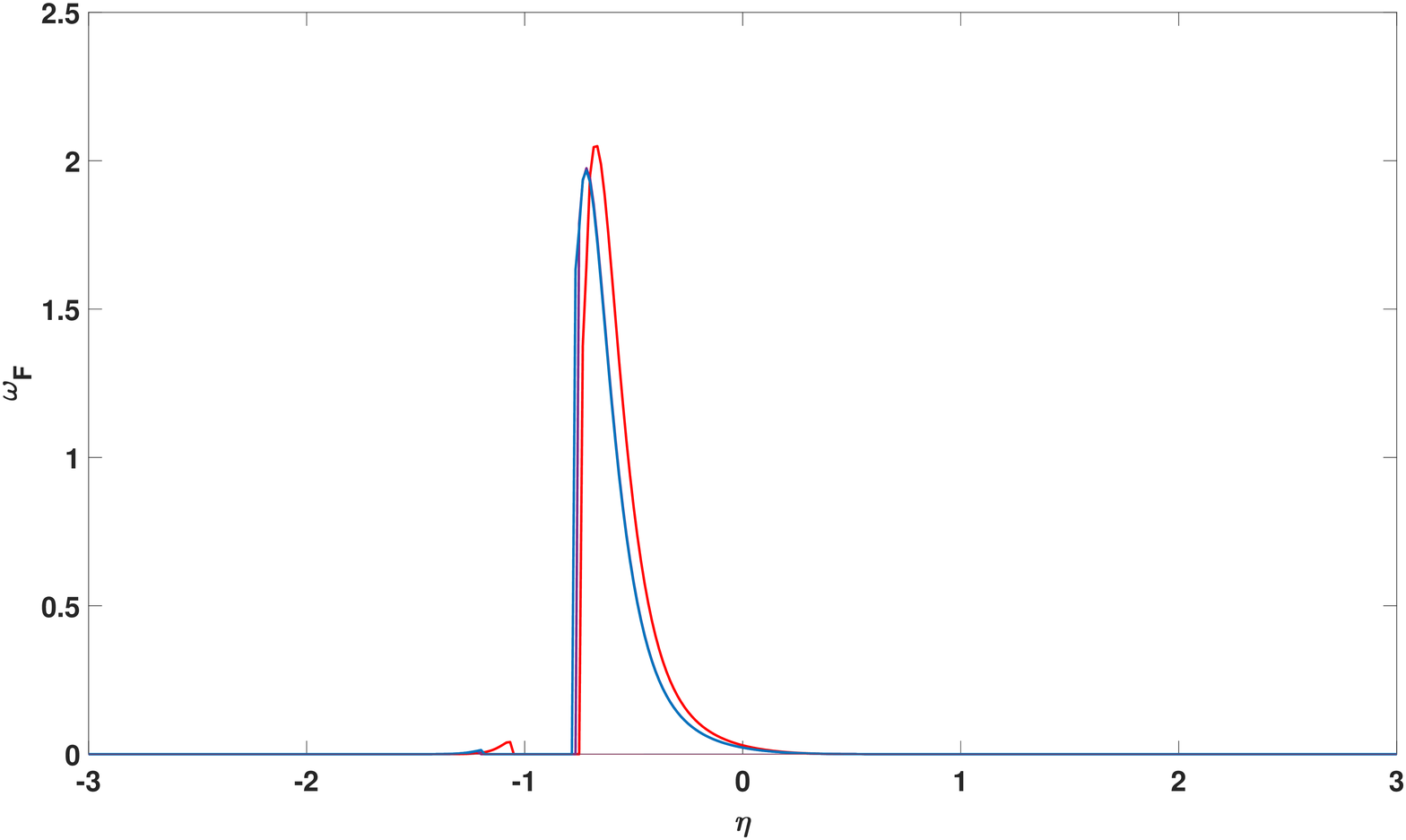}}
  \vspace{-0.1cm}
  \caption{Diffusion flame with varying vorticity and Damk\"{o}hler number. $S_1 =0.75 ;  \; S_2 = 0.25$. dash red, $K= 0.195, \;  \omega_{\kappa} = 0$; solid red, $ K= 0.196,  \; \omega_{\kappa} = 0$;  solid blue, $K= 0.196, \; \omega_{\kappa} = 1.0$;  solid purple, $ K= 0.185, \; \omega_{\kappa} = 1.0$;  dash blue, $K= 0.180, \; \omega_{\kappa} = 1.0$.  The $K=0.195, \; \omega_{\kappa}=1.0$ case is covered by the solid blue curve, except in subfigure d where it is given by the dash purple curve. }
  \label{WeakDiffFlame1}
\end{figure}

\begin{figure}
  \centering
 \subfigure[mass flux per area, $f=\rho u_{\chi}$ ]{
  \includegraphics[height = 4.8cm, width=0.45\linewidth]{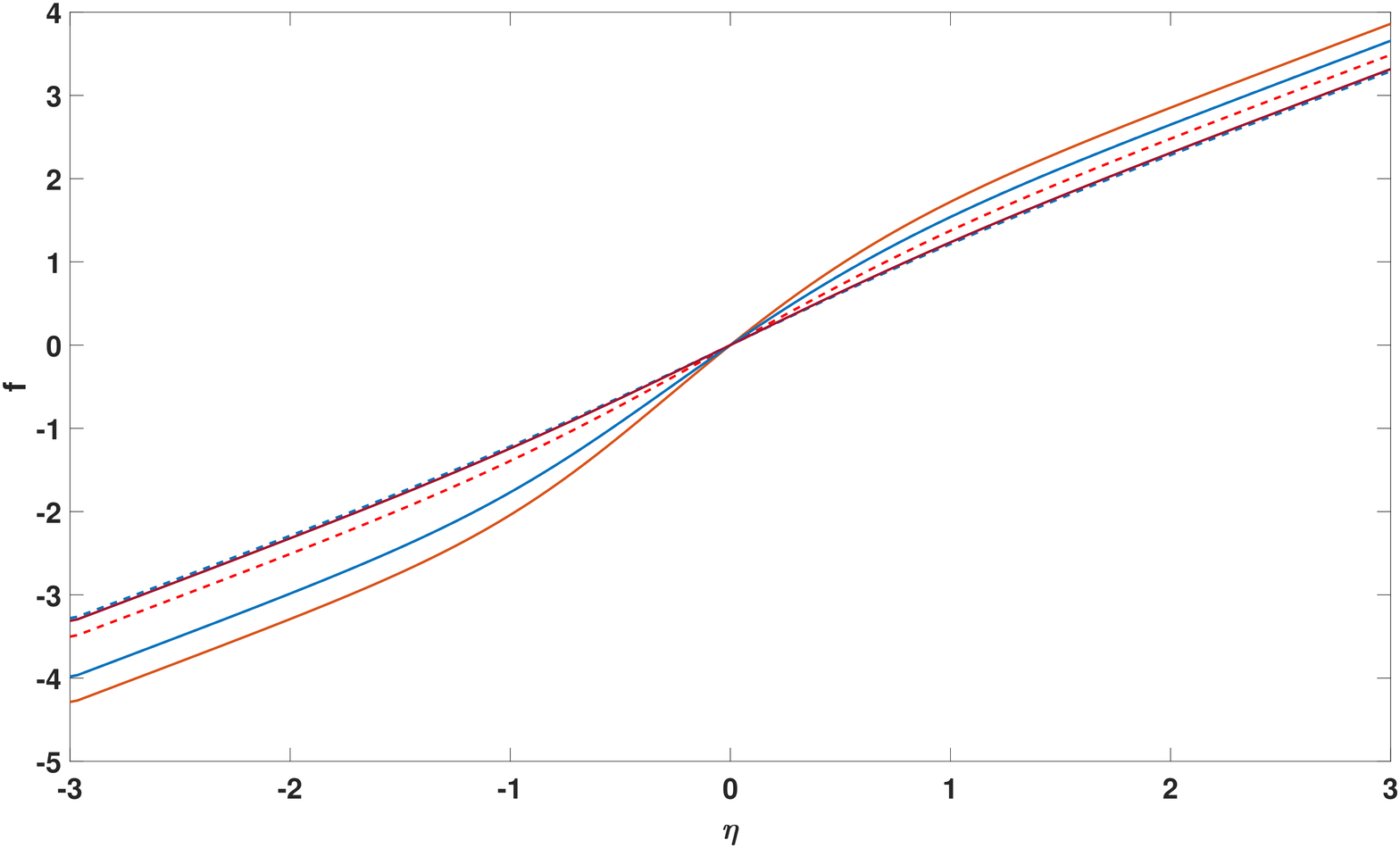}}
  \subfigure[velocity component, $f_1' = u_{\xi}/(S_1 \xi)$]{
  \includegraphics[height = 4.8cm, width=0.45\linewidth]{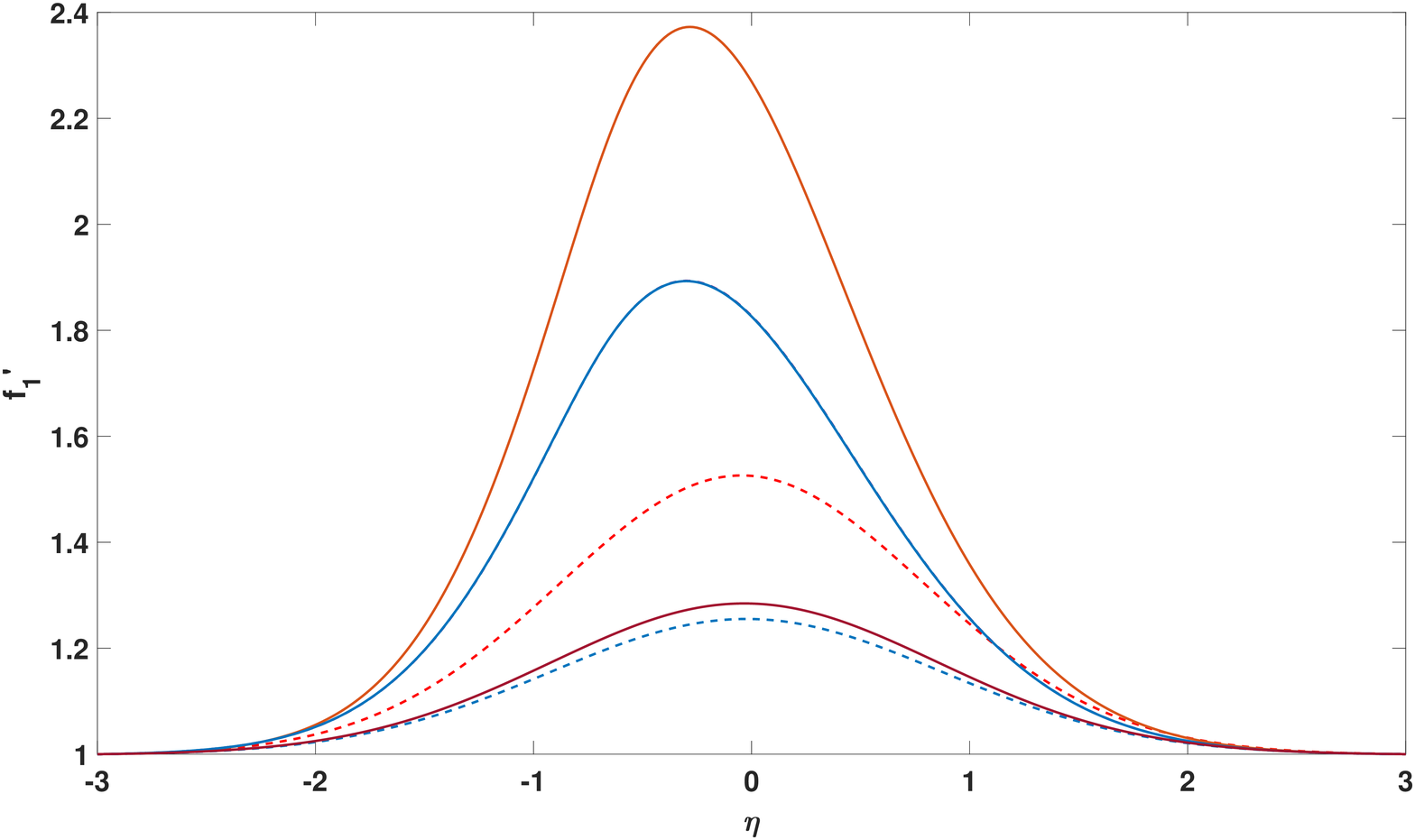}}     \\
  \vspace{0.2cm}
  \subfigure[ velocity component, $f_2' = w/(S_2z)$]{
  \includegraphics[height = 4.8cm, width=0.45\linewidth]{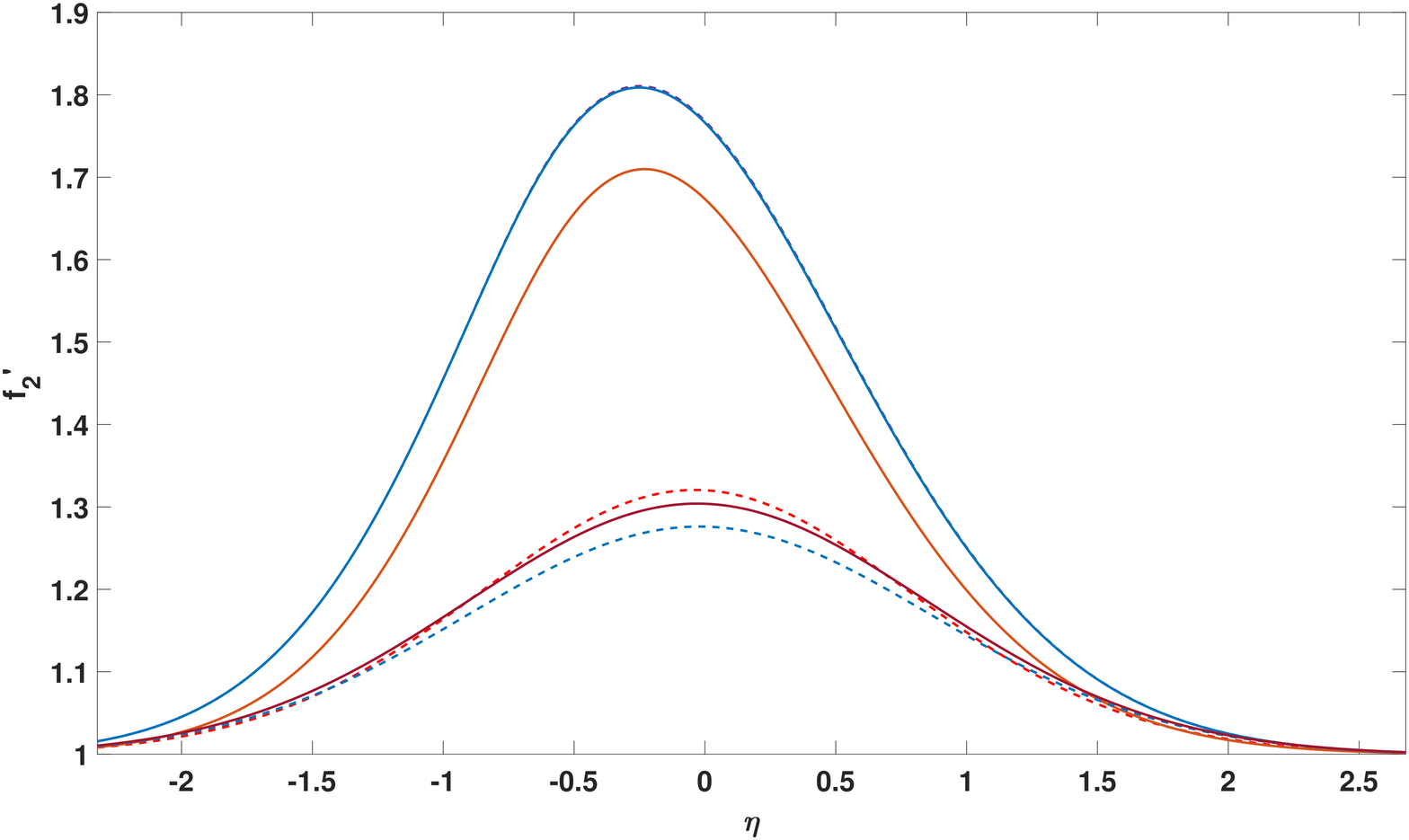}}
  \subfigure[velocity component, $u_{\chi}$]{
  \includegraphics[height = 4.8cm, width=0.45\linewidth]{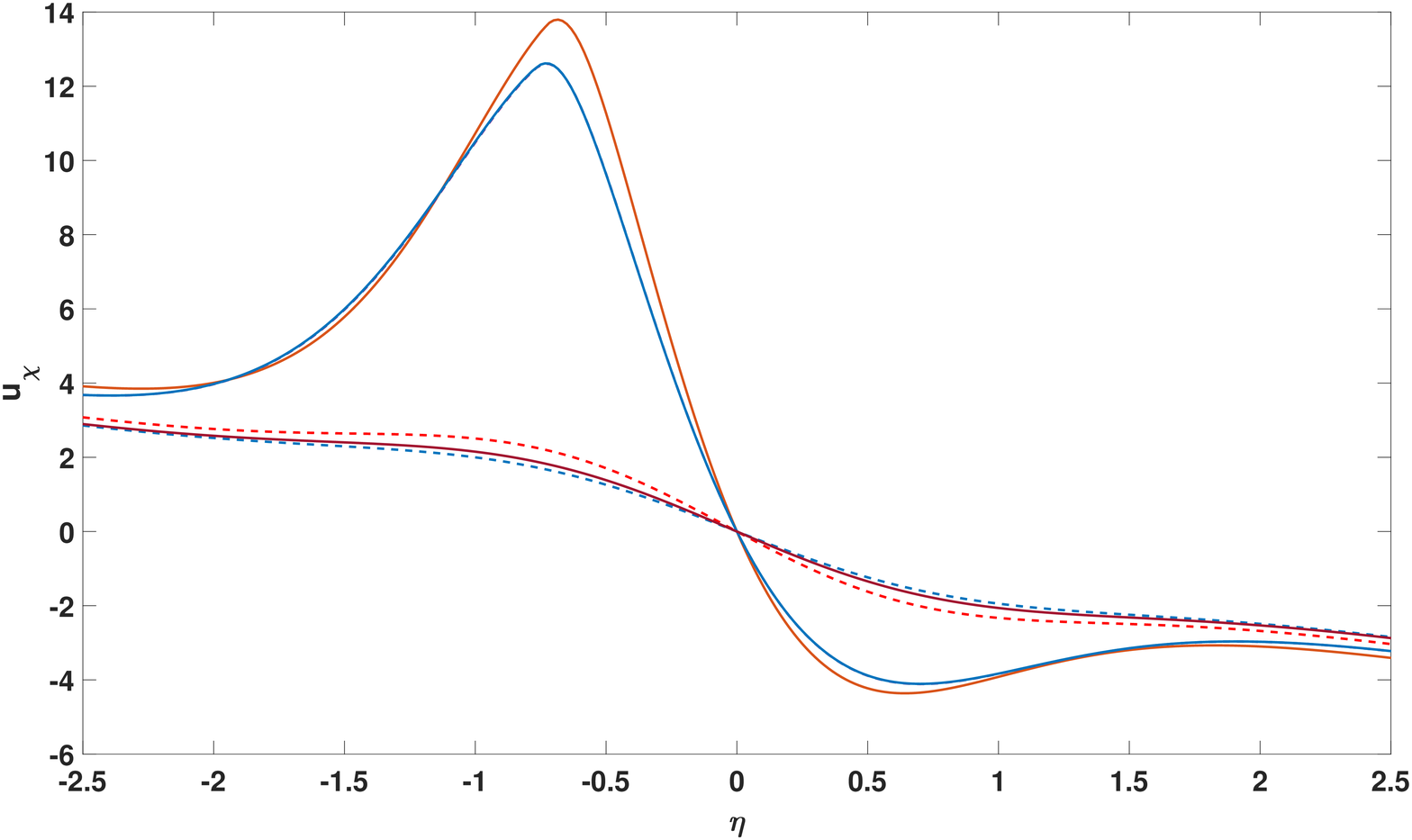}}     \\
  \vspace{0.2cm}
  \caption{Diffusion flame with varying vorticity and Damk\"{o}hler number. $S_1 =0.75 ; \;  S_2 = 0.25$. dash red, $K= 0.195,  \;   \omega_{\kappa} = 0$; solid red,   \; $ K= 0.196,  \; \omega_{\kappa} = 0$;  solid blue, $K= 0.196, \; \omega_{\kappa} = 1.0$;  solid purple, \;  $ K= 0.185, \;   \omega_{\kappa} = 1.0$;  dash blue, $K= 0.180, \;  \omega_{\kappa} = 1.0$.  The $K=0.195, \;  \omega_{\kappa}= 1.0$ case is covered by the solid blue curve. }
  \label{WeakDiffFlame2}
\end{figure}

Now, we treat a situation with a  three-dimensional diffusion-flame structure at the sub-grid level. Figures \ref{WeakDiffFlame1} and  \ref{WeakDiffFlame2} show the influence of vorticity on the flamelet stability near the extinction limit. The rotation of the flamelet due to vorticity causes a centrifugal effect on the counterflow velocity and thereby on the residence time in the vicinity of the reaction zone. The dash red and solid red curves represent $K$ values of 0.195 and 0.196, respectively, both without vorticity. Despite the very modest difference in $K$, one survives as a strong flame and the other is basically extinguished. By applying $K=0.195$ and $\omega_{\kappa} =1.0$ and therefore rotational speed of value $d\theta / dt =0.5$  , the solid blue curve is obtained, indicating a strong flame with regard to both reaction rate and peak temperature or enthalpy is induced by the rotation. Actually,  the $K=0.196, \;\omega_{\kappa}= 1.0$ curve falls right on the solid blue curve as well, except in Figure \ref{WeakDiffFlame1}d where it appears as a dash purple curve. The rotation causes a lower mass flux $f$ than was obtained without rotation. The inflow rate is diminished because the centrifugal effect creates a more adverse pressure gradient. However, the slower flow rate allows a longer residence time and modifies the extinction limit.   As $K$ is lowered to values of 0.180 or 0.185, the flame is essentially extinguished in spite of the increased residence time due to the rotational effects.

The rotation also causes a decrease in $f_1'$ and therefore in the $\xi$-component of velocity as shown in Figure \ref{WeakDiffFlame2}. There is an associated  increase in $f_2'$ and therefore an increase  in the $z$-component of velocity. Basically,  the low-density products of combustion find it easier to flow in the $z$-direction wherein no centrifugal effect is applied. This gives some physical explanation to the findings of  \cite{Nomura1993} where, for reacting flows but not for non-reacting flows, the major extensional axis tends to align parallel  with the vorticity vector.  Specifically, when reaction and energy release occur, the outflow for the counterflow configuration will have the greater extensional strain rate in the principal direction aligned with the vorticity. The outflow for the counterflow configuration might still experience a higher magnitude for extensional strain rate in the principal direction orthogonal to the vorticity than the inflow compressive-strain-rate magnitude due to the density decrease in the flow.  However, its increase will be less profound than for the axis aligned with the vorticity.

The asymptotic value as $\eta \rightarrow \infty$ in subfigure \ref{WeakDiffFlame1}d gives the integrated burning rate in the sub-grid volume. Realize that the factor $2b$ from Equation (\ref{rateintegral}) has not yet been factored into the integral result in the figures; it will reduce the values by an order of magnitude in order to give the average over the volume.   It is seen that rotation affects burning and burning rate is negligible in several cases. The decrease in mass flux rate with increasing rotation rate (i.e., vorticity) results in a decreased burning rate through the sub-grid volume.

\begin{figure}
  \centering
 \subfigure[enthalpy, $h/h_{\infty}$ ]{
  \includegraphics[height = 4.6cm, width=0.45\linewidth]{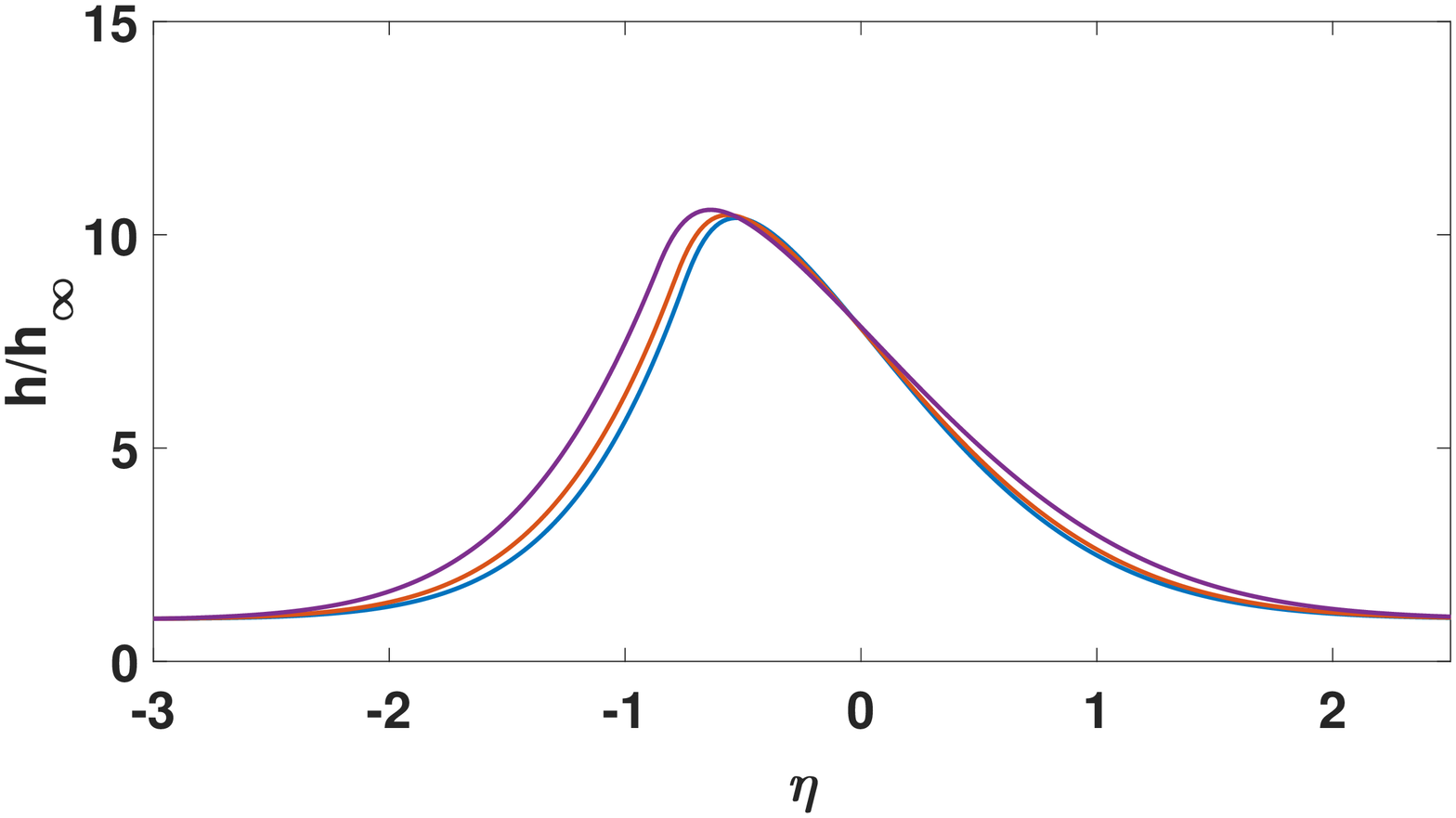}}
  \subfigure[fuel mass fraction, $Y_F$]{
  \includegraphics[height = 4.6cm, width=0.45\linewidth]{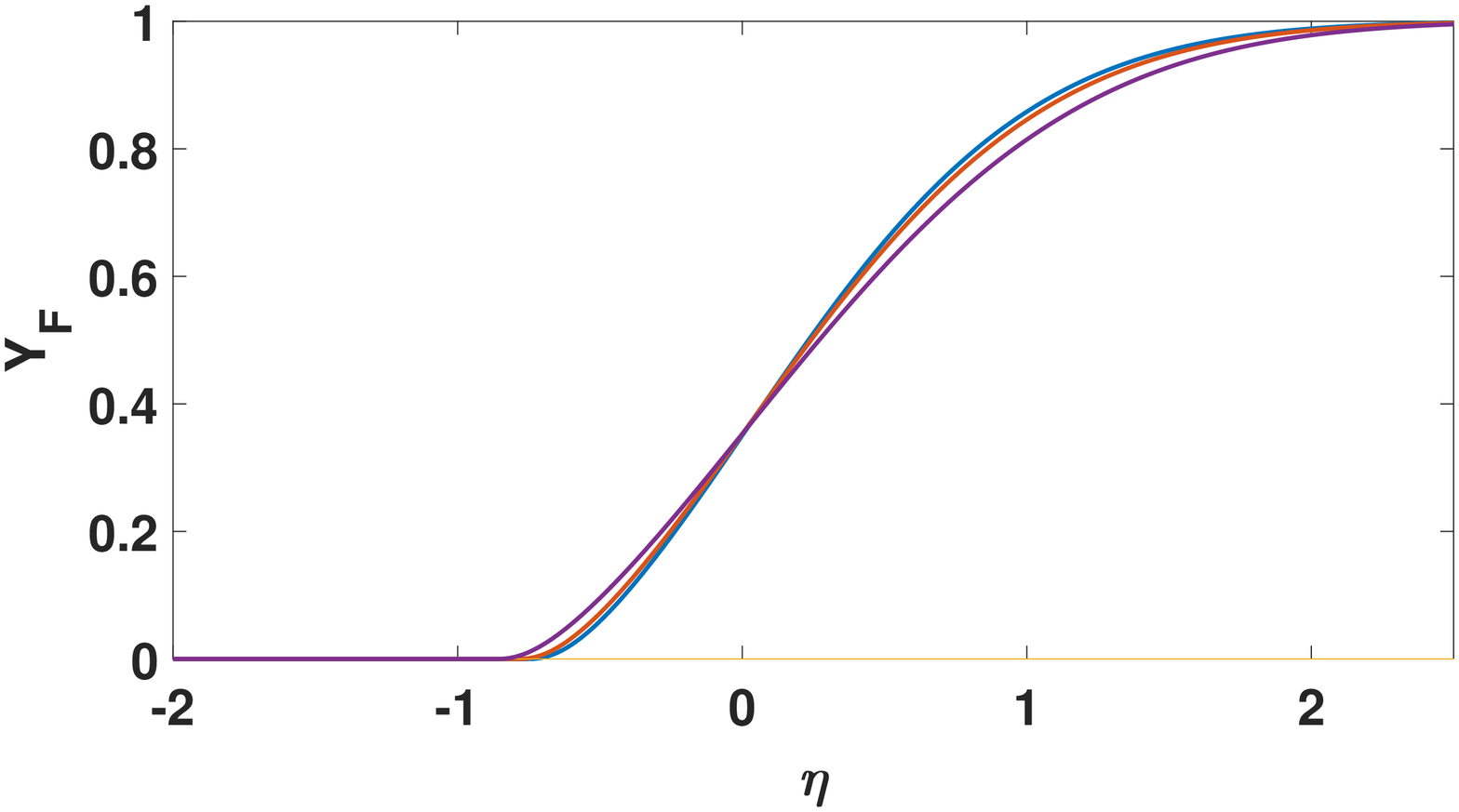}}     \\
  \vspace{0.2cm}
  \subfigure[ mass ratio x oxygen mass fraction, $\nu Y_O$]{
  \includegraphics[height = 4.6cm, width=0.45\linewidth]{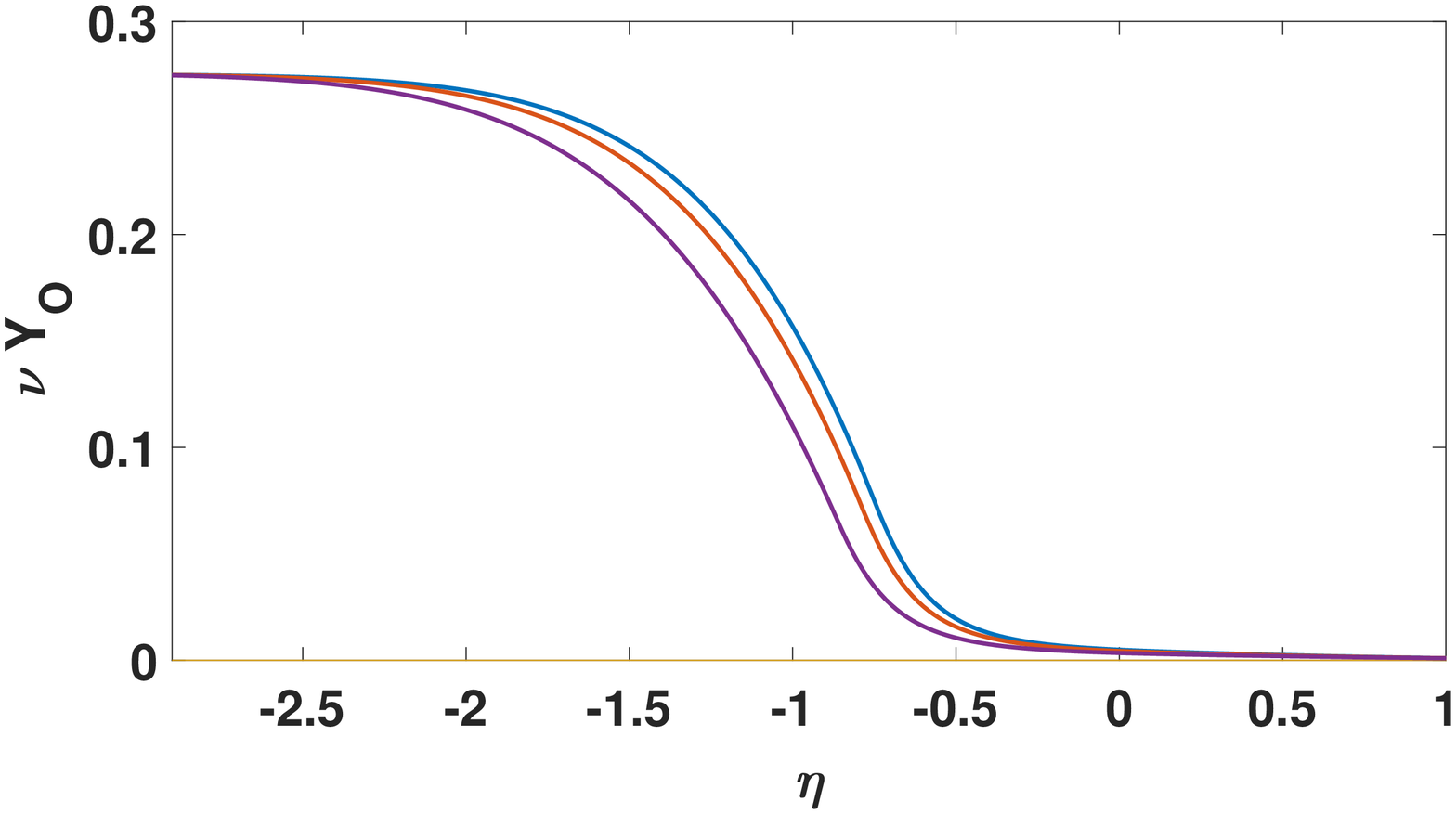}}
  \subfigure[integral of reaction rate,  $\int \dot{\omega}_F d \eta$]{
  \includegraphics[height = 4.6cm, width=0.45\linewidth]{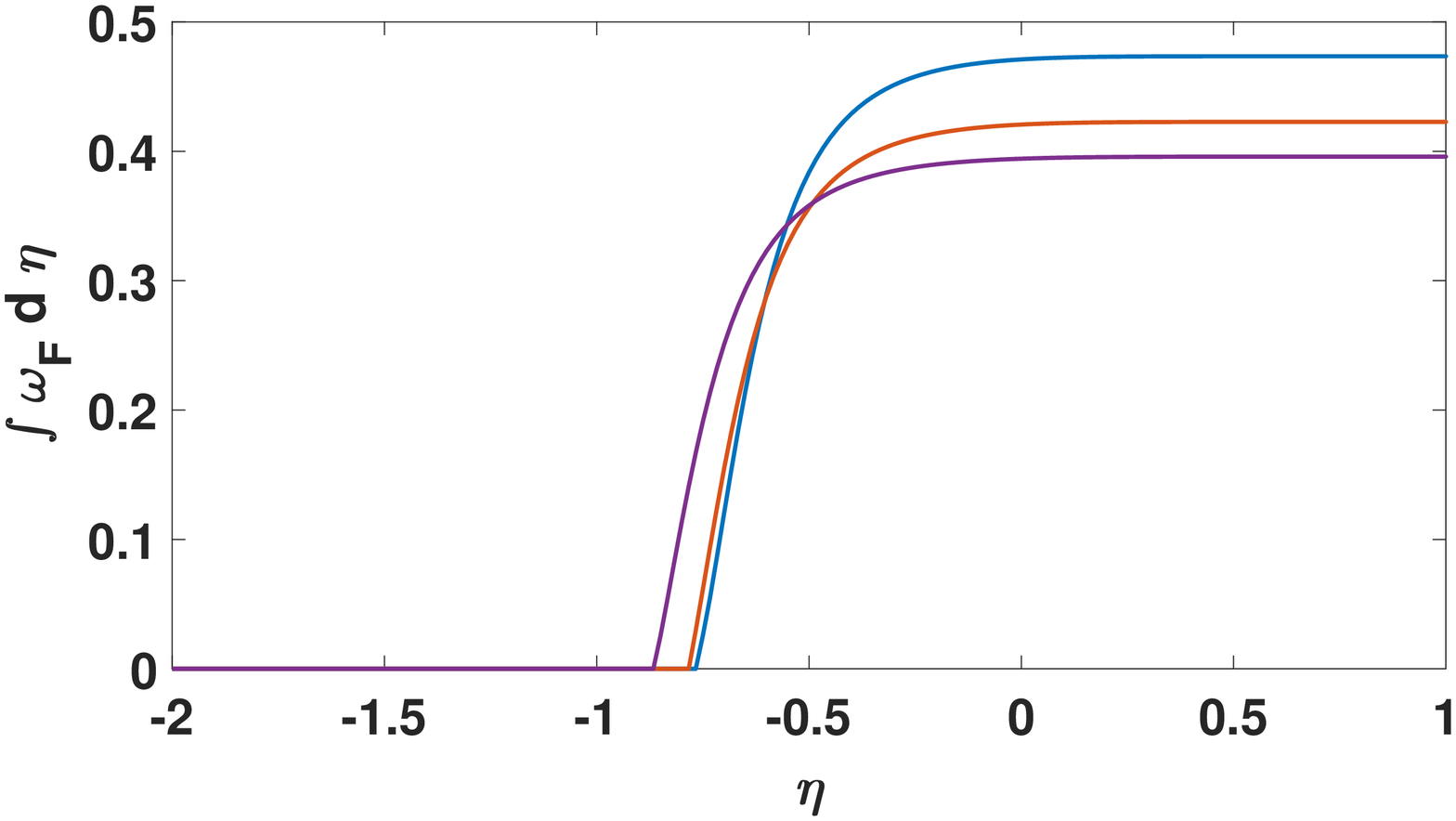}}     \\
  \vspace{0.2cm}
  \subfigure[reaction rate, $\dot{\omega}_F $]{
  \includegraphics[height = 4.6cm, width=0.45\linewidth]{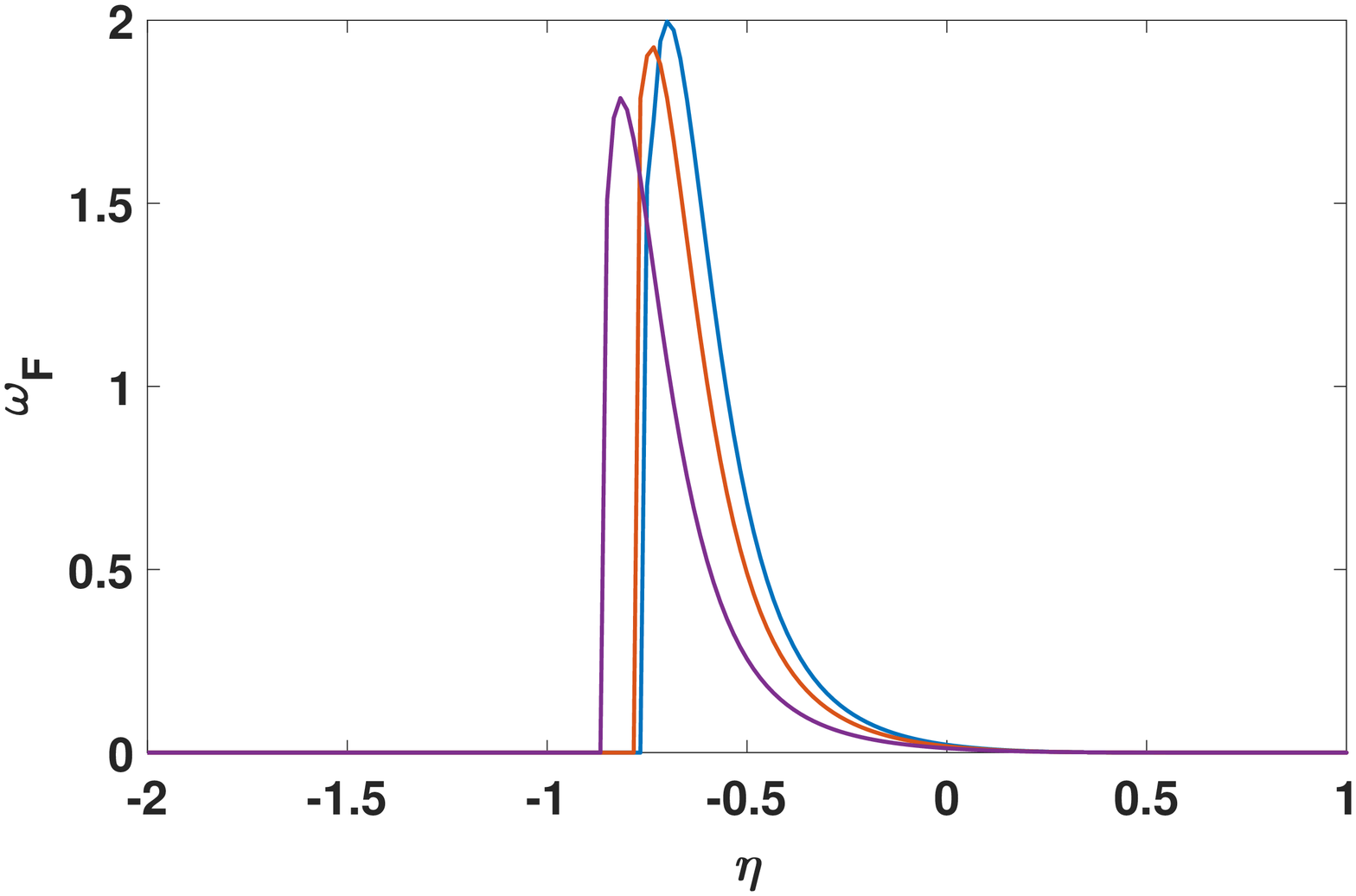}}
  \vspace{-0.1cm}
  \caption{Diffusion flame with varying strain rate.   $K= 0.195, \;  \omega_{\kappa} = 1.0$. blue $S_1 =0.750, \; S_2 = 0.250$;   red $S_1 =0.500, \;  S_2 = 0.500$ ;   purple $S_1 =0.333, \;  S_2 = 0.667$  .              }
  \label{S1DiffFlame1}
\end{figure}

\begin{figure}
  \centering
 \subfigure[mass flux per area, $f= \rho u_{\chi}$ ]{
  \includegraphics[height = 4.8cm, width=0.45\linewidth]{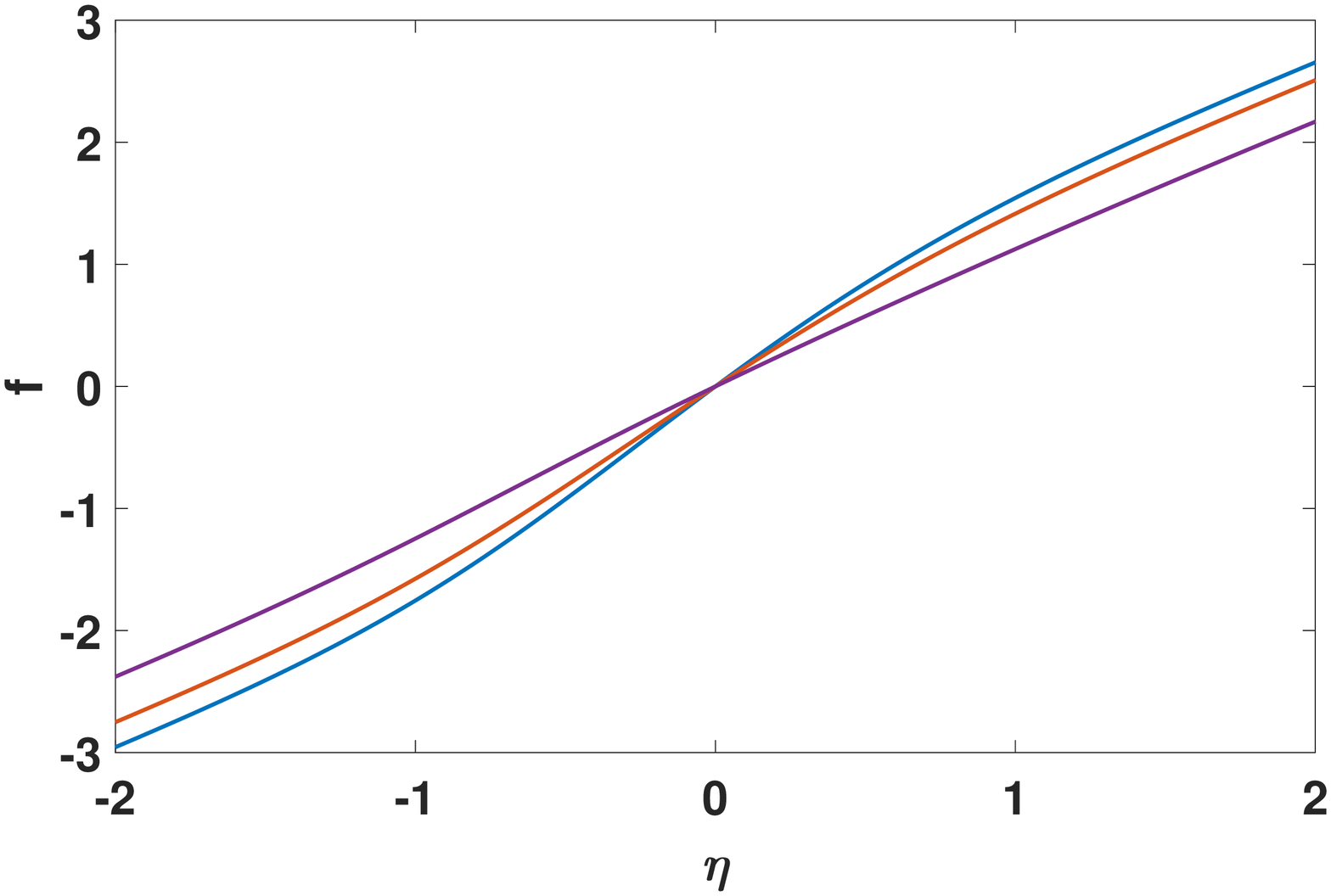}}
  \subfigure[velocity component, $f_1' = u_{\xi}/(S_1 \xi)$]{
  \includegraphics[height = 4.8cm, width=0.45\linewidth]{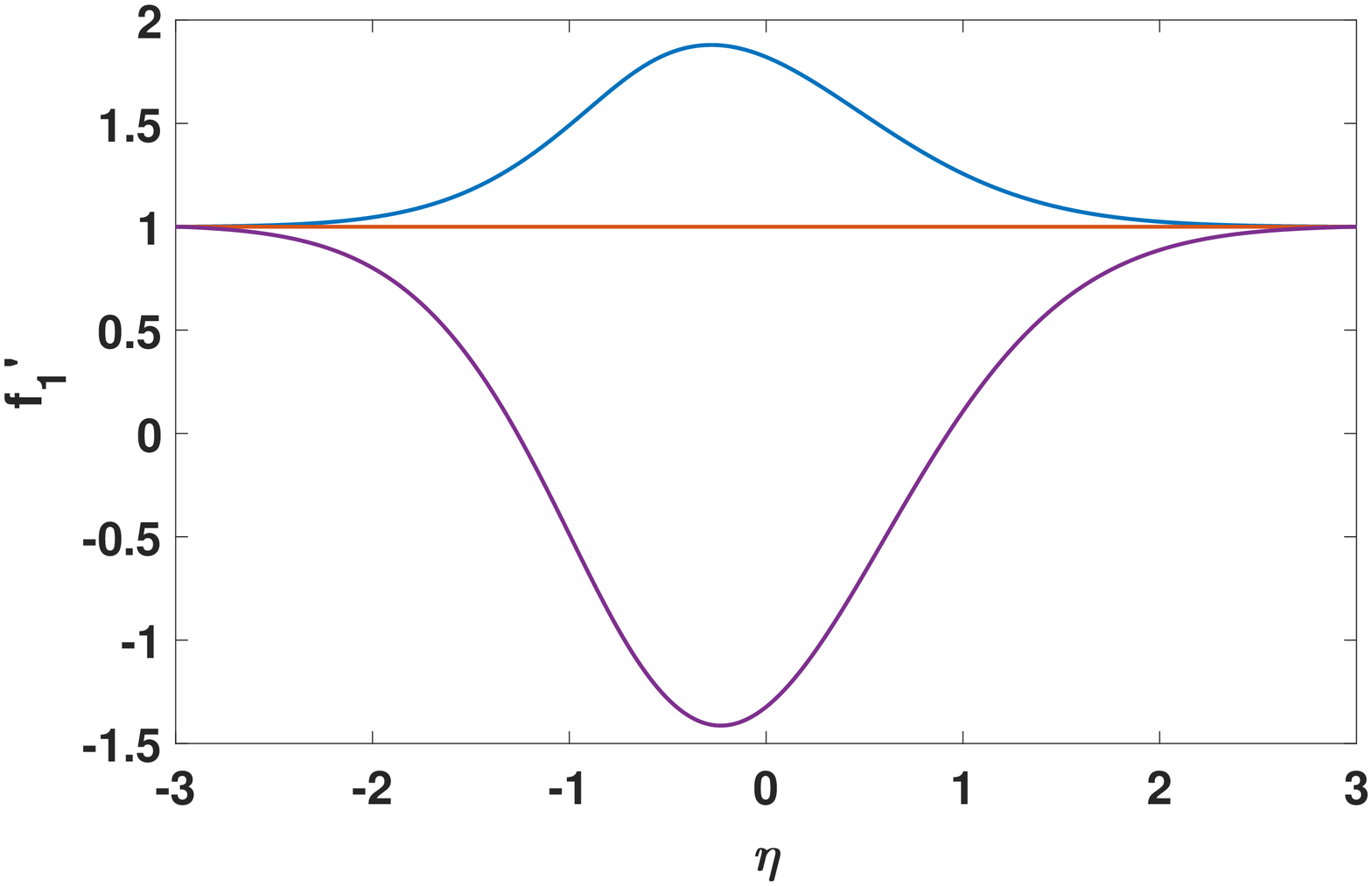}}     \\
  \vspace{0.2cm}
  \subfigure[ velocity component, $f_2' = w/(S_2 z)$]{
  \includegraphics[height = 4.8cm, width=0.45\linewidth]{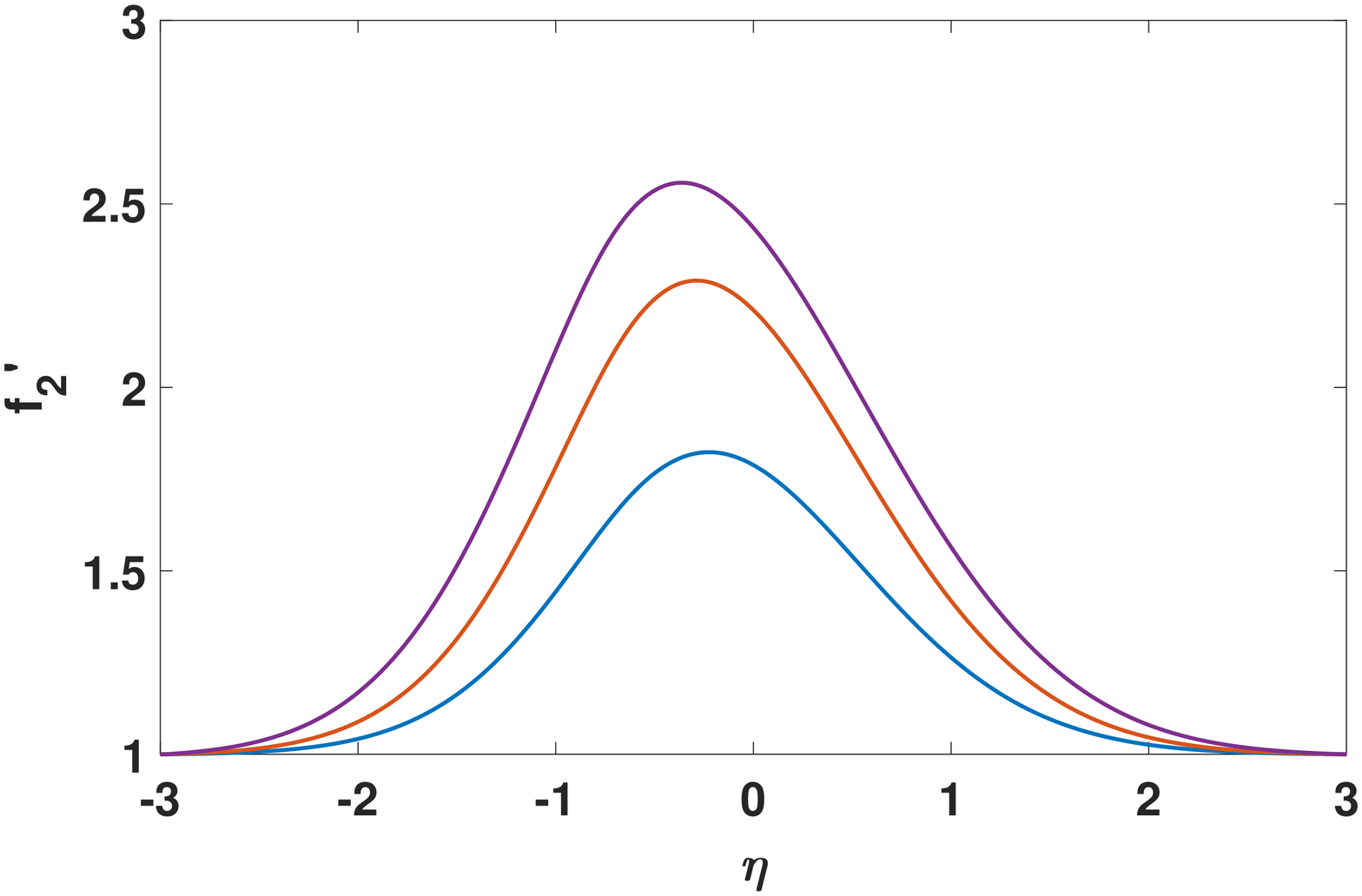}}
  \subfigure[velocity component, $u_{\chi}$]{
  \includegraphics[height = 4.8cm, width=0.45\linewidth]{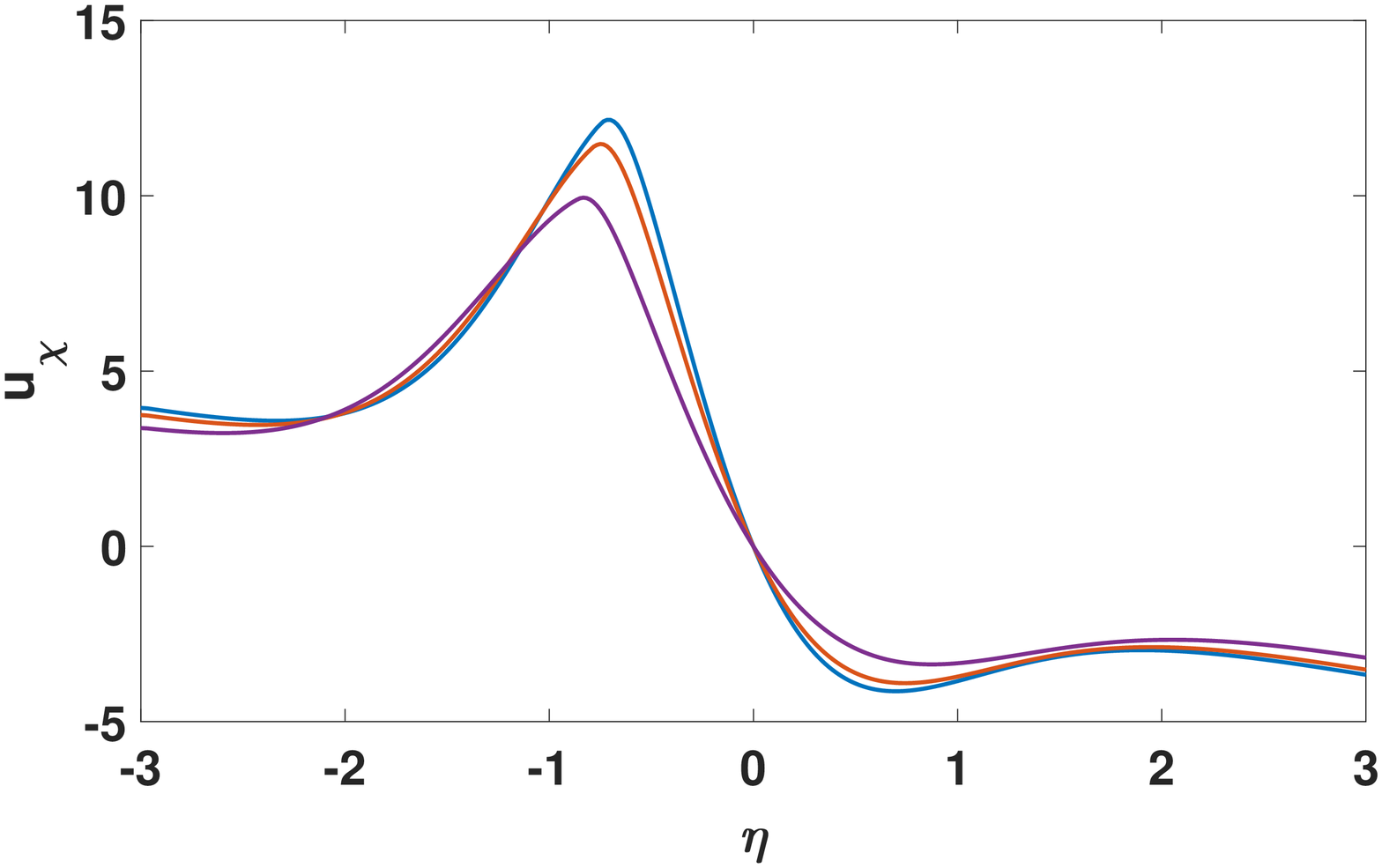}}     \\
  \vspace{0.2cm}
  \caption{Diffusion flame with varying strain rate.   $K= 0.195, \;  \omega_{\kappa} = 1.0$. blue $S_1 =0.750, \;  S_2 = 0.250$;   red $S_1 =0.500, \;  S_2 = 0.500$ ;   purple $S_1 =0.333, \;  S_2 = 0.667$  .   }
  \label{S1DiffFlame2}
\end{figure}

The  values of $S_1$ and $S_2$ do have some consequence on the behavior. In Figures \ref{S1DiffFlame1} and \ref{S1DiffFlame2}, $S_1$ varies between 0.750 and 0.333. $S_2 = 1 -S_1$ and varies accordingly.  As $S_1 $ decreases and $S_2$ increases, the flame zone moves slightly, the integrated burning rate decreases, and the normalized mass flux $f$ through the counterflow decreases. Very interestingly, as $S_1$ decreases, both the $u_{\chi}$ velocity component and $f_1'$ decrease while the $w$ velocity component and $f_2'$ increase.  $u_{\chi}$ also decreases.  As the $S_1$ value moves from 0.500 to 0.333, some reversal of the $u_{\xi}$ velocity occurs in the region of highest temperature and lowest density. In that region, there is inflow (compressive strain) in two directions with outflow (tensile strain) only in the $z$-direction. This implies that, for $S_1 =0.333$, an inflowing particle of material enters at first with decreasing magnitude of $\chi$ and increasing values of $\xi$ and $z$ in Lagrangian time. Then, it changes to decreasing values of both $\chi$ and $\xi$ in that time but remaining with an increasing value for $z$. Note that a case with $K =0.195, \;  \omega_{\kappa} = 1.0,$ and $ S_1 = 0.250$ is not plotted here. It resulted in extinction of the flame.

As seen from Figures \ref{WeakDiffFlame2} and \ref{S1DiffFlame2}, because of density gradients, velocity gradients  exist in the new rotating coordinate system such that vorticity components result in the $z$ and $\xi$  direction from $f_1''$ and $f_2''$ , respectively. However, those velocity components and thereby the associated vorticity components have an antisymmetry. Thus, the induced circulation around the flamelet due to density gradients is zero. Only the circulation due to the imposed vorticity $\omega_{\kappa}$ will exist. These findings are consistent with   the results of \cite{Sirignano2019b, Sirignano2019a}.

Clearly, the combination of fluid rotation, variable-density, and three-dimensional structure have major consequences for flamelet behavior.

\newpage

\subsection{Uncoupled Premixed Flame Calculations}



The premixed flame is treated with consideration of a mixture that is 12 per cent propane by mass and 43.5 per cent oxygen by mass (stoichiometric proportion), with the remainder an inert gas.   This flame structure requires a much larger Damk\"{o}hler number to exist when compared to the diffusion flame. It extinguishes under high strain rate (i.e., low residence time in the counterflow).

\begin{figure}
  \centering
 \subfigure[enthalpy, $h/h_{\infty}$ ]{
  \includegraphics[height = 4.6cm, width=0.45\linewidth]{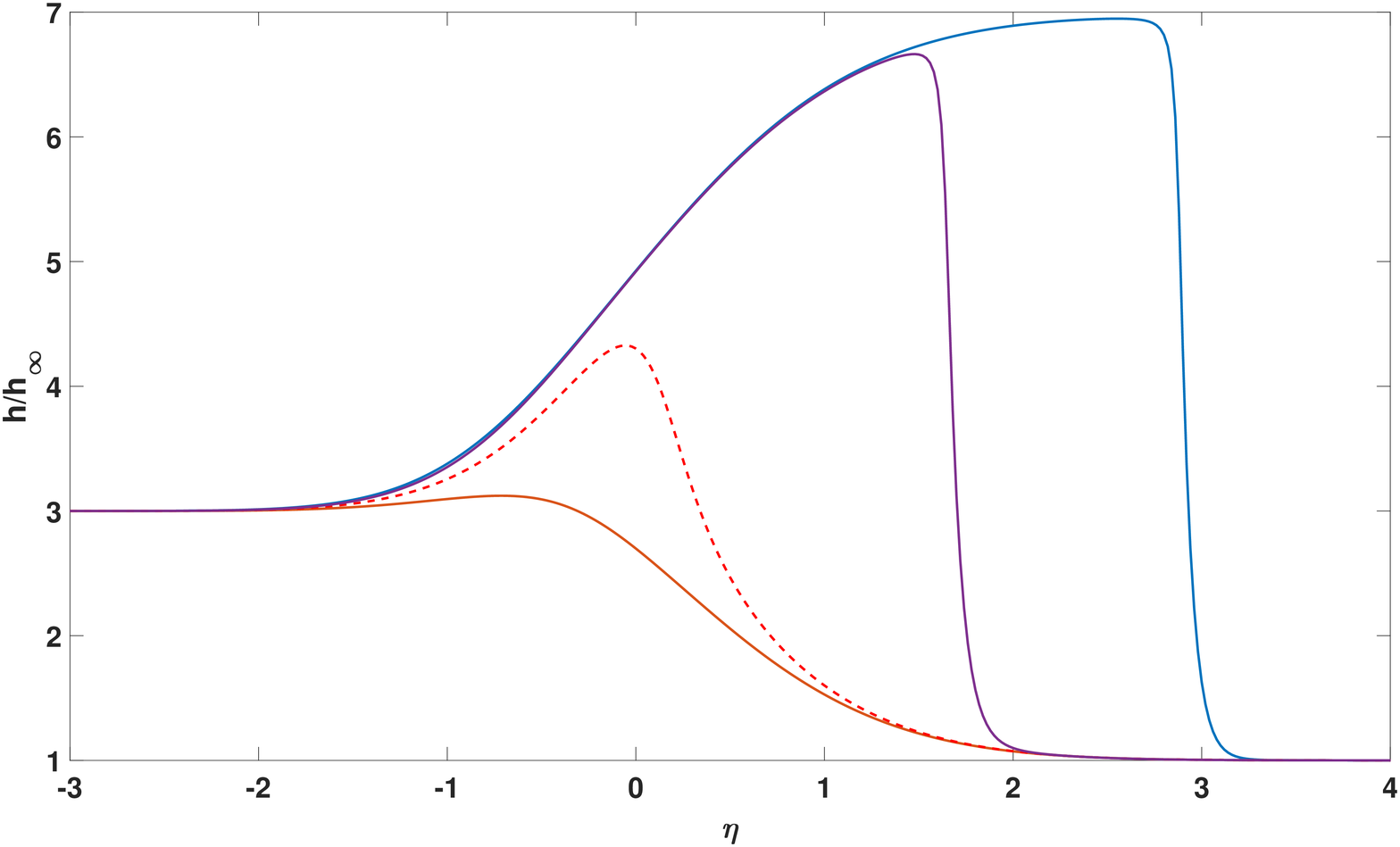}}
  \subfigure[fuel mass fraction, $Y_F$]{
  \includegraphics[height = 4.6cm, width=0.45\linewidth]{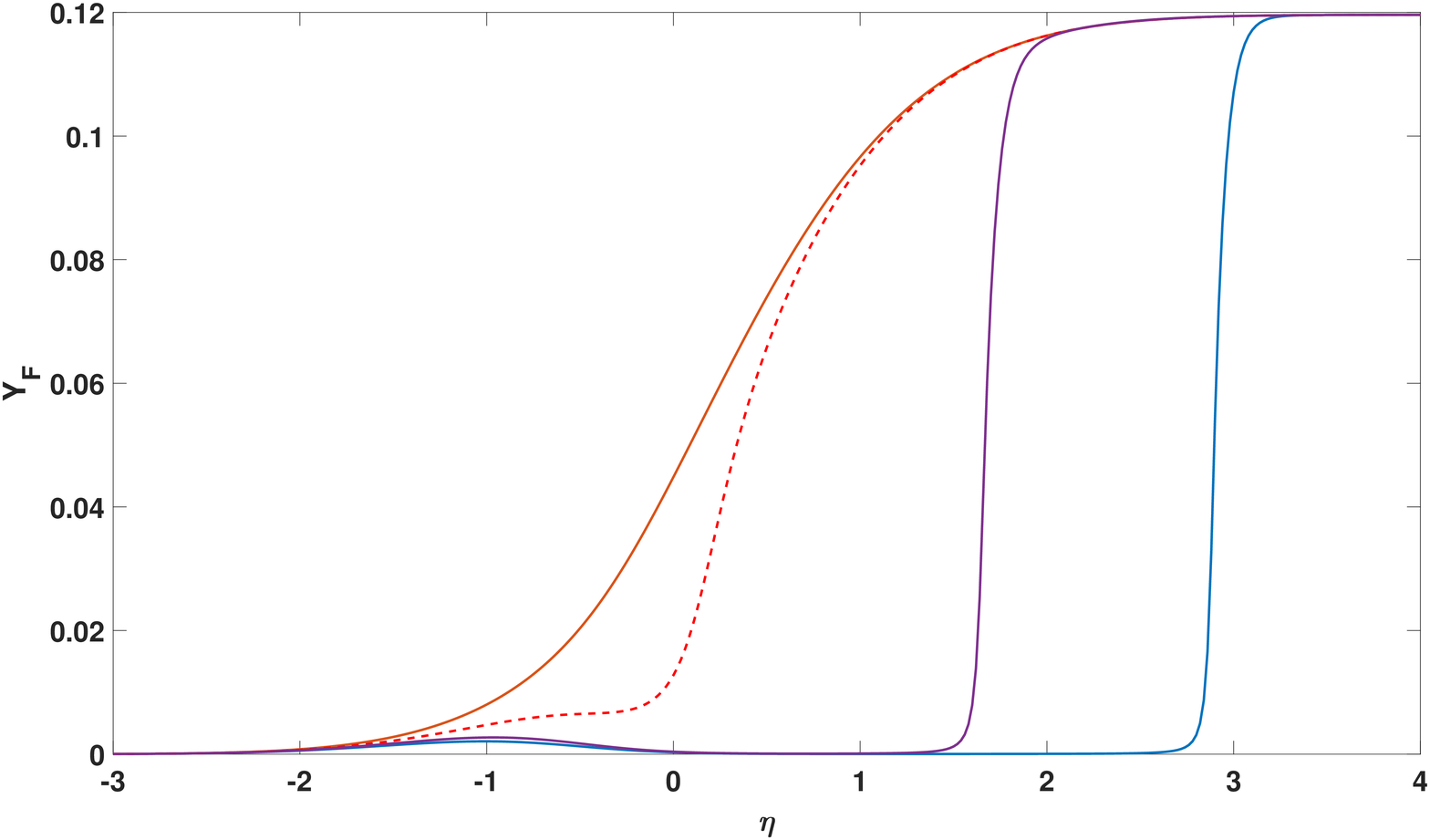}}     \\
  \vspace{0.2cm}
  \subfigure[ mass ratio x oxygen mass fraction, $\nu Y_O$]{
  \includegraphics[height = 4.6cm, width=0.45\linewidth]{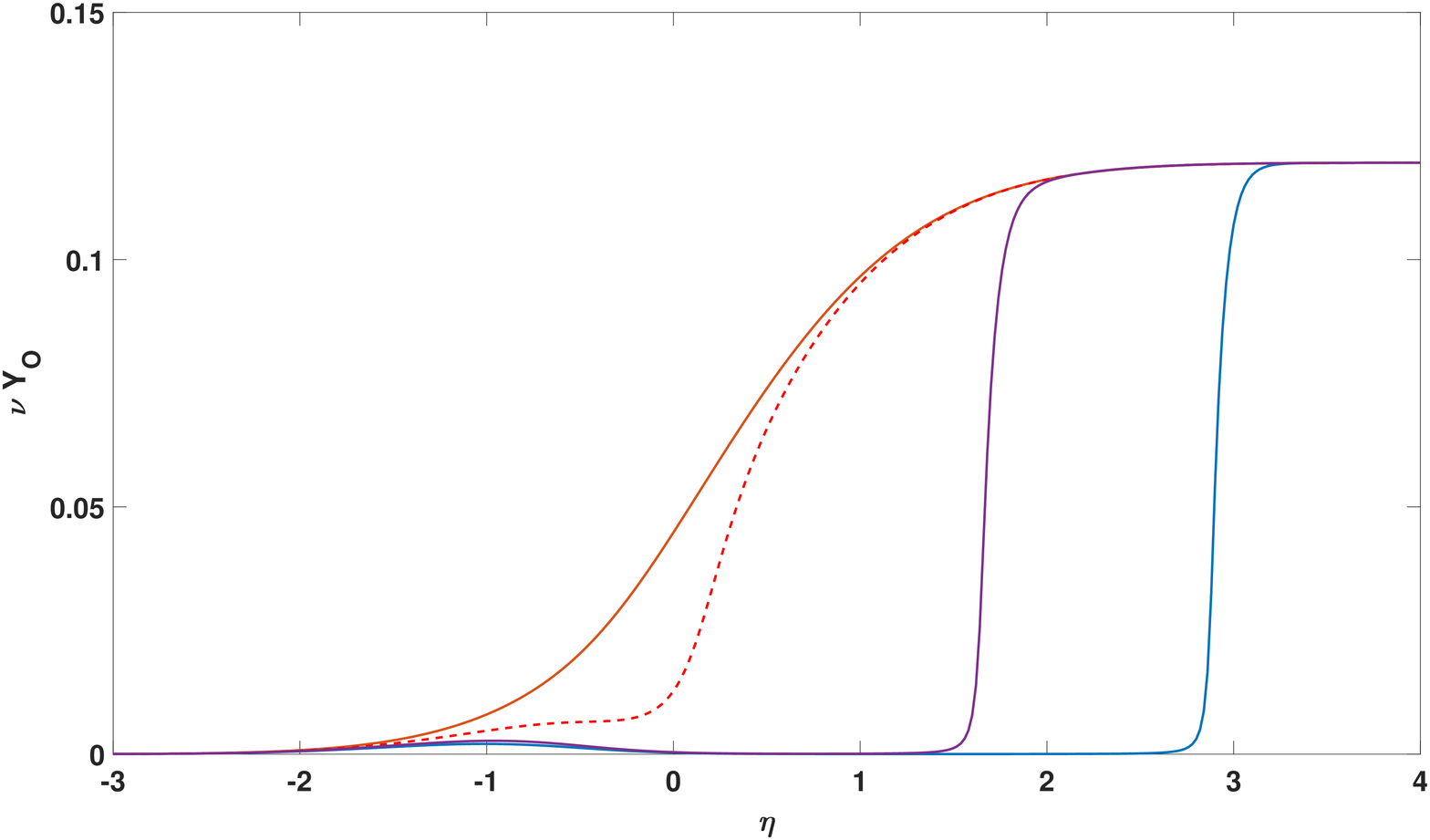}}
  \subfigure[integral of reaction rate, $\int \dot{\omega}_F d \eta$]{
  \includegraphics[height = 4.6cm, width=0.45\linewidth]{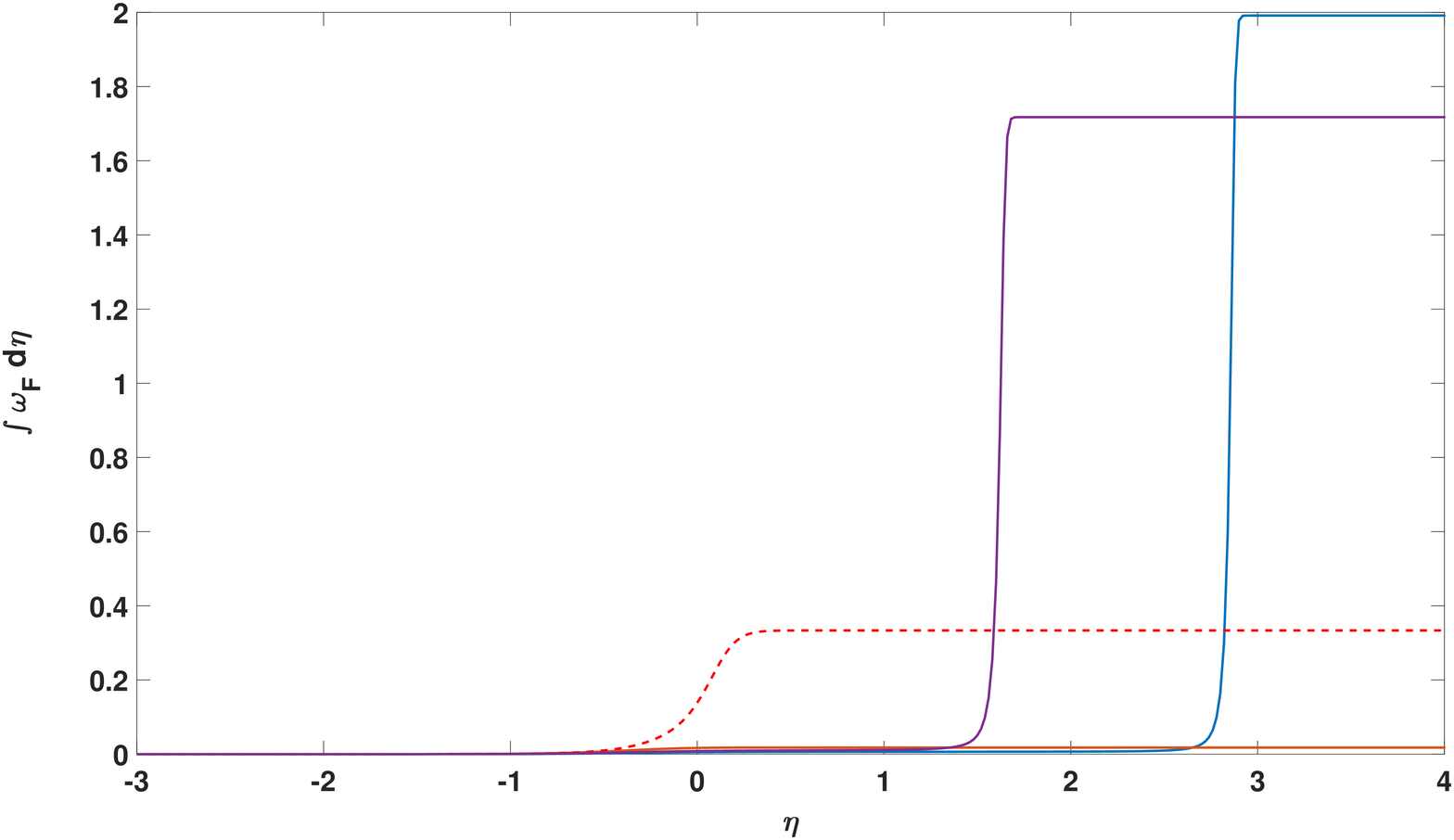}}     \\
  \vspace{0.2cm}
  \subfigure[reaction rate, $\dot{\omega}_F $]{
  \includegraphics[height = 4.6cm, width=0.45\linewidth]{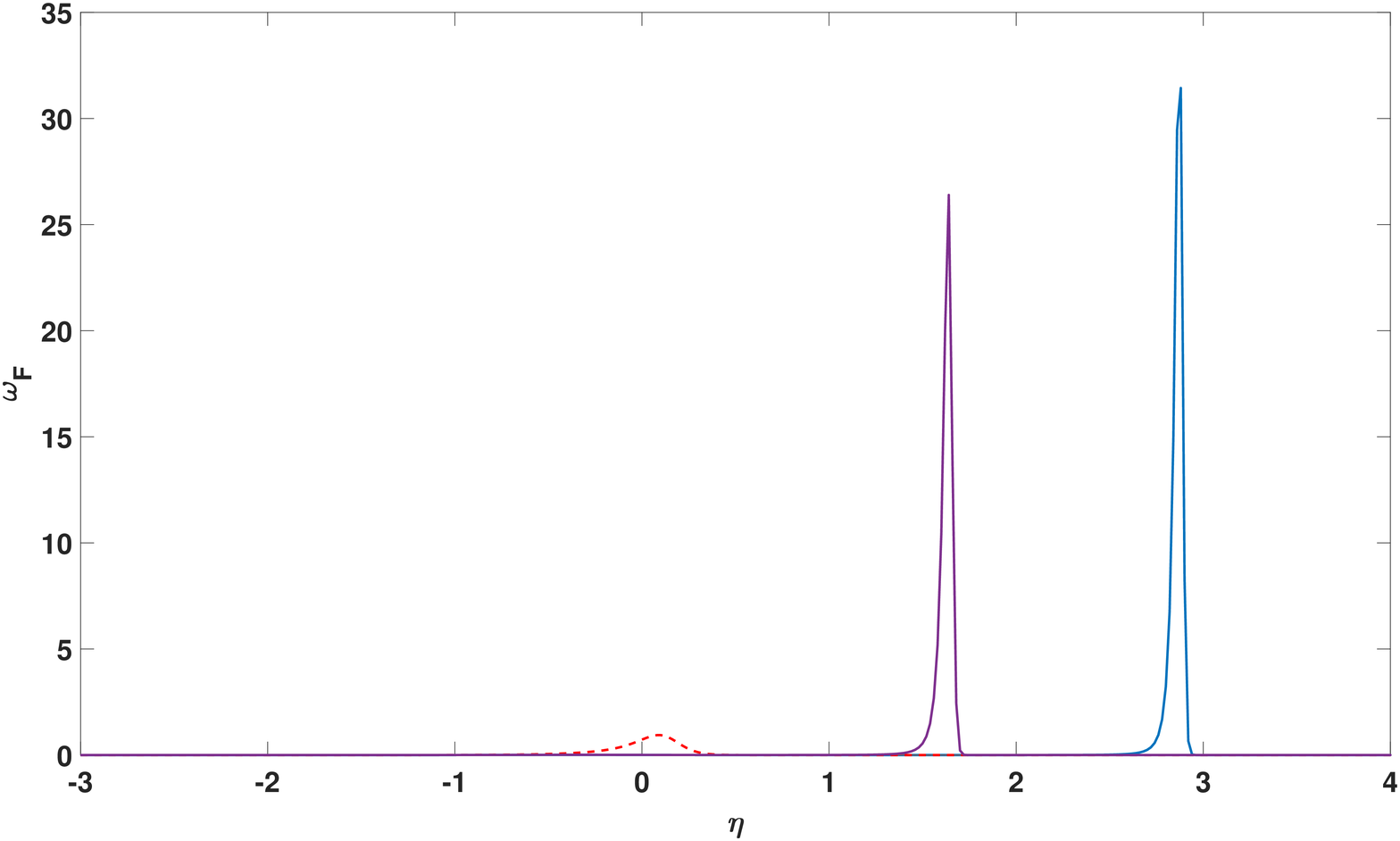}}
  \vspace{-0.1cm}
  \caption{Premixed flame: influence of  Damk\"{o}hler number.  $S_1 = 0.750 , \; S_2 = 0.250, \; \omega_{\kappa} = 1.0$. $K= 235$, solid red; $K =238$, dash red; $K =239, $ purple; $K=240$, blue.           }
  \label{PremixedFlame1}
\end{figure}

\begin{figure}
  \centering
 \subfigure[mass flux per area, $f= \rho u_{\chi}$ ]{
  \includegraphics[height = 4.8cm, width=0.45\linewidth]{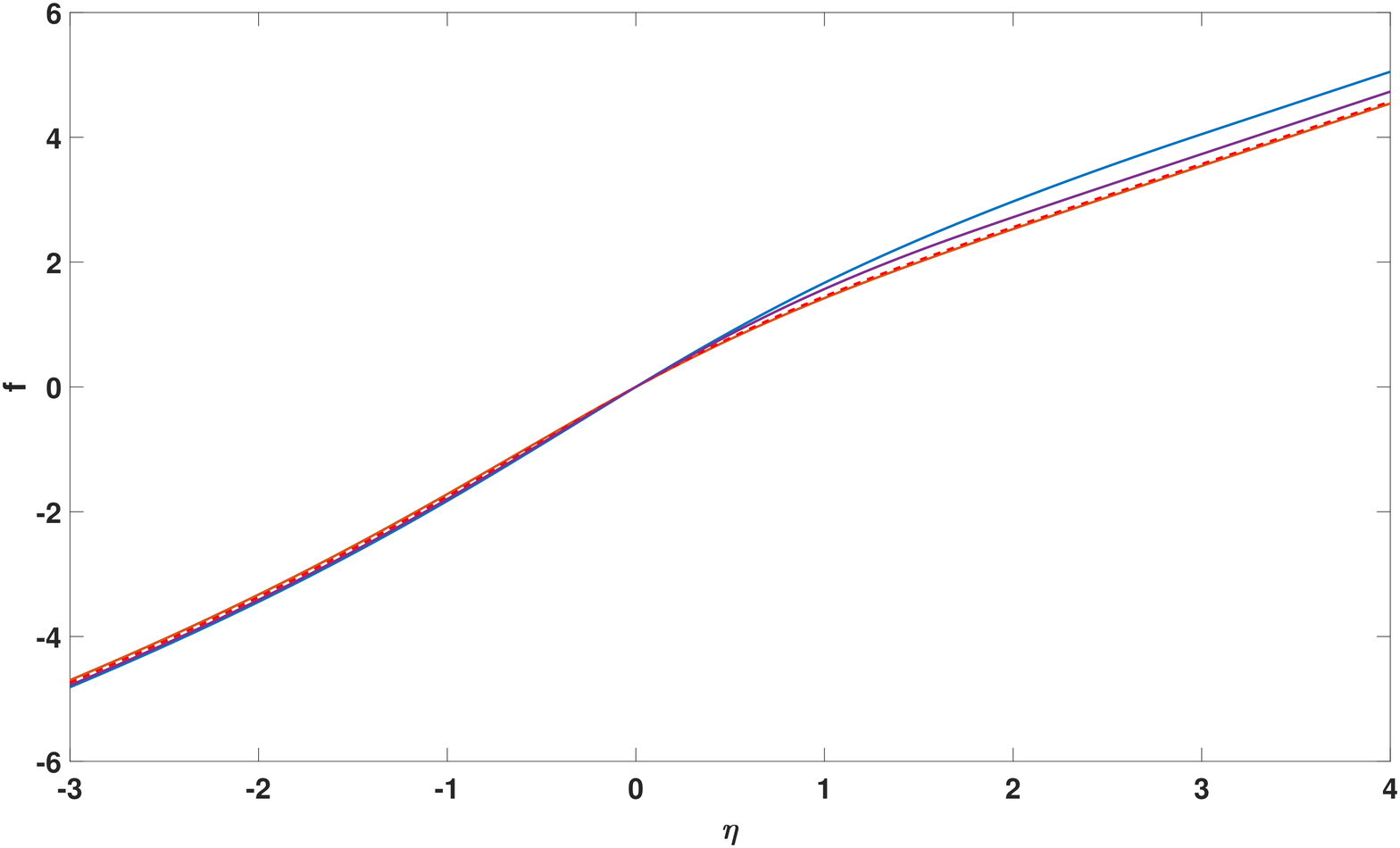}}
  \subfigure[velocity component, $f_1' = u_{\xi}/(S_1 \xi)$]{
  \includegraphics[height = 4.8cm, width=0.45\linewidth]{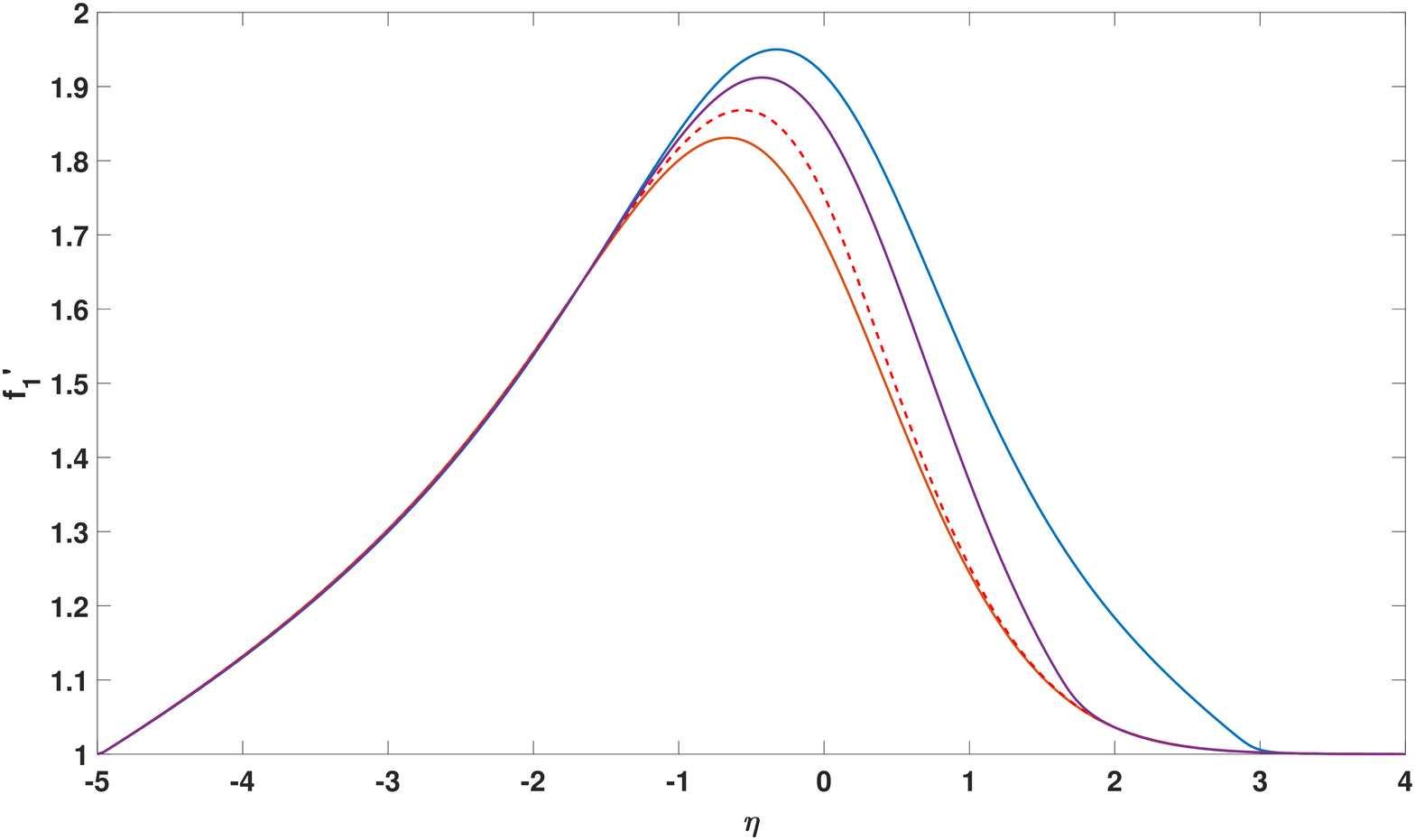}}     \\
  \vspace{0.2cm}
  \subfigure[ velocity component, $f_2' = w/(S_2 z)$]{
  \includegraphics[height = 4.8cm, width=0.45\linewidth]{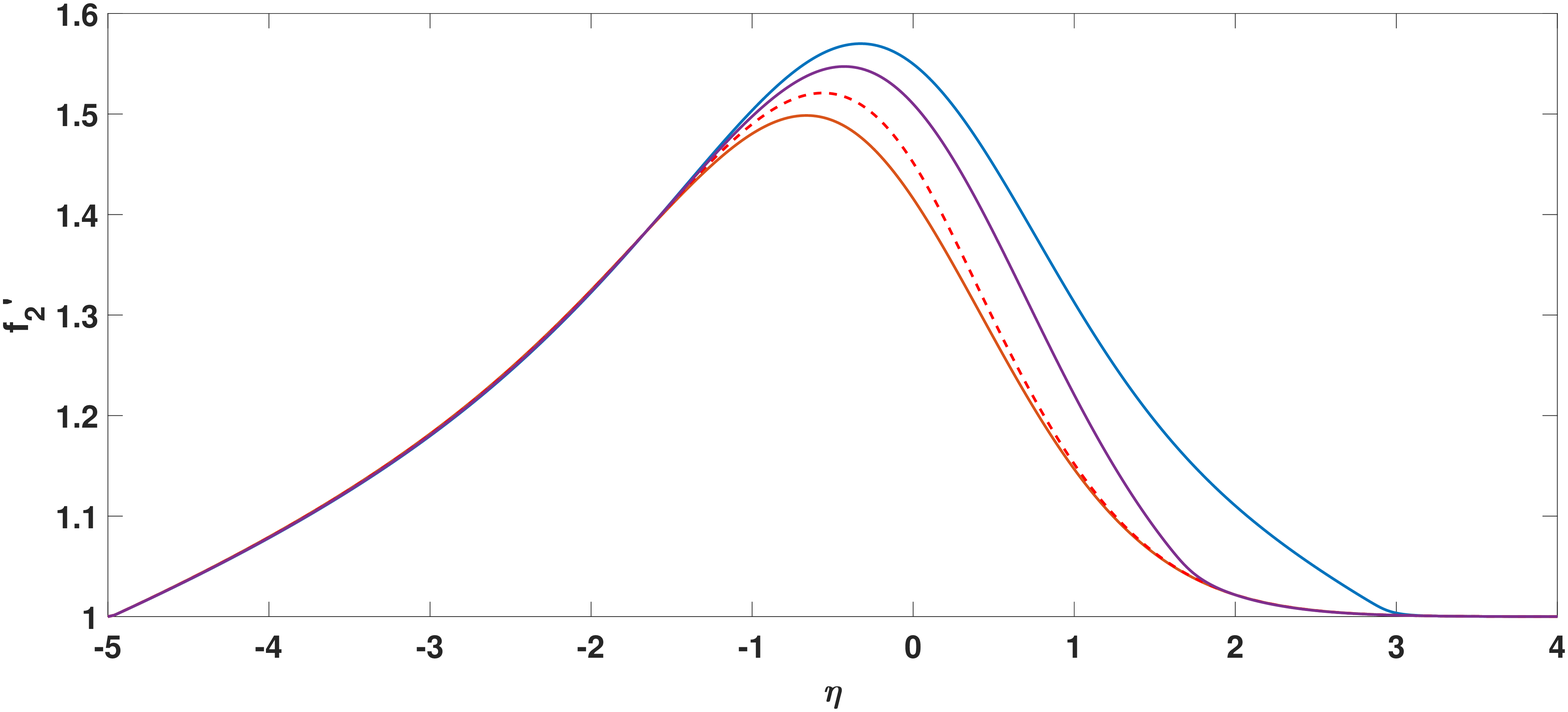}}
  \subfigure[velocity component, $u_{\chi}$]{
  \includegraphics[height = 4.8cm, width=0.45\linewidth]{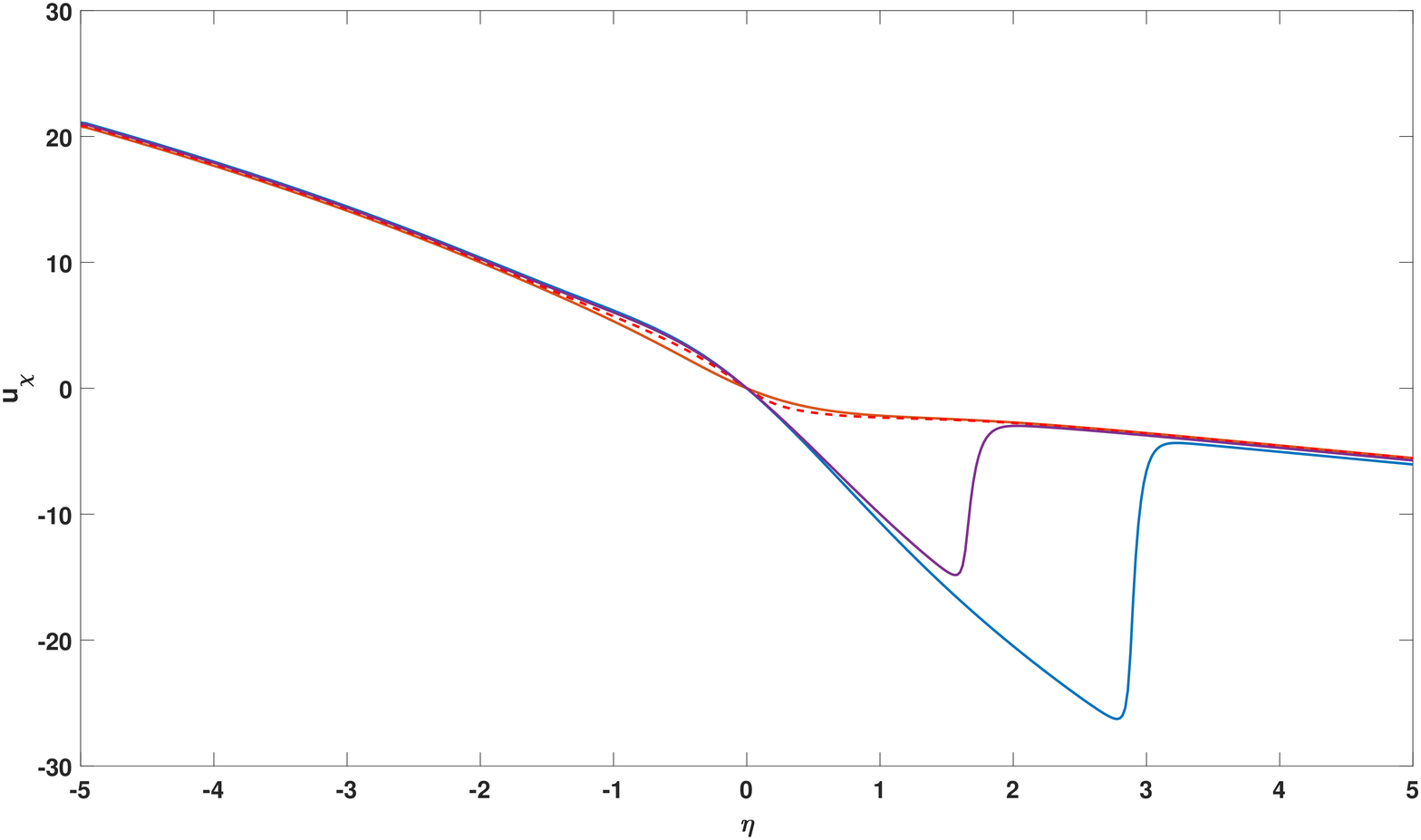}}     \\
  \vspace{0.2cm}
  \caption{Premixed flame: influence of Damk\"{o}hler number.   $S_1 = 0.750 ,\; S_2 = 0.250, \;\omega_{\kappa} = 1.0$. $K= 235$, solid red; $K =238$, dash red ; $K =239$,   purple; $K=240$, blue.            }
  \label{PremixedFlame2}
\end{figure}

Figures \ref{PremixedFlame1}  shows that small changes for the Damk\"{o}hler number $Da$ (i.e., in the third significant digit and reflected through the coefficient  $K$)    can result in large differences in peak enthalpy and temperature, flame velocity, and burning rate. As $Da$ increases, the flame shifts to a position in the counterflow with higher incoming velocity and mass flux and the flame thickness narrows. All three components of  the velocity at each position in the counterflow actually increase as  $Da $ increases; see Figure  \ref{PremixedFlame2}. It can be viewed as an increase in the chemical reaction rate leading to an increase in flame speed. At the margin, the flame extinguishes as the reaction rate decreases because in the spatially varying velocity field, it cannot obtain the needed residence time. The greater energy release rate causes a higher temperature and a lower density in the flame region. The higher velocities result from the combination of an increase in mass burning rate  and a decrease in density. The $u_{\xi}$ velocity component is substantially higher then the $w$ component.

\begin{figure}
  \centering
 \subfigure[enthalpy, $h/h_{\infty}$ ]{
  \includegraphics[height = 4.6cm, width=0.45\linewidth]{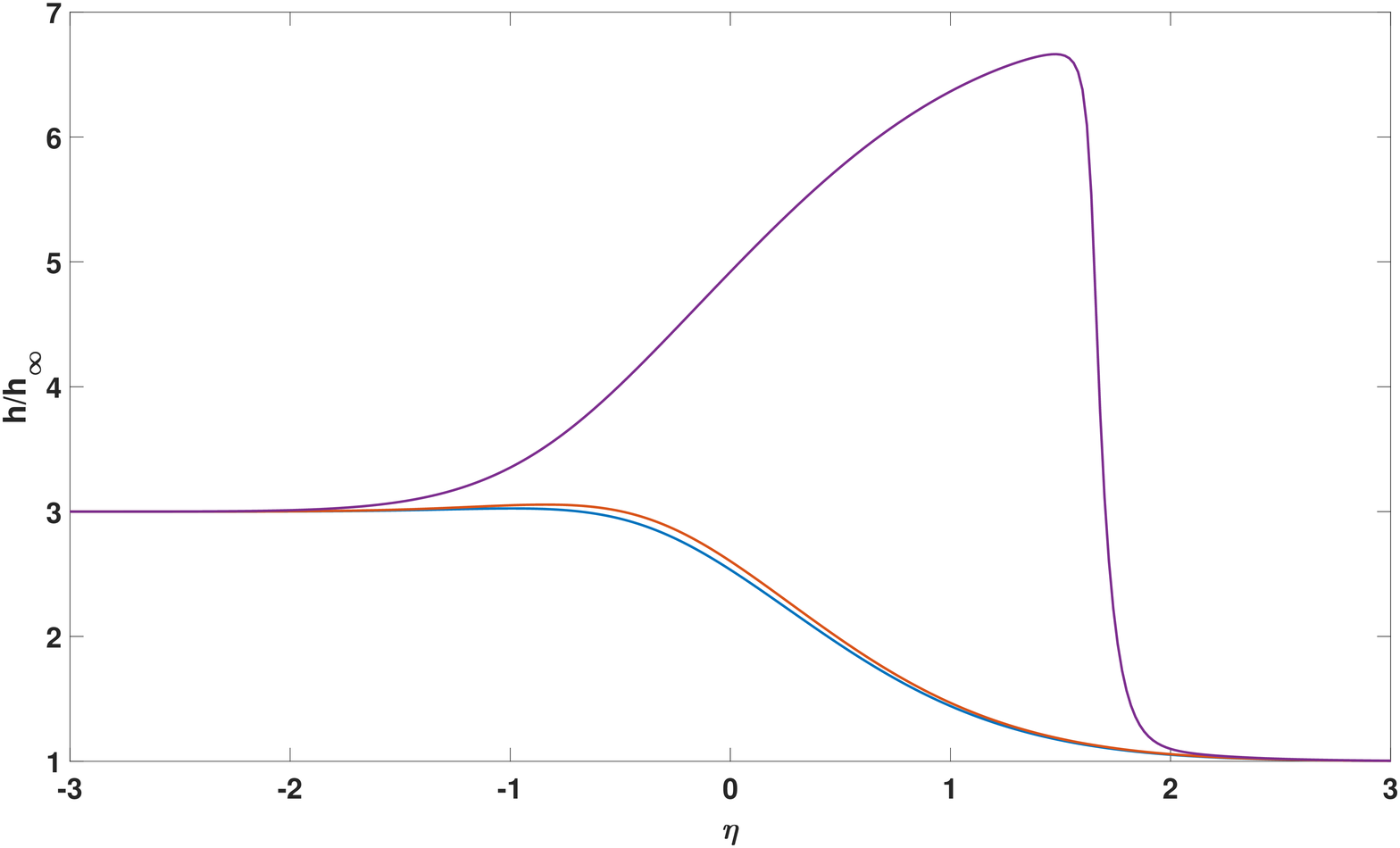}}
  \subfigure[fuel mass fraction, $Y_F$]{
  \includegraphics[height = 4.6cm, width=0.45\linewidth]{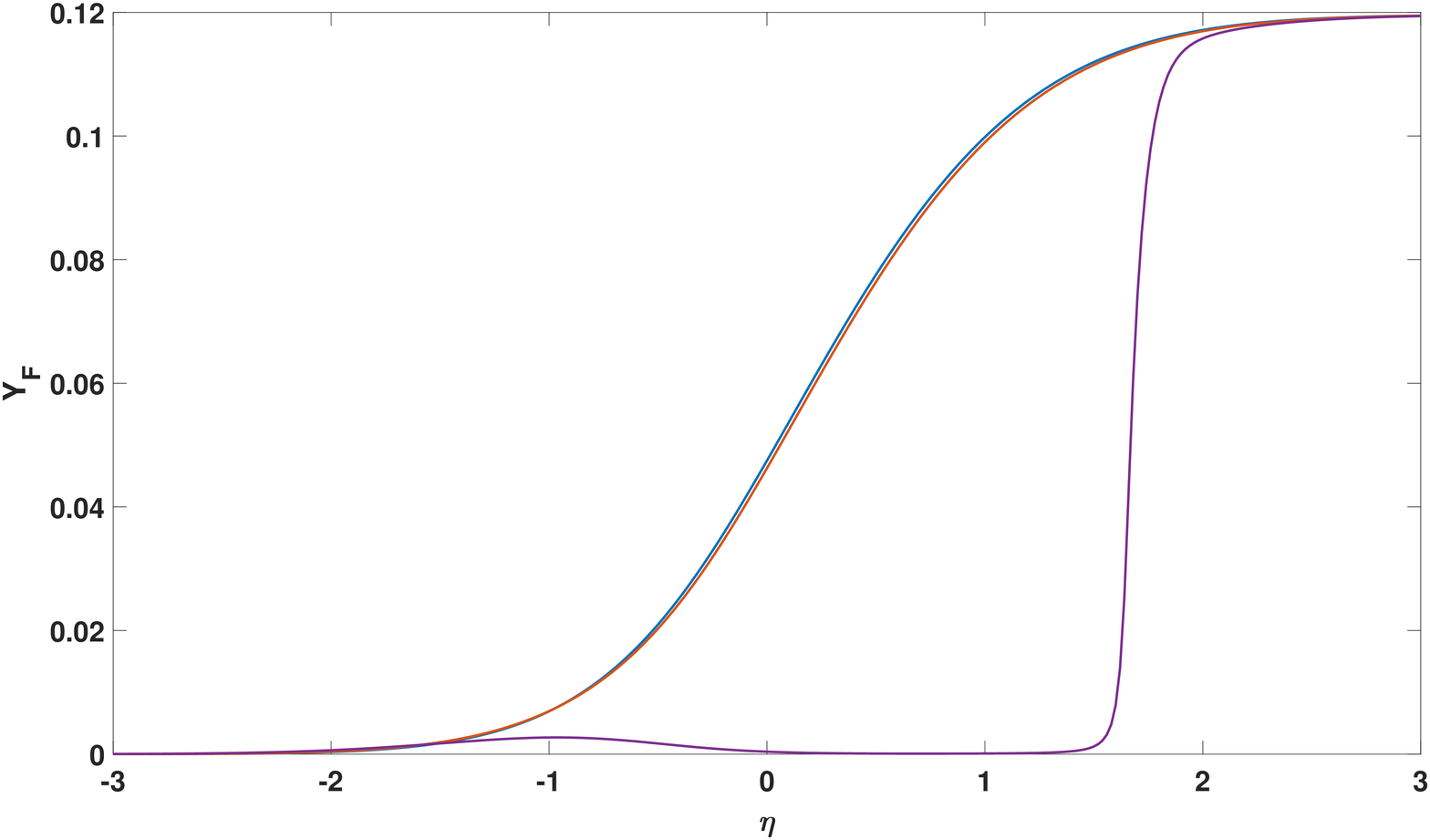}}     \\
  \vspace{0.2cm}
  \subfigure[ mass ratio x oxygen mass fraction, $\nu Y_O$]{
  \includegraphics[height = 4.6cm, width=0.45\linewidth]{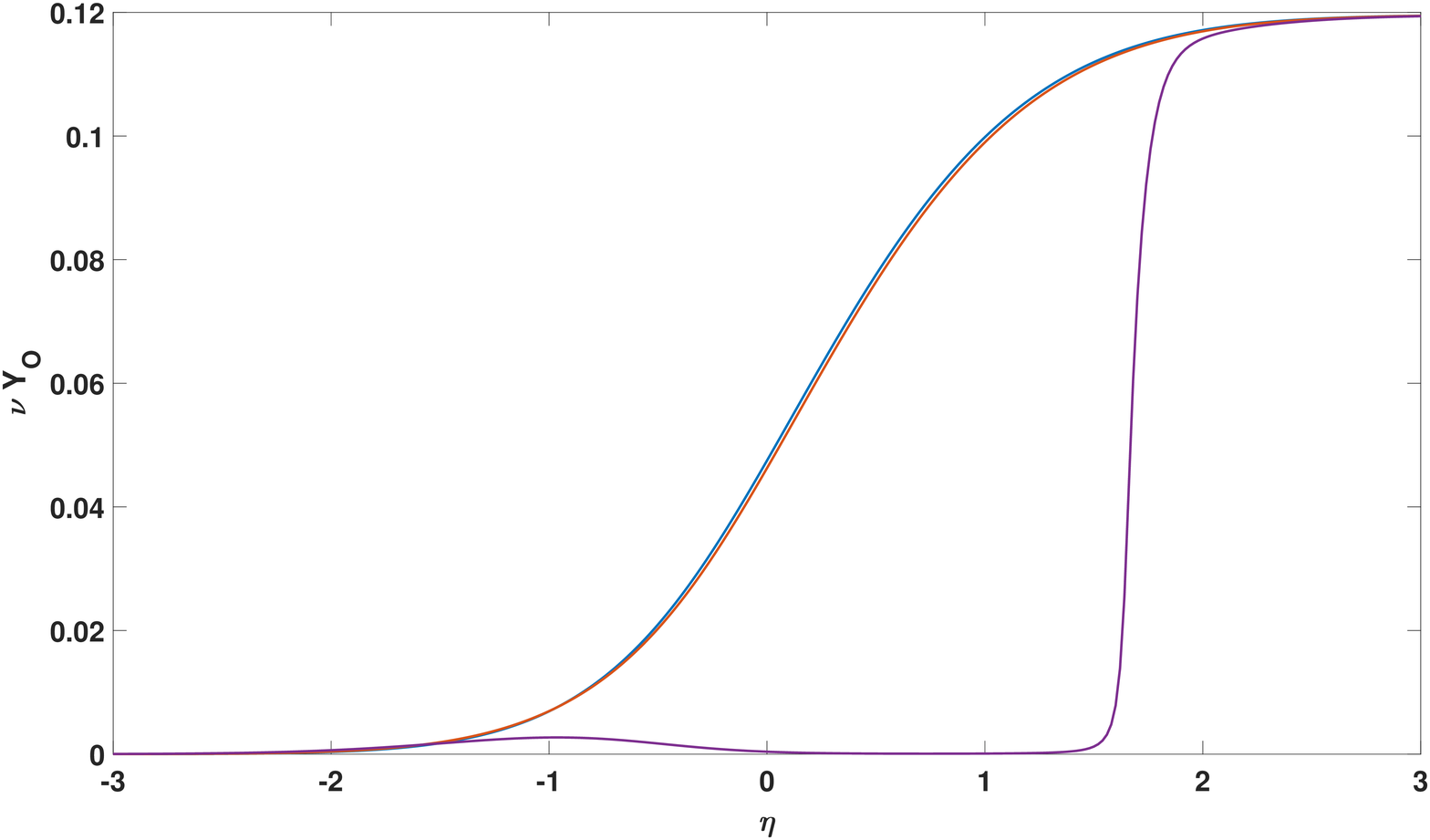}}
  \subfigure[integral of reaction rate, $\int \dot{\omega}_F d \eta$]{
  \includegraphics[height = 4.6cm, width=0.45\linewidth]{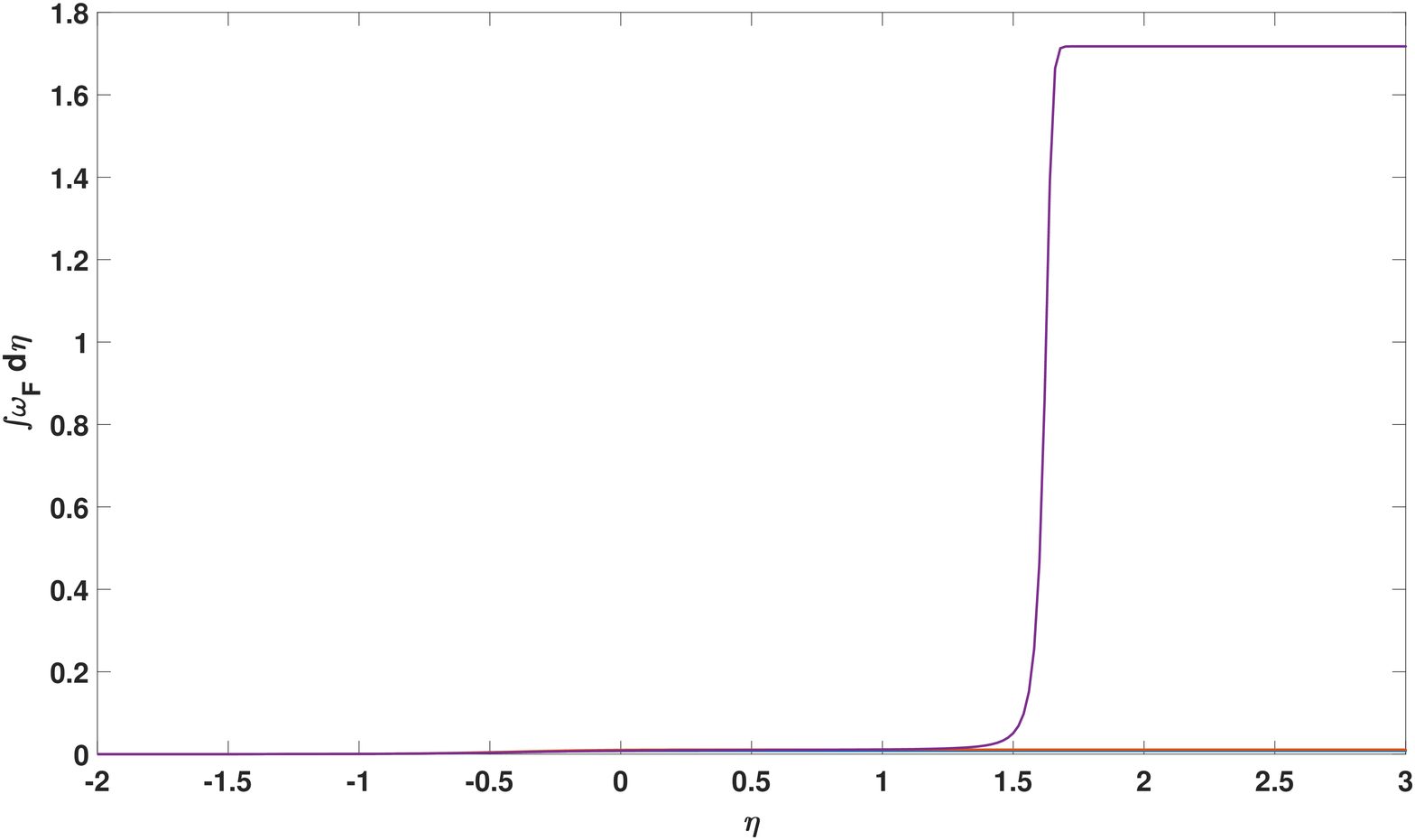}}     \\
  \vspace{0.2cm}
  \subfigure[reaction rate, $\dot{\omega}_F $]{
  \includegraphics[height = 4.6cm, width=0.45\linewidth]{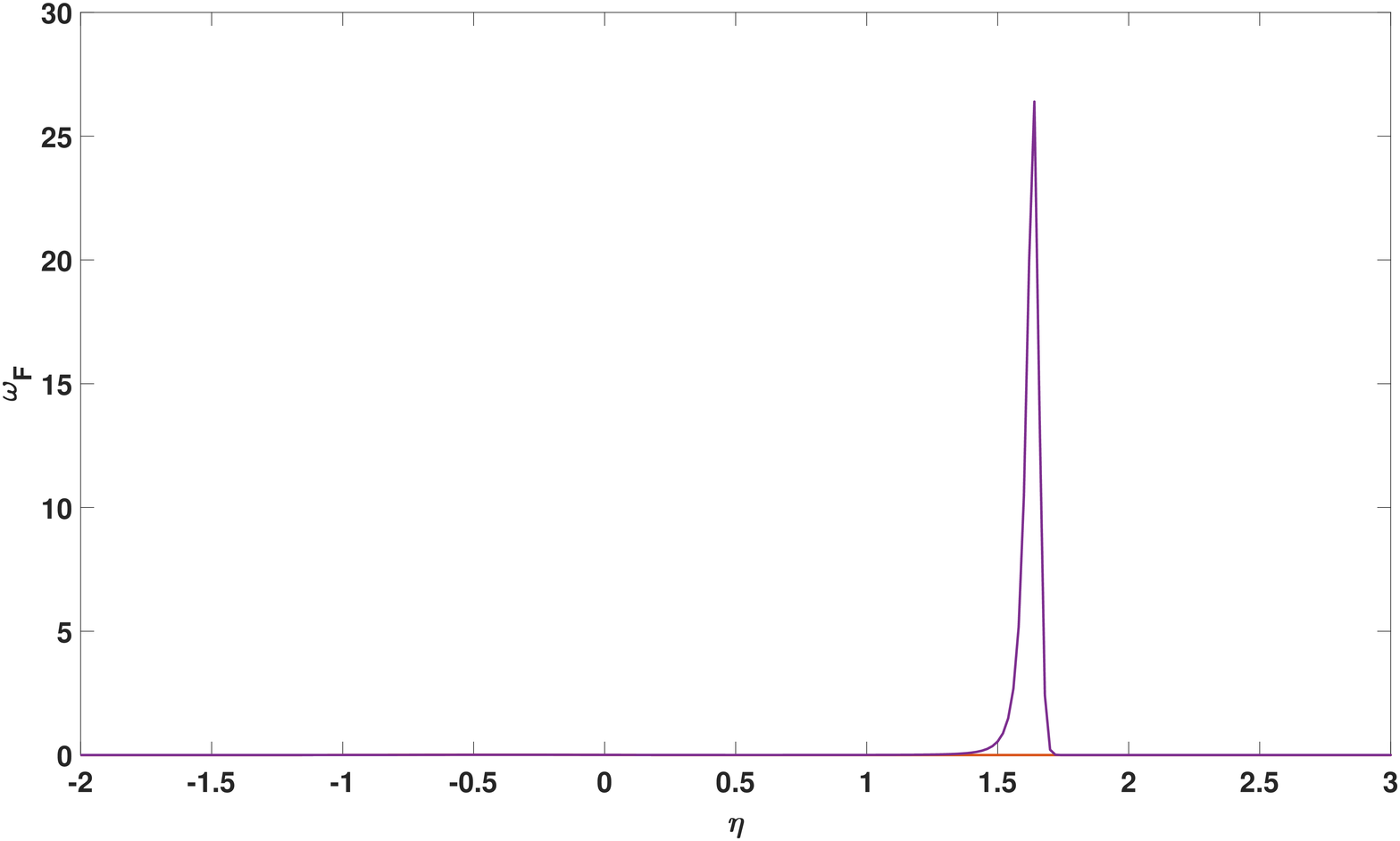}}
  \vspace{-0.1cm}
  \caption{Premixed flame: influence of vorticity.   $K= 239, S_1 =0.750, S_2 = 0.250$.  $\omega_{\kappa} = 0$, blue ; $\omega_{\kappa} = 0.5$,  red  ;   $\omega_{\kappa} = 1.0$,  purple.              }
  \label{PremixedFlame3}
\end{figure}

\begin{figure}
  \centering
 \subfigure[mass flux per area, $f= \rho u_{\chi}$ ]{
  \includegraphics[height = 4.8cm, width=0.45\linewidth]{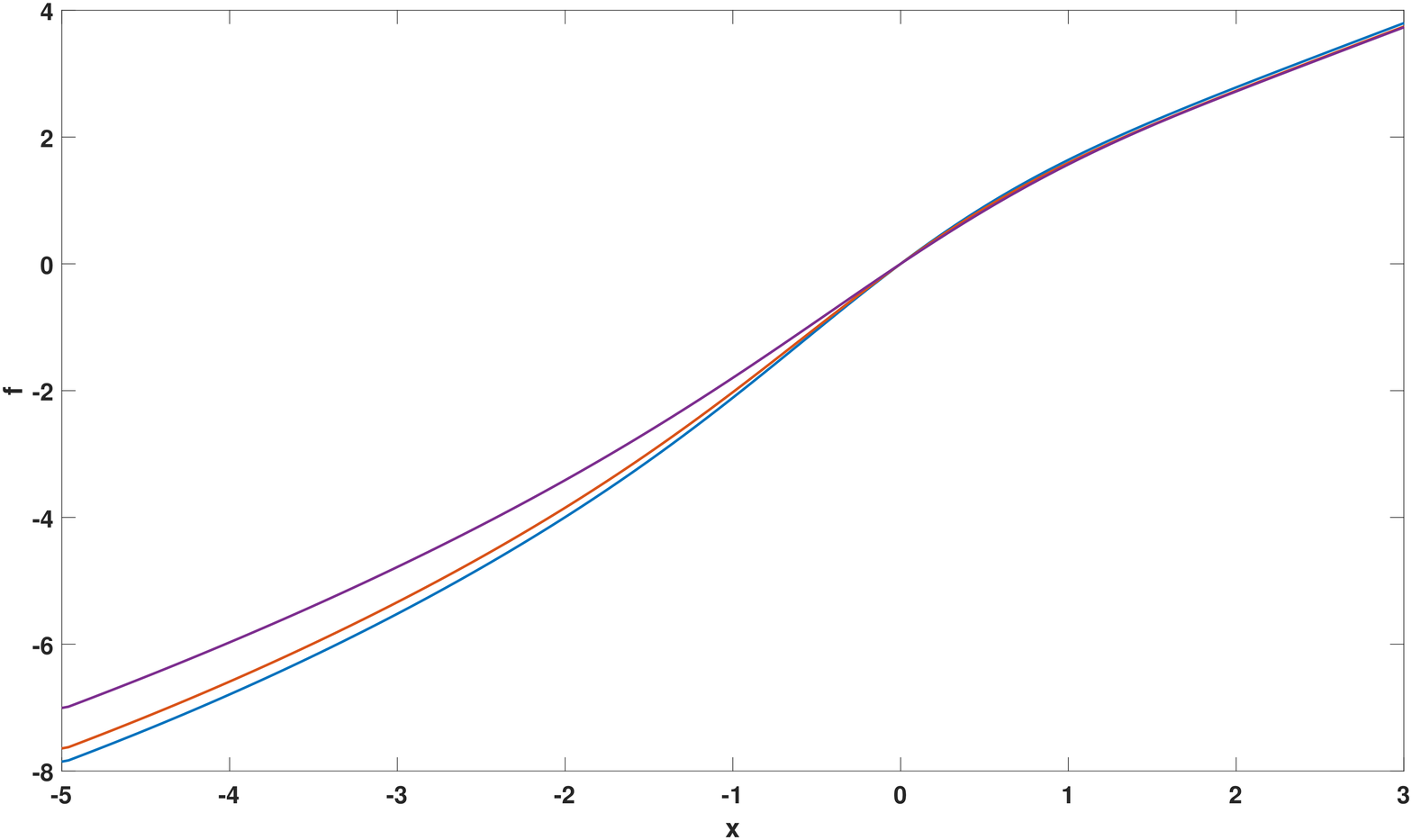}}
  \subfigure[velocity component, $f_1' = u_{\xi}/(S_1 \xi)$]{
  \includegraphics[height = 4.8cm, width=0.45\linewidth]{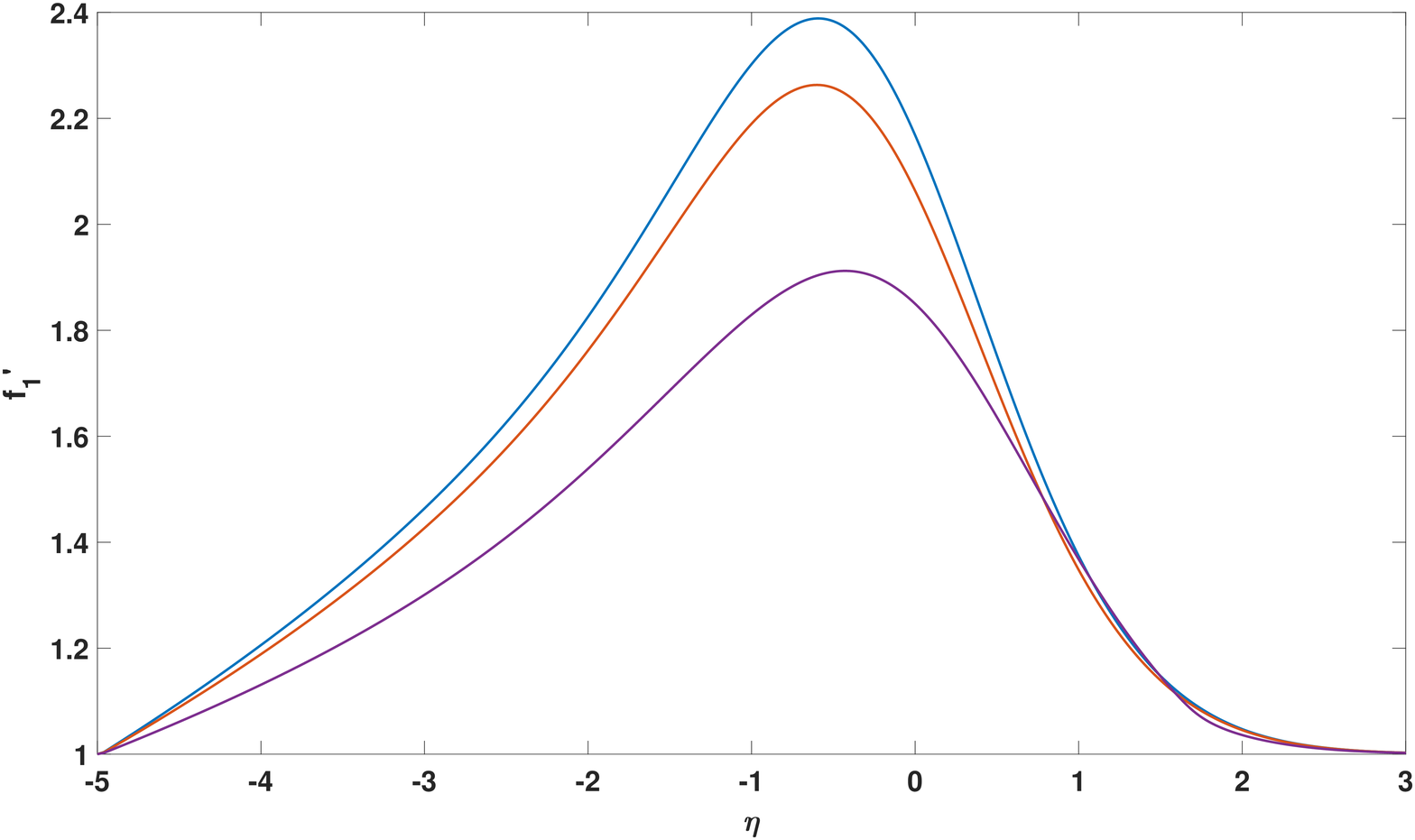}}     \\
  \vspace{0.2cm}
  \subfigure[ velocity component, $f_2' = w/(S_2 z)$]{
  \includegraphics[height = 4.8cm, width=0.45\linewidth]{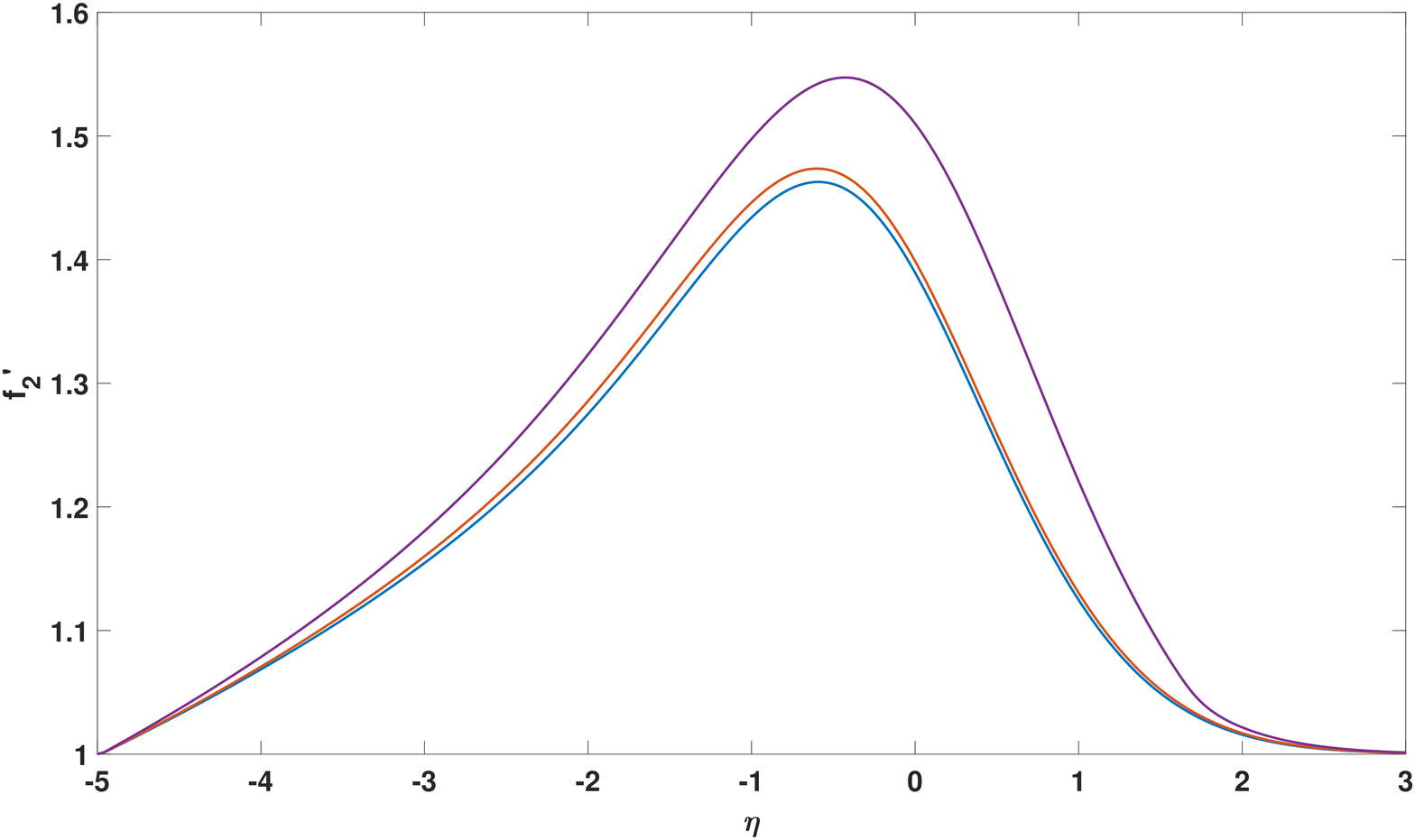}}
  \subfigure[velocity component, $u_{\chi}$]{
  \includegraphics[height = 4.8cm, width=0.45\linewidth]{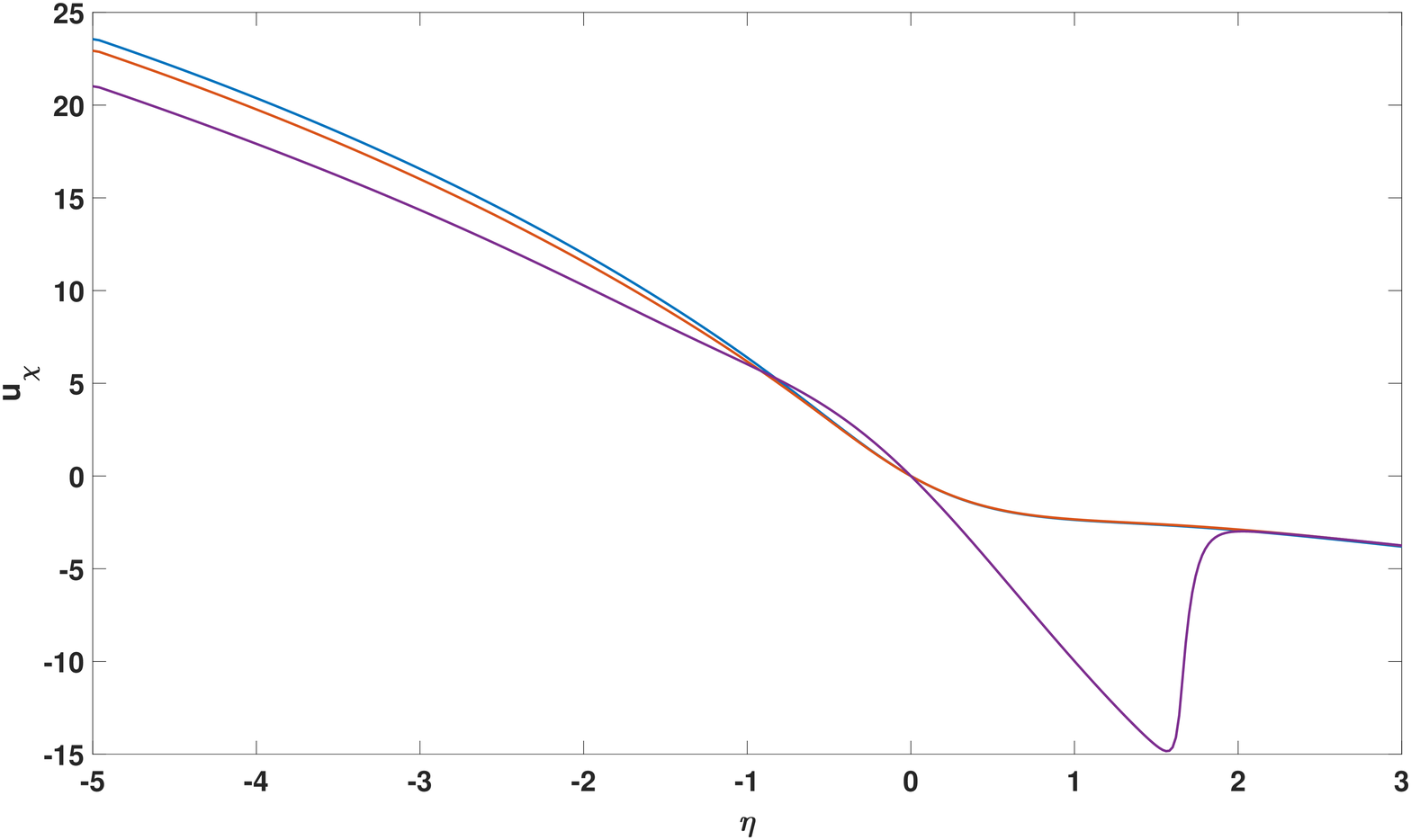}}     \\
  \vspace{0.2cm}
  \caption{Premixed flame influence of vorticity.   $K= 239, S_1 =0.750, S_2 = 0.250$.  $\omega_{\kappa} = 0$, blue ; $\omega_{\kappa} = 0.5$,  red  ;   $\omega_{\kappa} = 1.0$,  purple.         }
  \label{PremixedFlame4}
\end{figure}

In Figures   \ref{PremixedFlame1} and  \ref{PremixedFlame2}, the rate of strain  and vorticity values are held constant at $S_1 =0.750, S_2 = 0.250 ,$ and $\omega_{\kappa} = 1.0$ while $Da$ is varied. In Figures   \ref{PremixedFlame3} and  \ref{PremixedFlame4}, the same values of $S_1$ and $S_2$ apply  and $Da =239$ but $\omega_{\kappa}$ varies between $0$ and $1$. At lower values of the imposed vorticity, the flame extinguishes. Increased vorticity strengthens the centrifugal effect, increasing the efflux of the counterflow in the $z$-direction aligned with the vorticity vector.


The modification of the strain rate parameters $S_1$ and $S_2$  were studied with $\omega_{\kappa} = 1.0$ and $K = 239$.  $S_1 =0.500$ produced a higher peak temperature  and burning rate than was obtained with $S_1 = 0.750$. As  $S_1$ decreased, the $w$ component of velocity and $f_2'$ increased while $u_{\xi}$ and $f_1'$ decreased.

\newpage

\subsection{Uncoupled Multibranched Flamelet Calculations}\label{MultiFlame}

\begin{figure}
  \centering
 \subfigure[enthalpy, $h/h_{\infty}$ ]{
  \includegraphics[height = 4.6cm, width=0.45\linewidth]{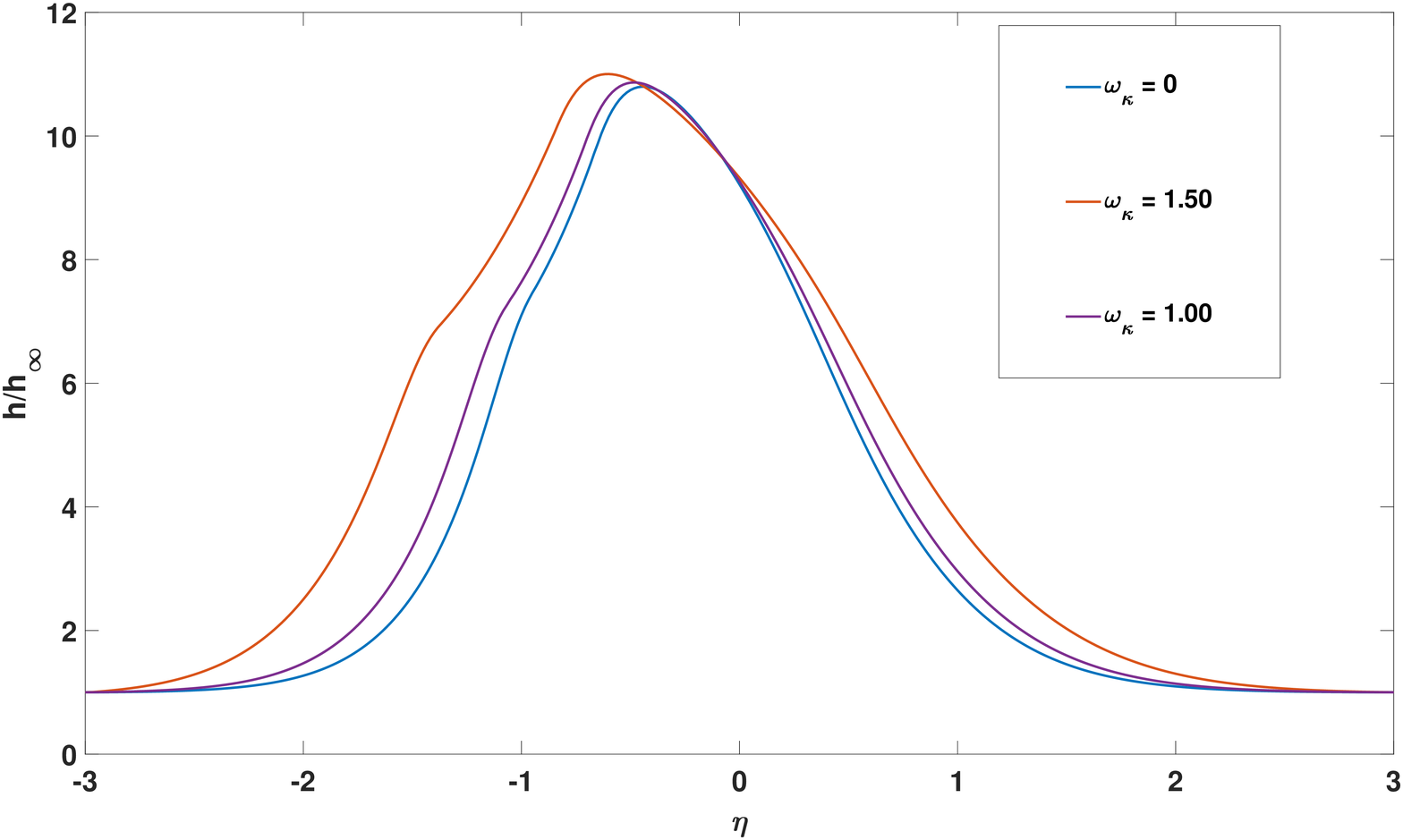}}
  \subfigure[fuel mass fraction, $Y_F$]{
  \includegraphics[height = 4.6cm, width=0.45\linewidth]{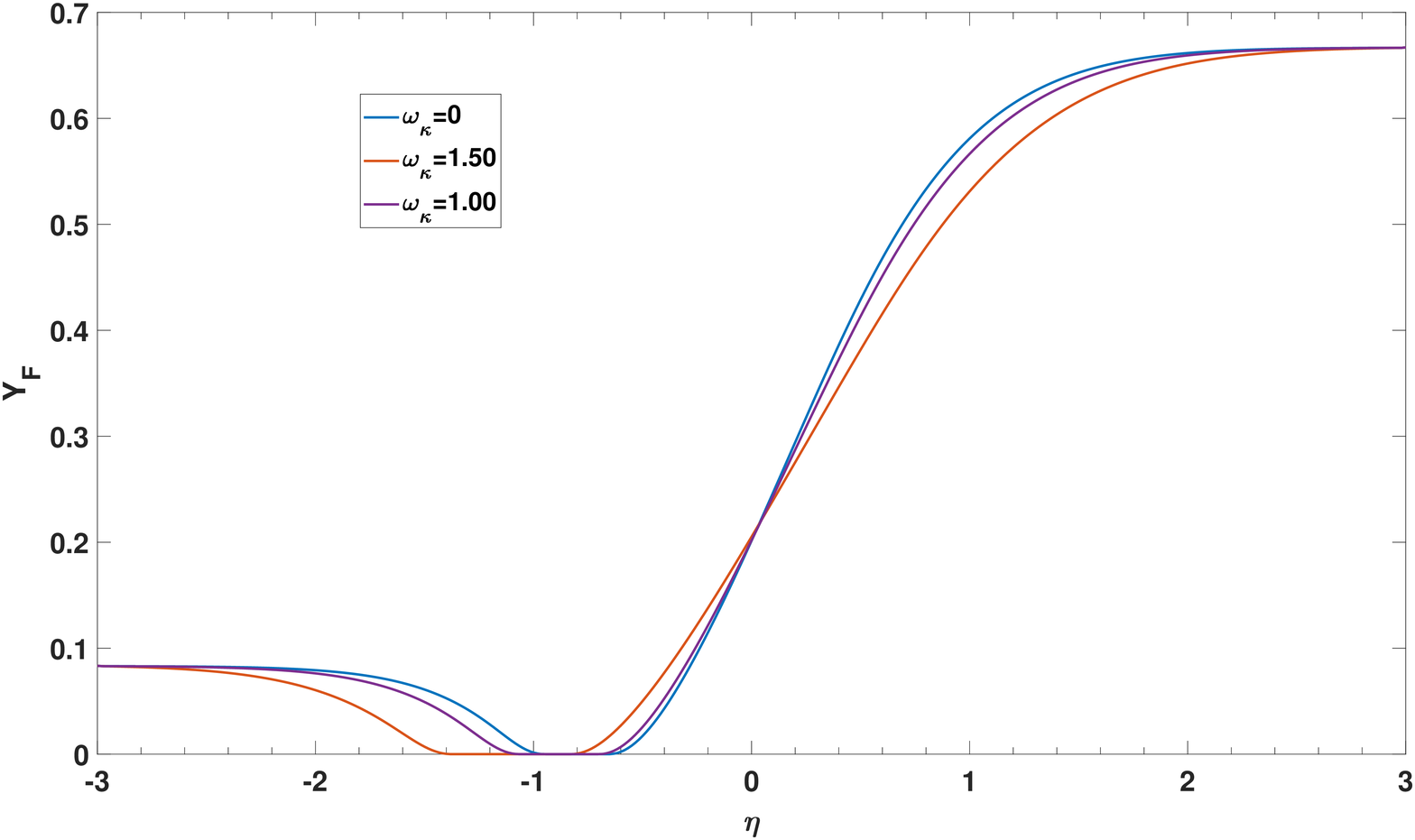}}     \\
  \vspace{0.2cm}
  \subfigure[ mass ratio x oxygen mass fraction, $\nu Y_O$]{
  \includegraphics[height = 4.6cm, width=0.45\linewidth]{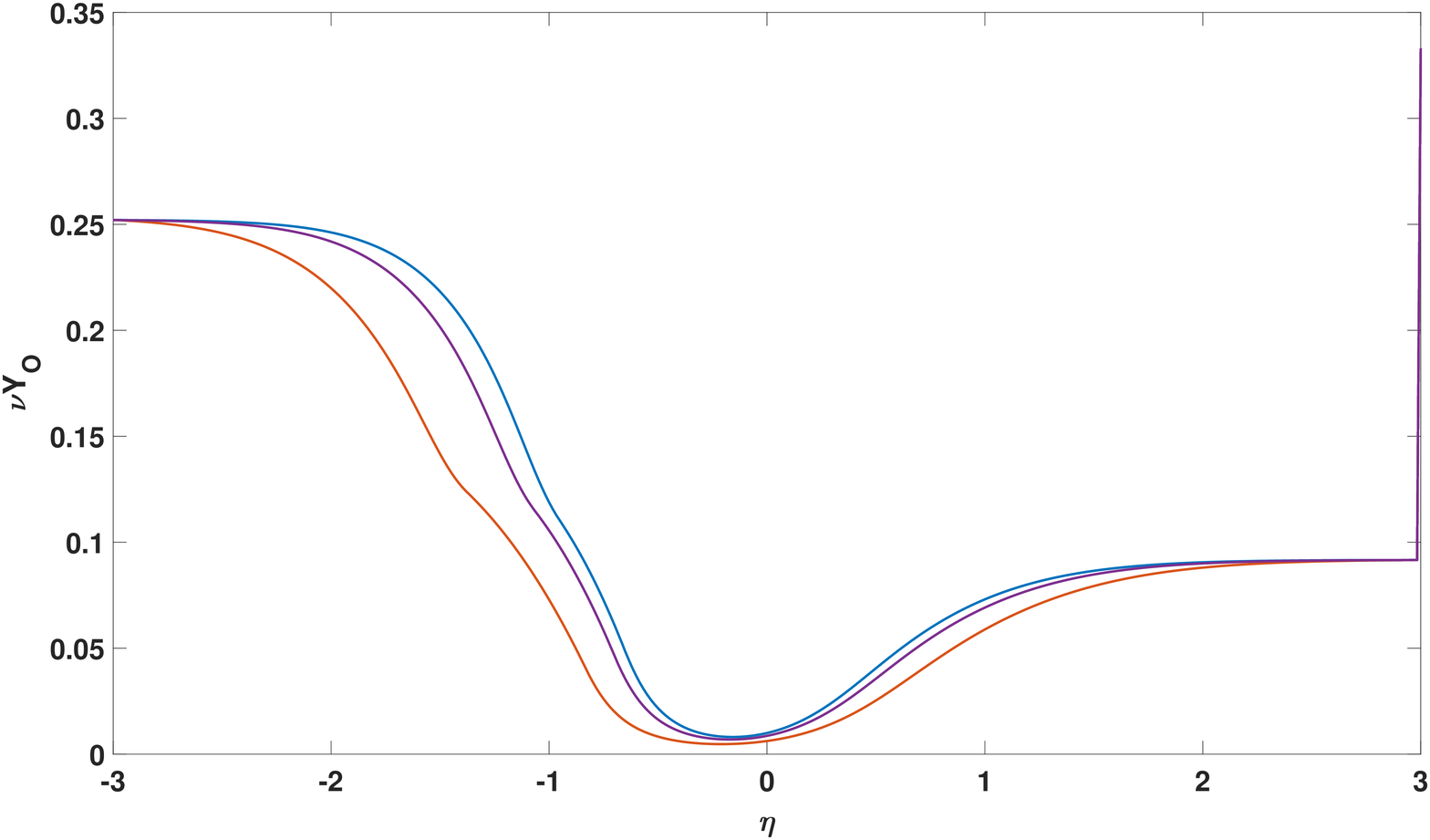}}
  \subfigure[integral of reaction rate, $\int \dot{\omega}_F d \eta$]{
  \includegraphics[height = 4.6cm, width=0.45\linewidth]{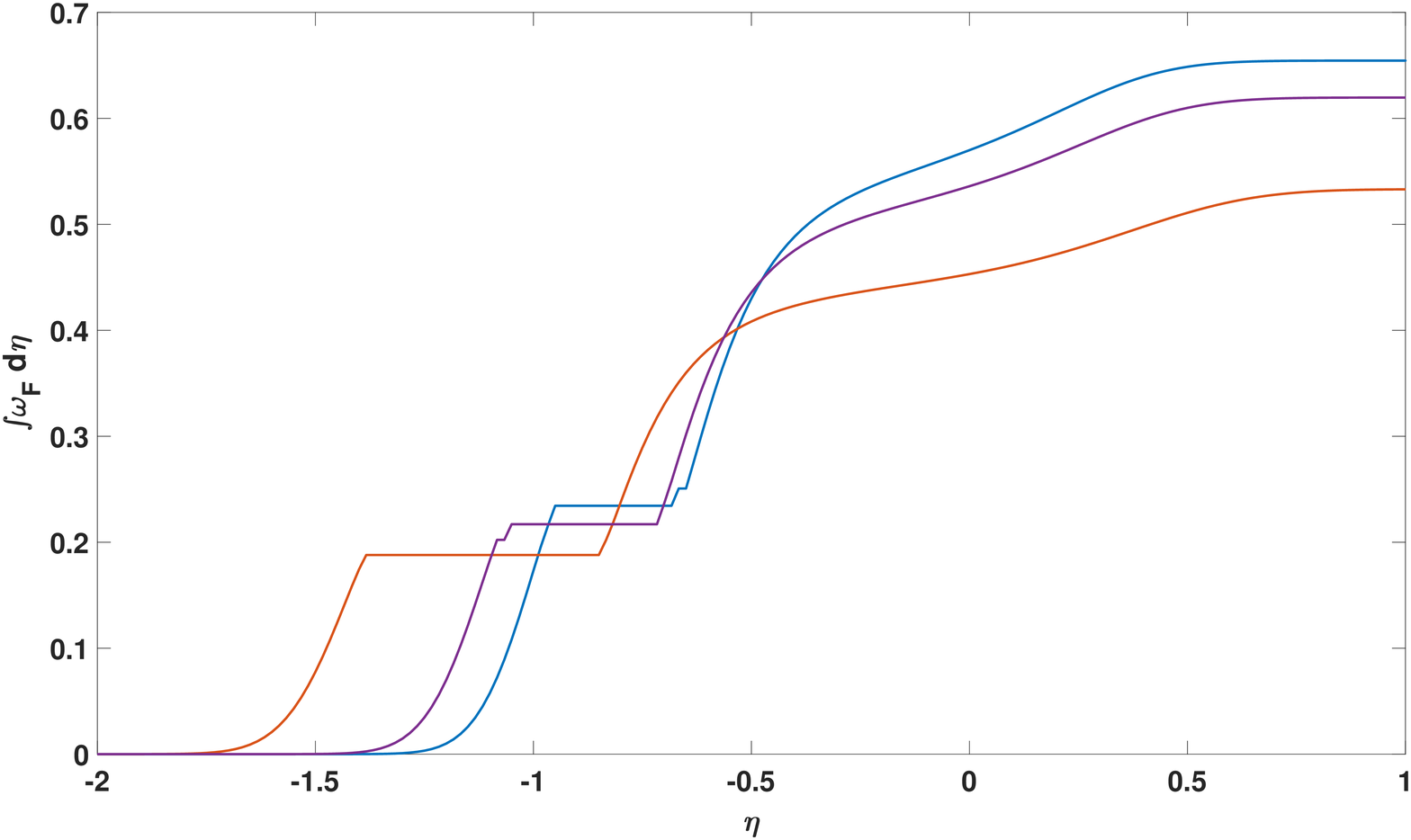}}     \\
  \vspace{0.2cm}
  \subfigure[reaction rate, $\dot{\omega}_F $]{
  \includegraphics[height = 4.6cm, width=0.45\linewidth]{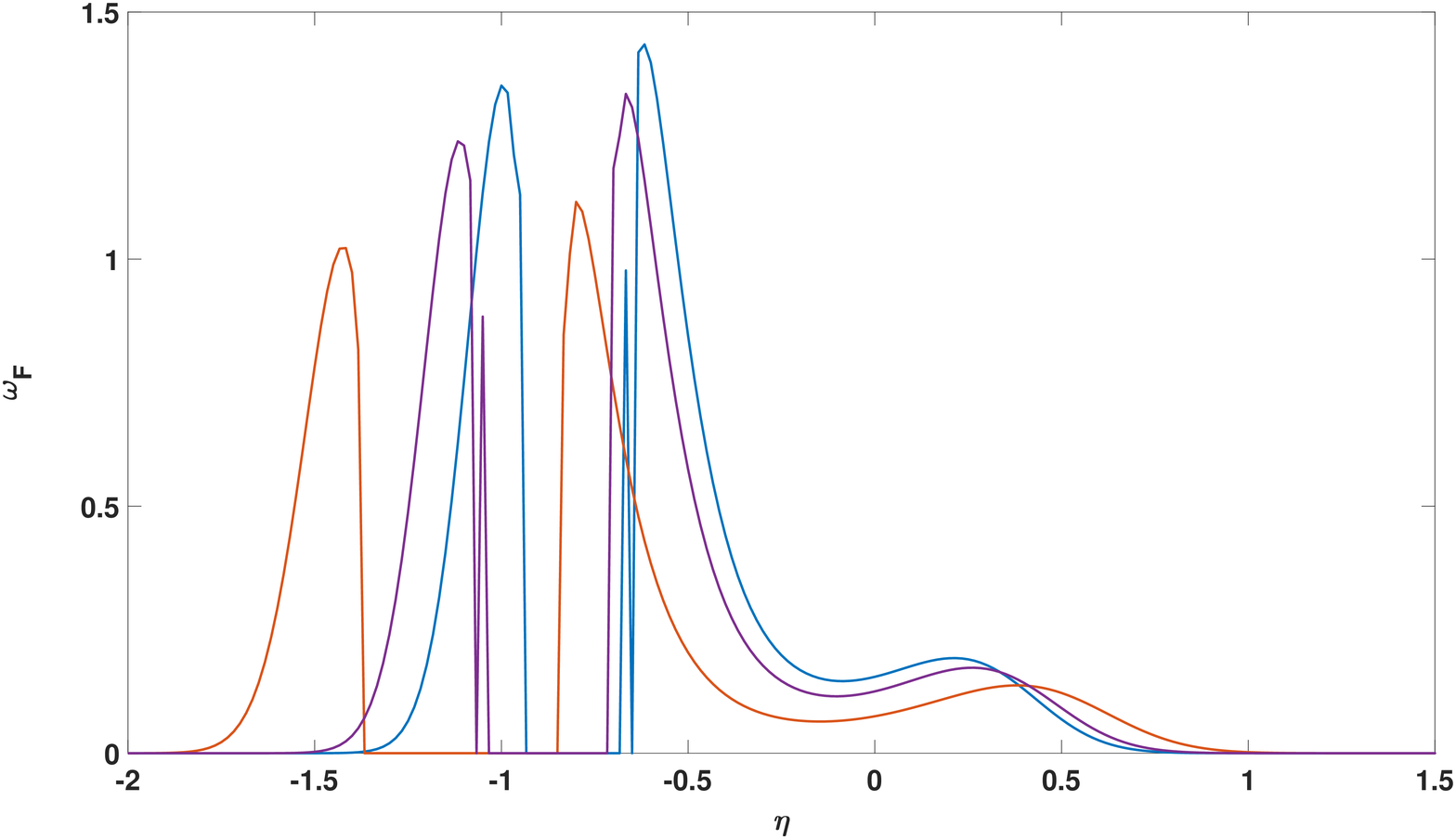}}
  \vspace{-0.1cm}
  \caption{Multi-branched flame with $K =0.200$ and varying vorticity. $S_1 =0.75 ; S_2 = 0.25$. blue,  $\omega_{\kappa} = 0$; red, $ \omega_{\kappa} = 1.50$;  purple, $\omega_{\kappa} = 1.0$. }
  \label{TriFlame1}
\end{figure}

\begin{figure}
  \centering
 \subfigure[mass flux per area, $f= \rho u_{\chi}$ ]{
  \includegraphics[height = 4.8cm, width=0.45\linewidth]{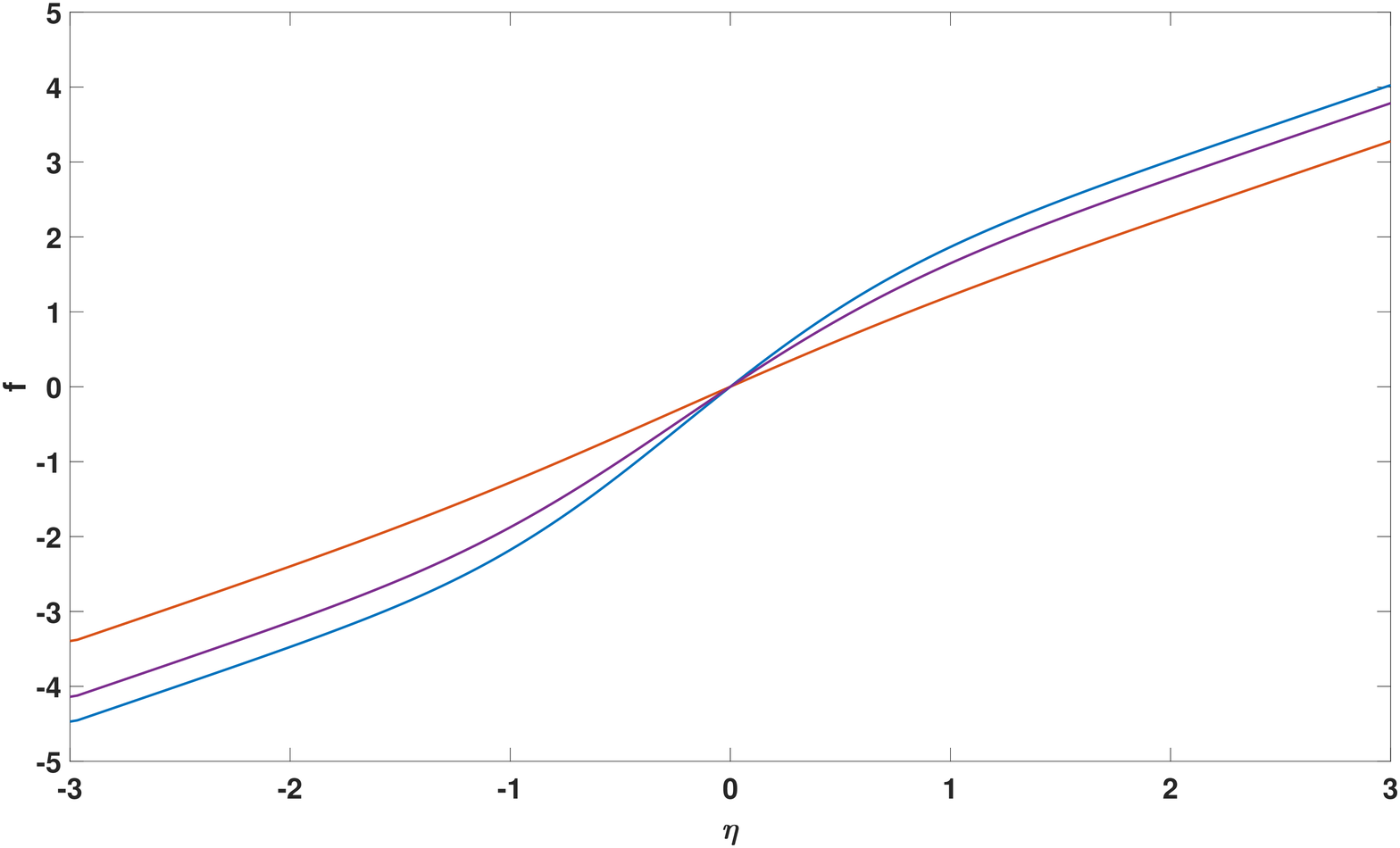}}
  \subfigure[velocity component, $f_1' = u_{\xi}/(S_1 \xi)$]{
  \includegraphics[height = 4.8cm, width=0.45\linewidth]{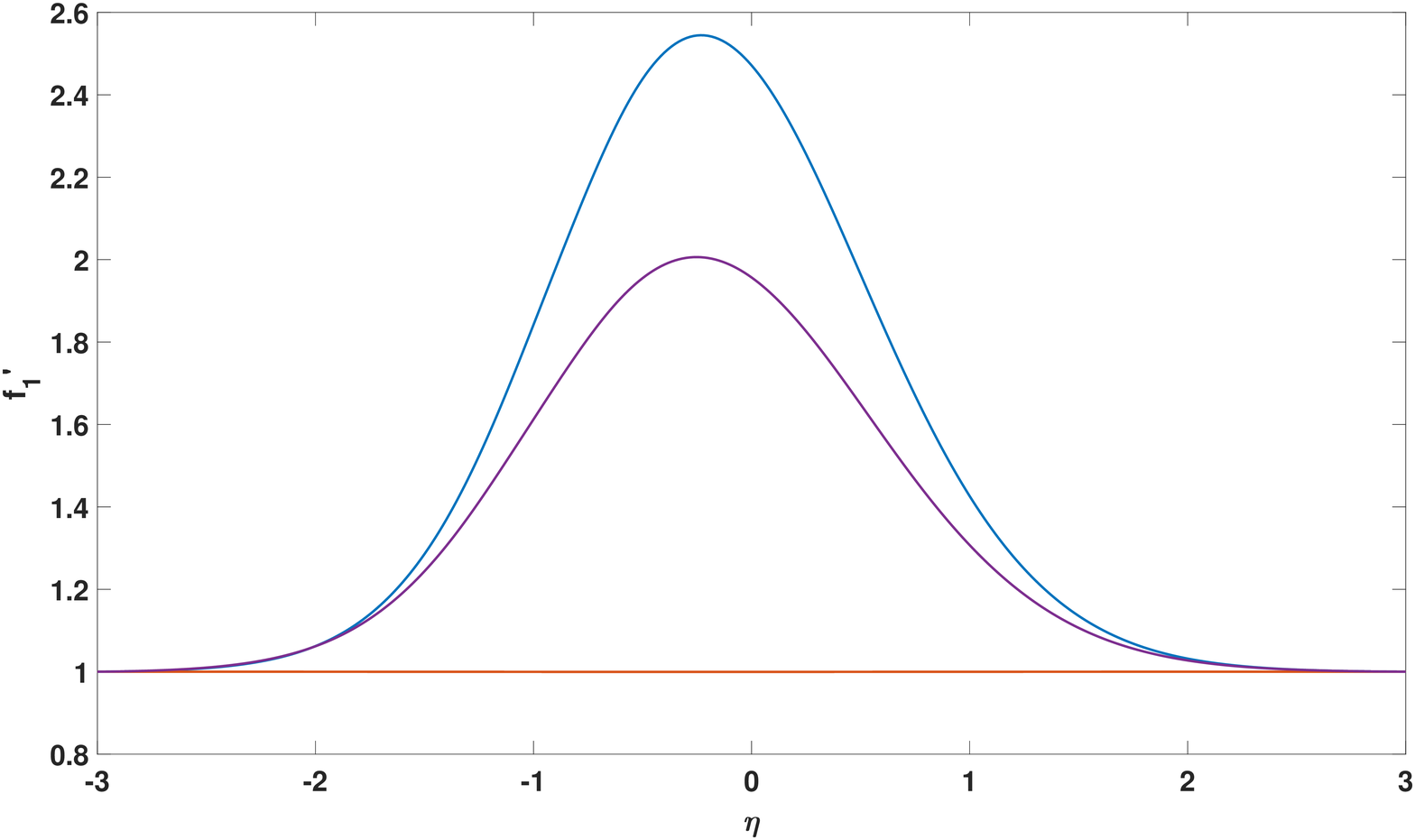}}     \\
  \vspace{0.2cm}
  \subfigure[ velocity component, $f_2' = w/(S_2z)$]{
  \includegraphics[height = 4.8cm, width=0.45\linewidth]{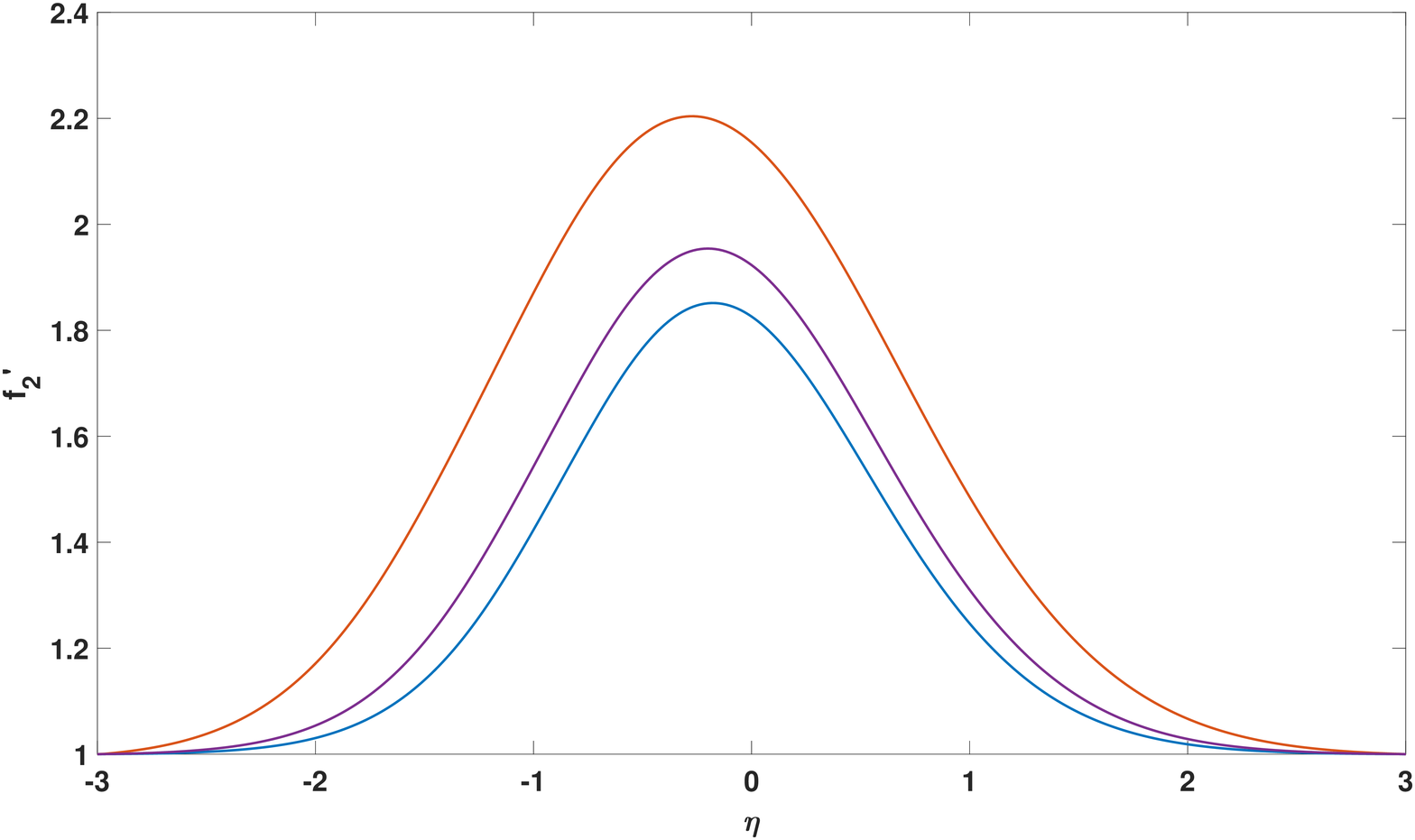}}
  \subfigure[velocity component, $u_{\chi}$]{
  \includegraphics[height = 4.8cm, width=0.45\linewidth]{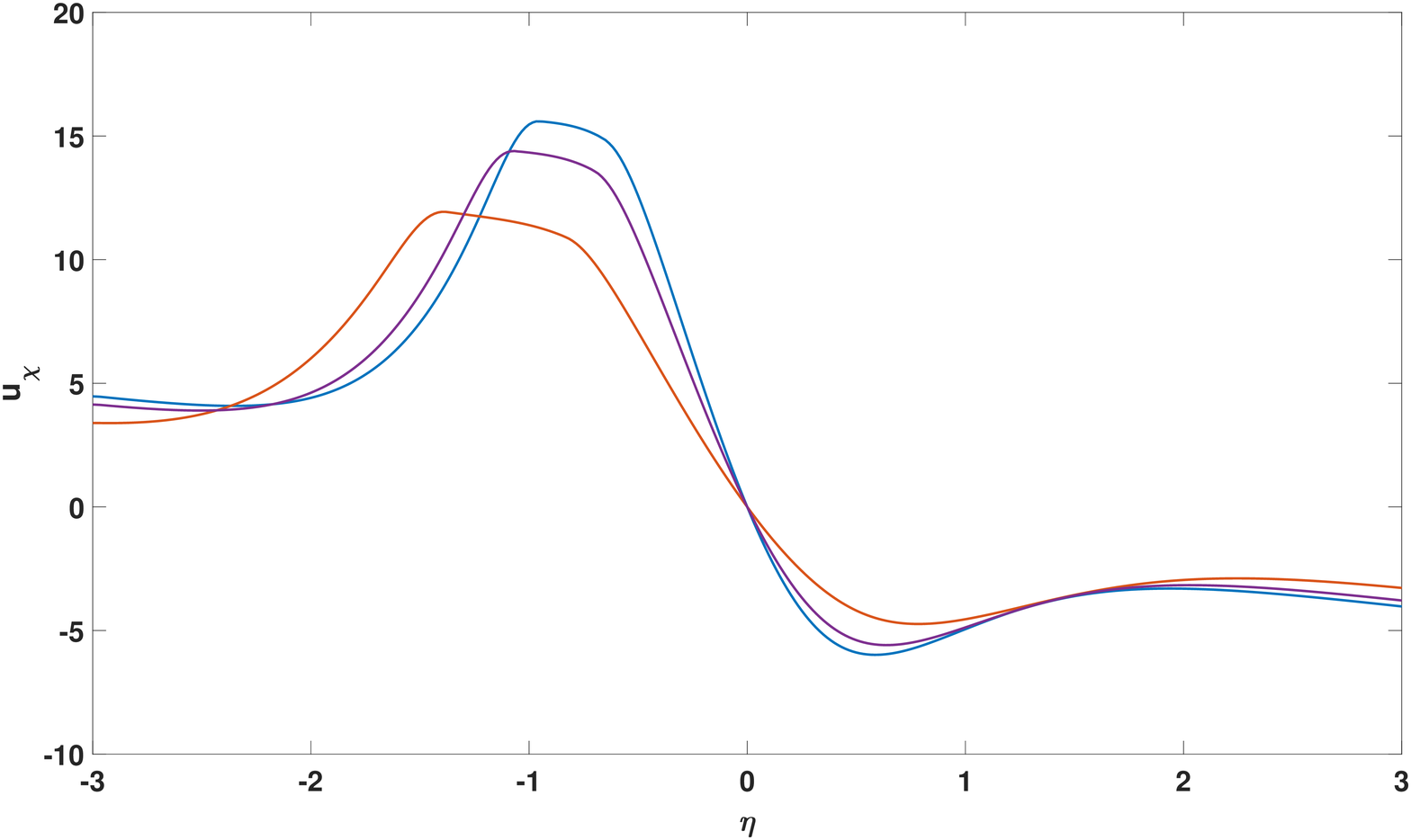}}
  \vspace{0.2cm}
  \caption{Multi-branched flame with $K =0.200$ and varying vorticity. $S_1 =0.75 ; S_2 = 0.25$.  blue,  $\omega_{\kappa} = 0$; red,  $\omega_{\kappa} = 1.50$;  purple, $\omega_{\kappa} = 1.0$. }
  \label{TriFlame2}
\end{figure}

In Figures \ref{TriFlame1} and \ref{TriFlame2}, results are shown for a multi-flame configuration with $K= 0.200 $ where a fuel-rich mixture with
$Y_F = 2/3$ and $Y_O = 1/3$
exists on one side of the counterflow flame and a fuel-lean mixture with
$Y_F =1/12$ and $Y_O = 11/12$
exists on the other side. This allows for the possibility of multi-flame structures as found by \cite{Sirignano2019b}, in contrast to the FPV approach which disallows multi-branched or premixed flame behavior by using a mixture fraction $Z$ domain that extends for $0 \leq Z \leq  1$.   The figures show that three flame structures appear based on this new theory: a fuel-lean premixed flame on the left, a diffusion flame in the center, and a fuel-rich premixed flame on the right. As noted by \cite{Sirignano2021}, the    two premixed flames are not independent structures but rather depend substantially on heat flux from the diffusion flame.  The figures also show that, as vorticity and rotation rate increase, the burning rate for each of the three flame branches and mass-flux through the counterflow decrease. The location of each of the flames shifts towards the fuel-lean side as rotation rate increases. Also, the fraction of the outflow in the direction aligned with the vorticity increases substantially with increasing rate of rotation.


For the multi-branched flame as well as the simple diffusion flame,  fluid rotation, variable-density, and three-dimensional structure combine to have major consequences for the behavior.

\begin{figure}
  \centering
 \subfigure[enthalpy, $h/h_{\infty}$ ]{
  \includegraphics[height = 4.6cm, width=0.45\linewidth]{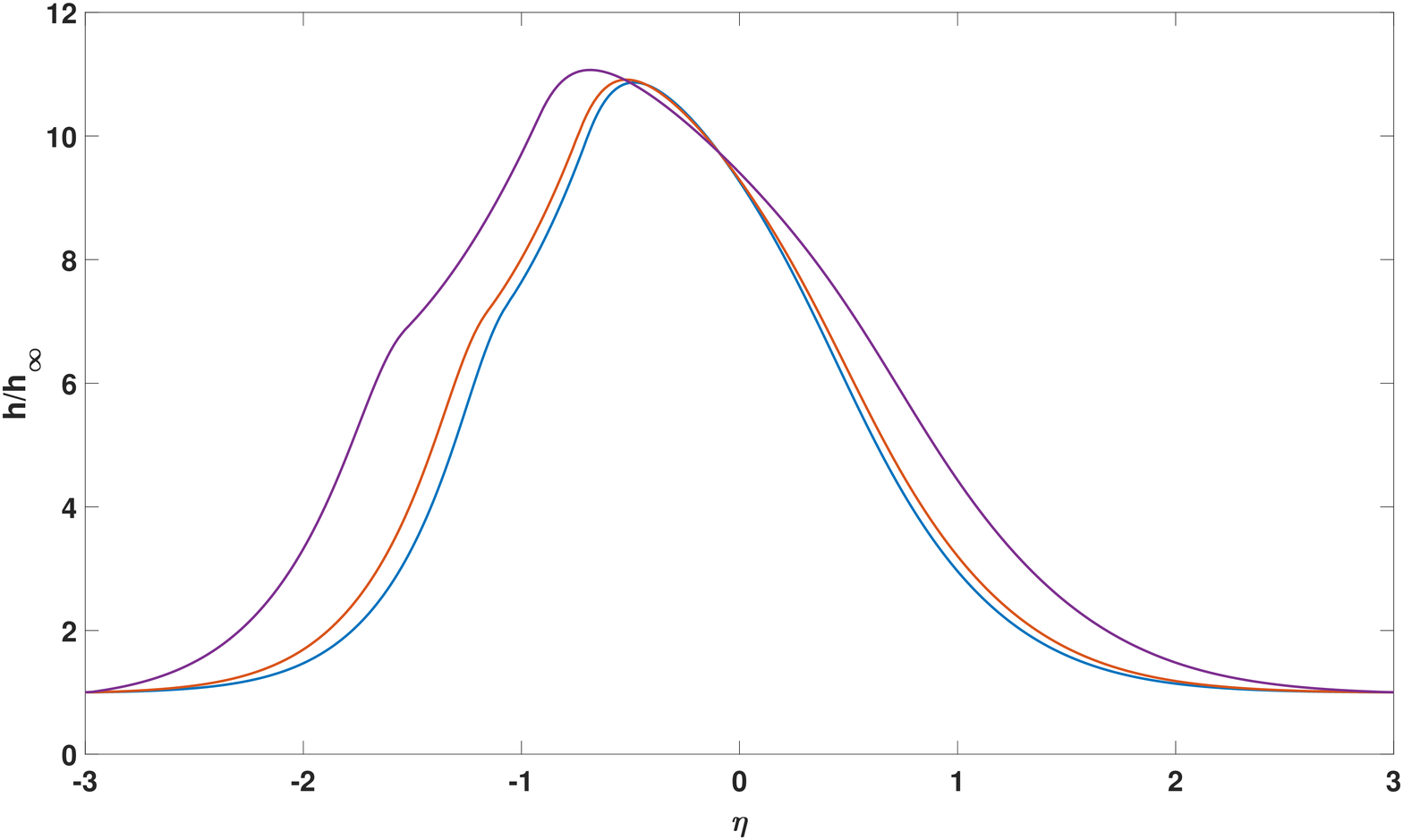}}
  \subfigure[fuel mass fraction, $Y_F$]{
  \includegraphics[height = 4.6cm, width=0.45\linewidth]{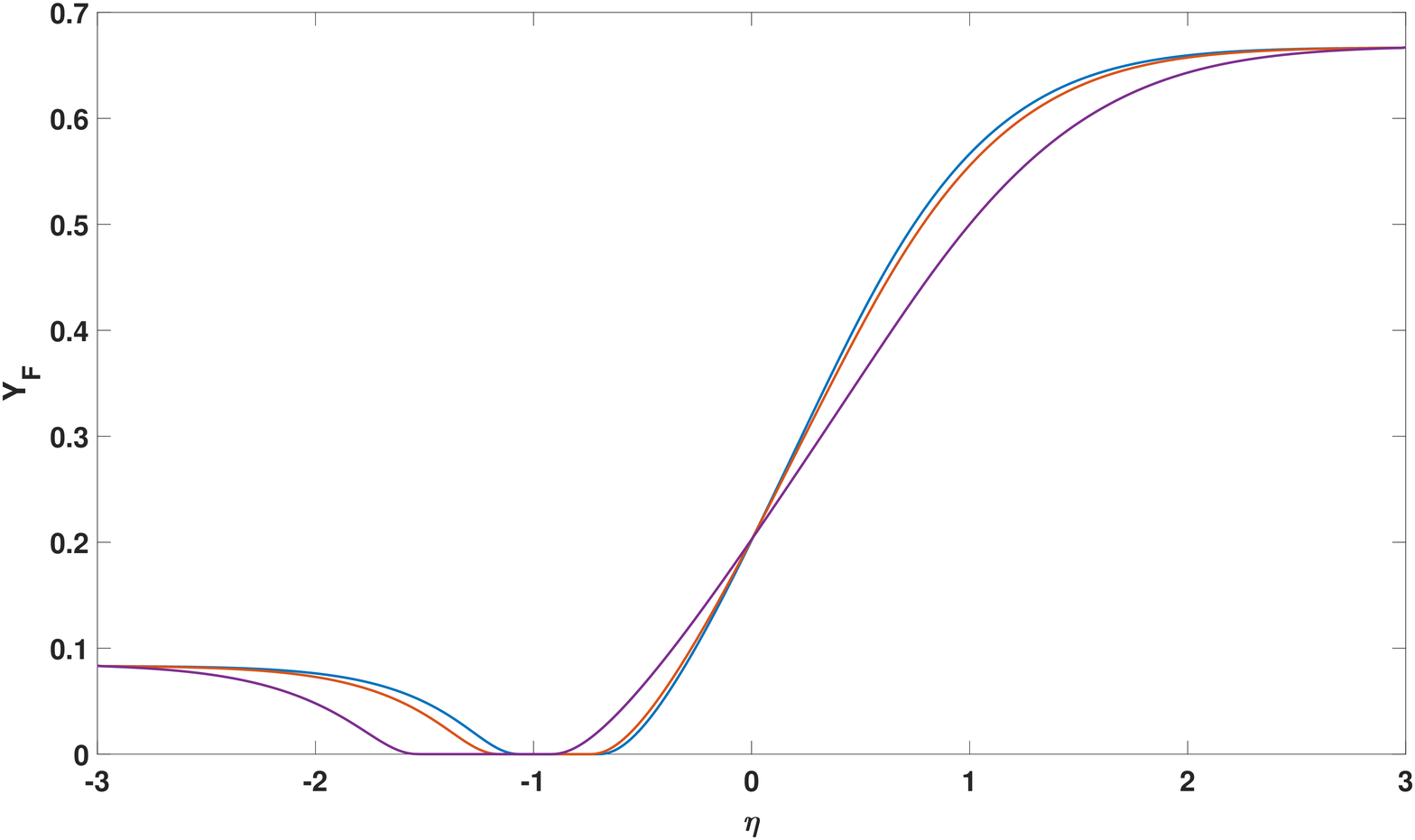}}     \\
  \vspace{0.2cm}
  \subfigure[ mass ratio x oxygen mass fraction, $\nu Y_O$]{
  \includegraphics[height = 4.6cm, width=0.45\linewidth]{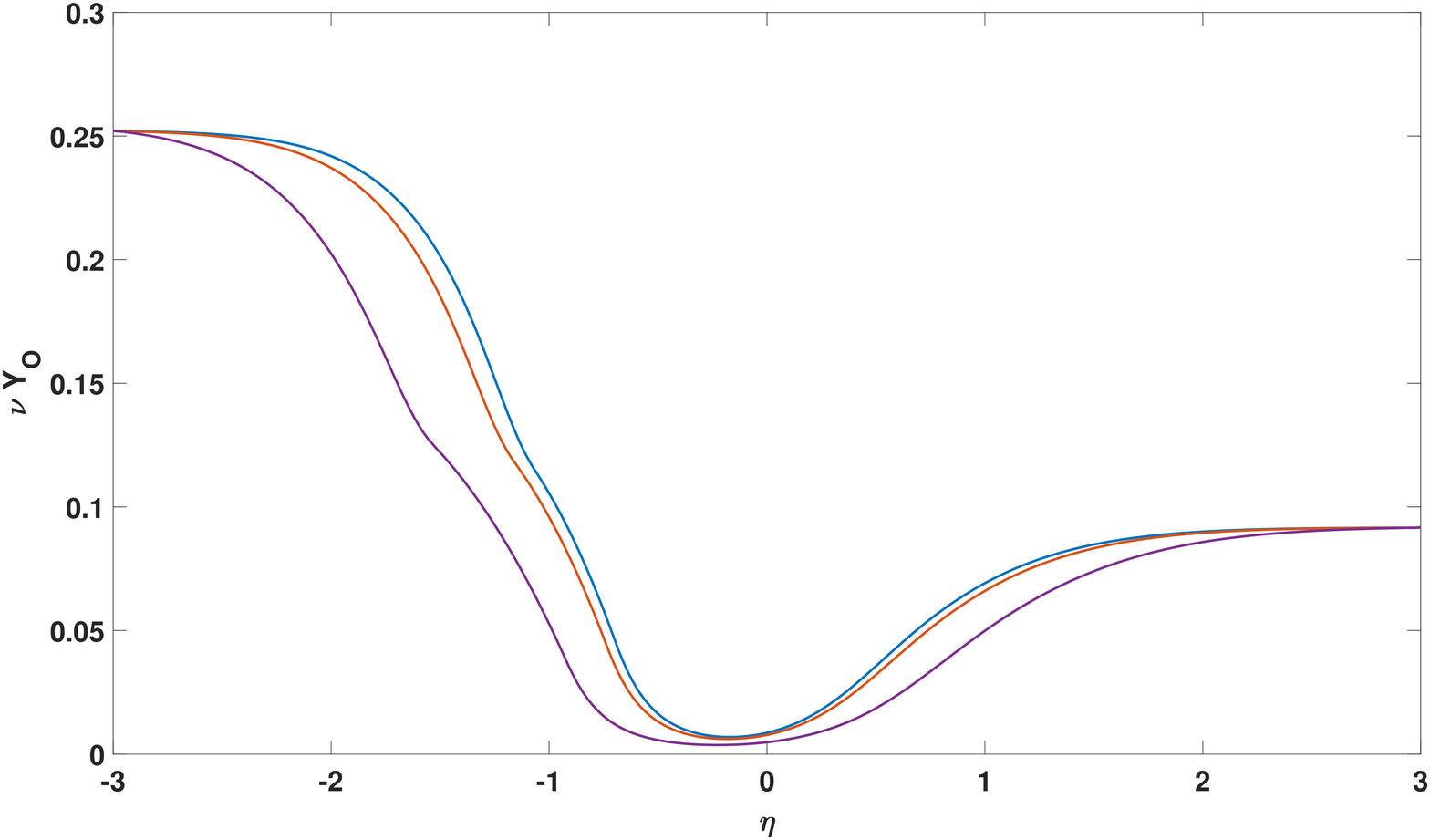}}
  \subfigure[integral of reaction rate, $\int \dot{\omega}_F d \eta$]{
  \includegraphics[height = 4.6cm, width=0.45\linewidth]{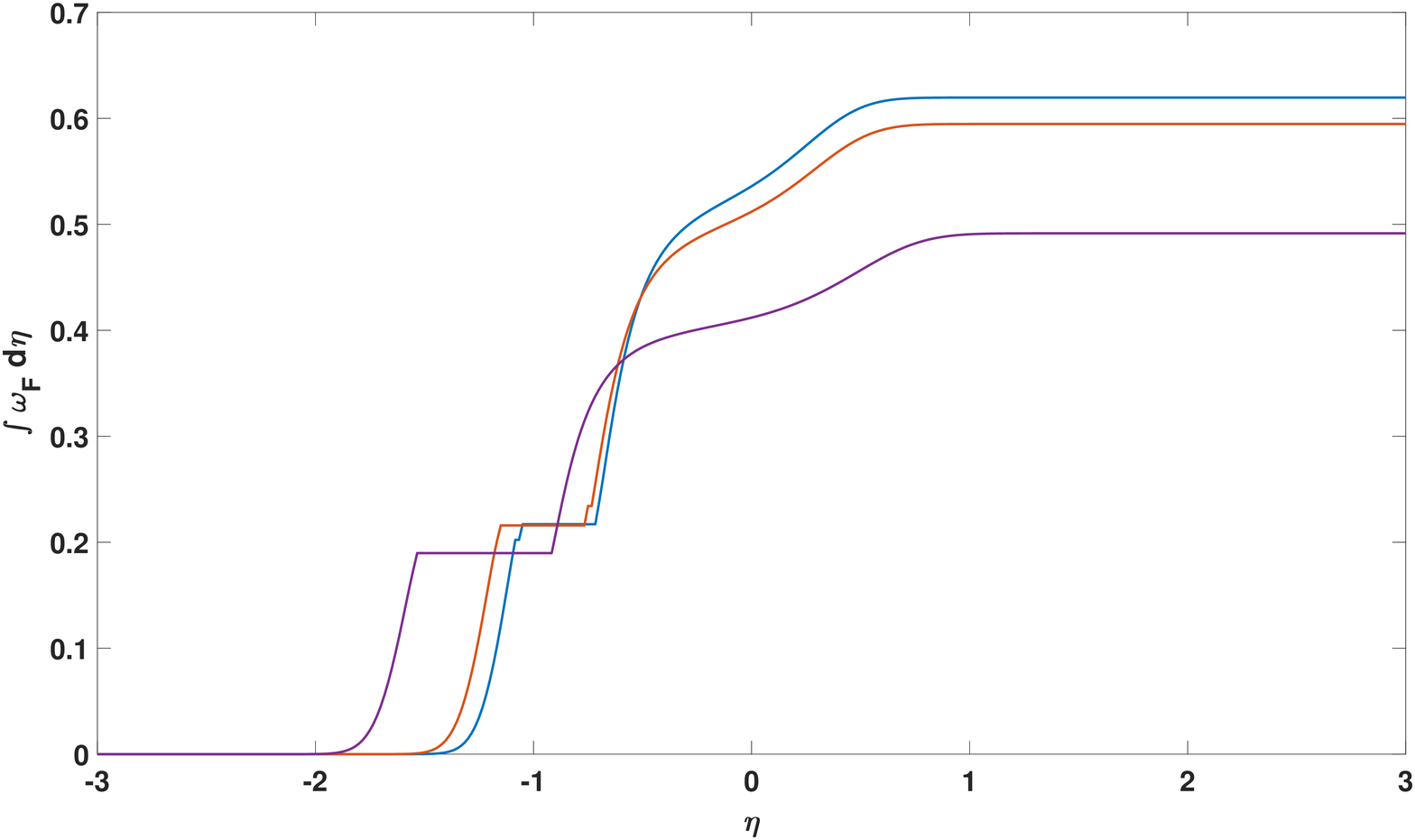}}     \\
  \vspace{0.2cm}
  \subfigure[reaction rate, $\dot{\omega}_F $]{
  \includegraphics[height = 4.6cm, width=0.45\linewidth]{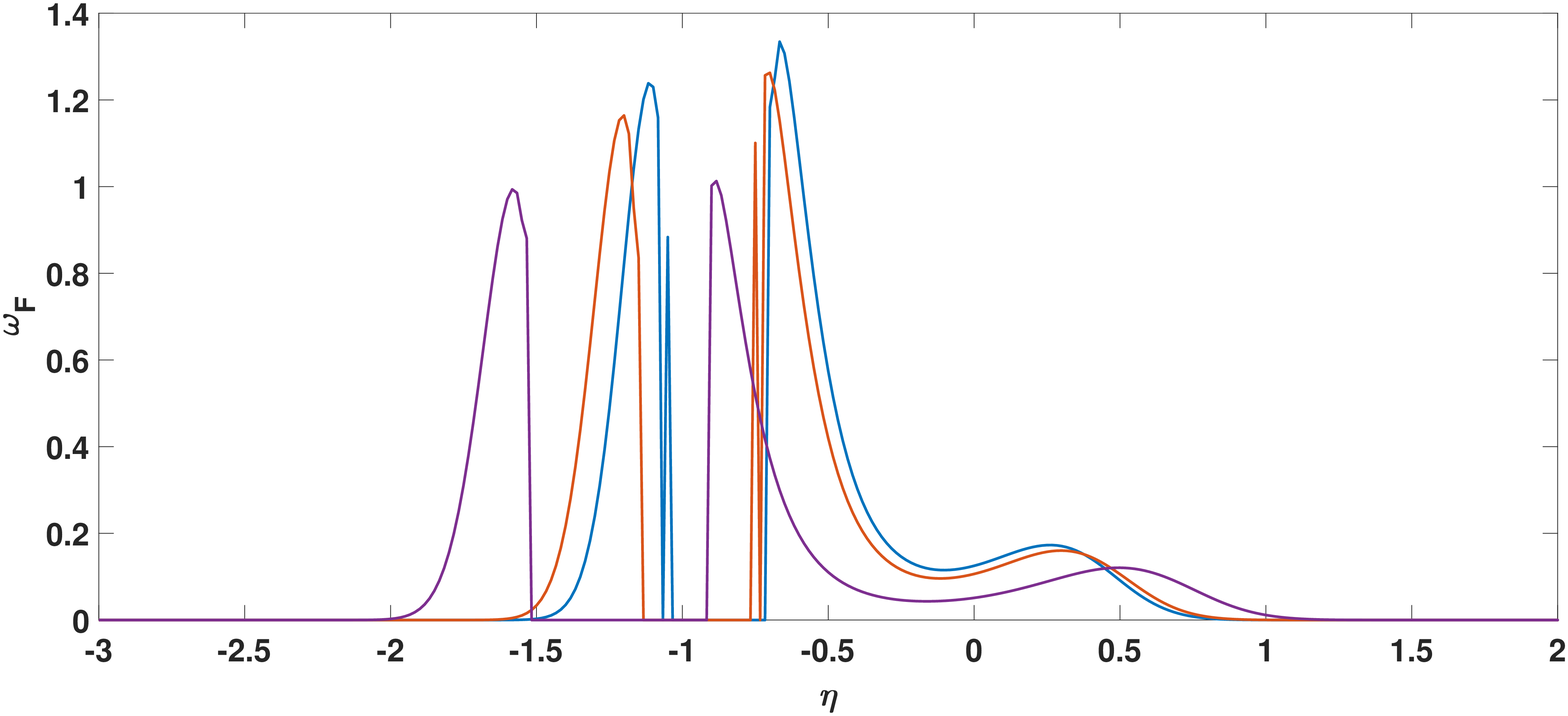}}
  \vspace{-0.1cm}
  \caption{Multi-branched flame with varying strain rate.   $K= 0.200,  \omega_{\kappa} = 1.0$. blue $S_1 =0.750, S_2 = 0.250$;   red $S_1 =0.500, S_2 = 0.500$ ;   purple $S_1 =0.333, S_2 = 0.667$  .              }
  \label{S1TripleFlame1}
\end{figure}

\begin{figure}
  \centering
 \subfigure[mass flux per area, $f= \rho u_{\chi}$ ]{
  \includegraphics[height = 4.8cm, width=0.45\linewidth]{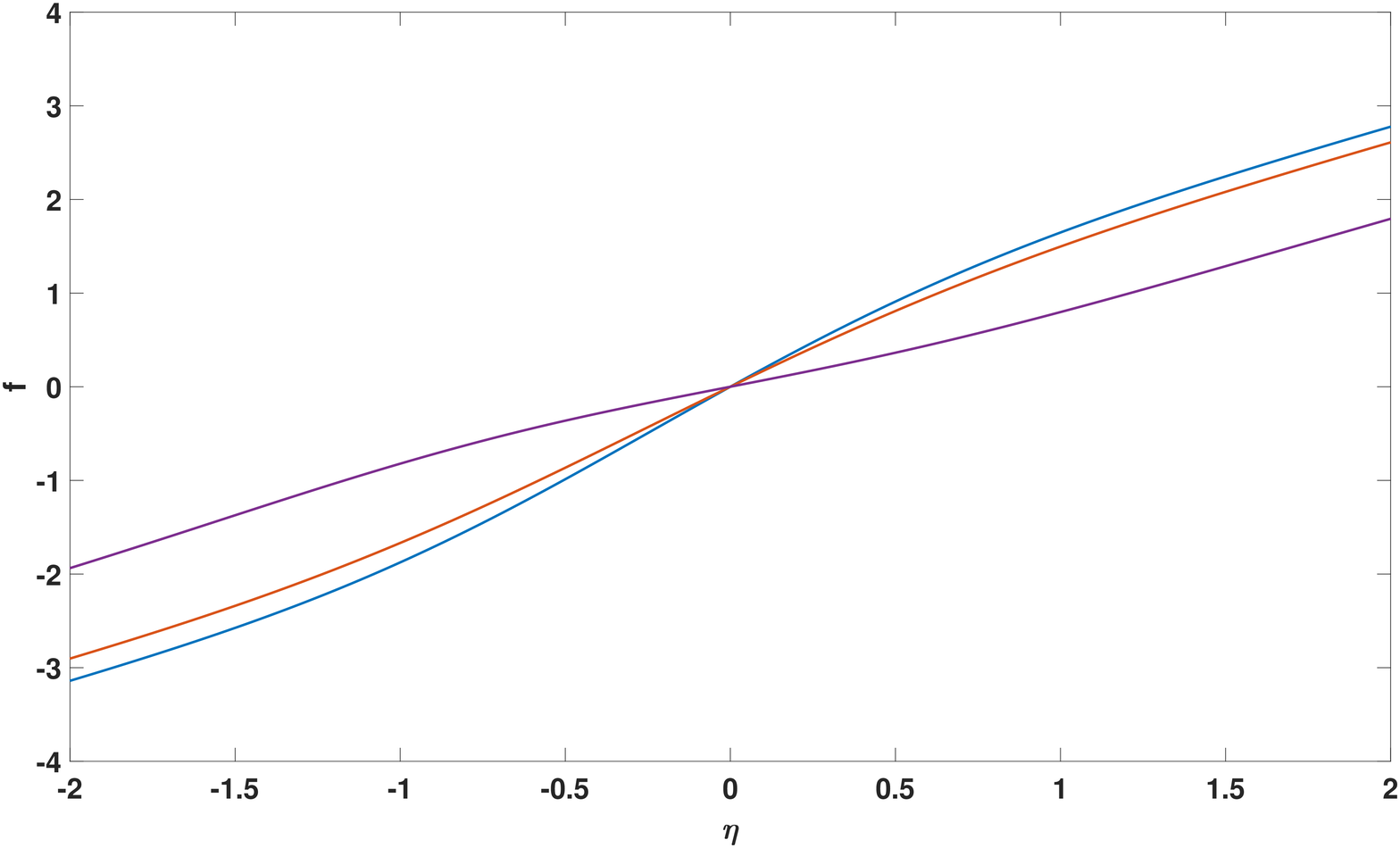}}
  \subfigure[velocity component, $f_1' = u_{\xi}/(S_1 \xi)$]{
  \includegraphics[height = 4.8cm, width=0.45\linewidth]{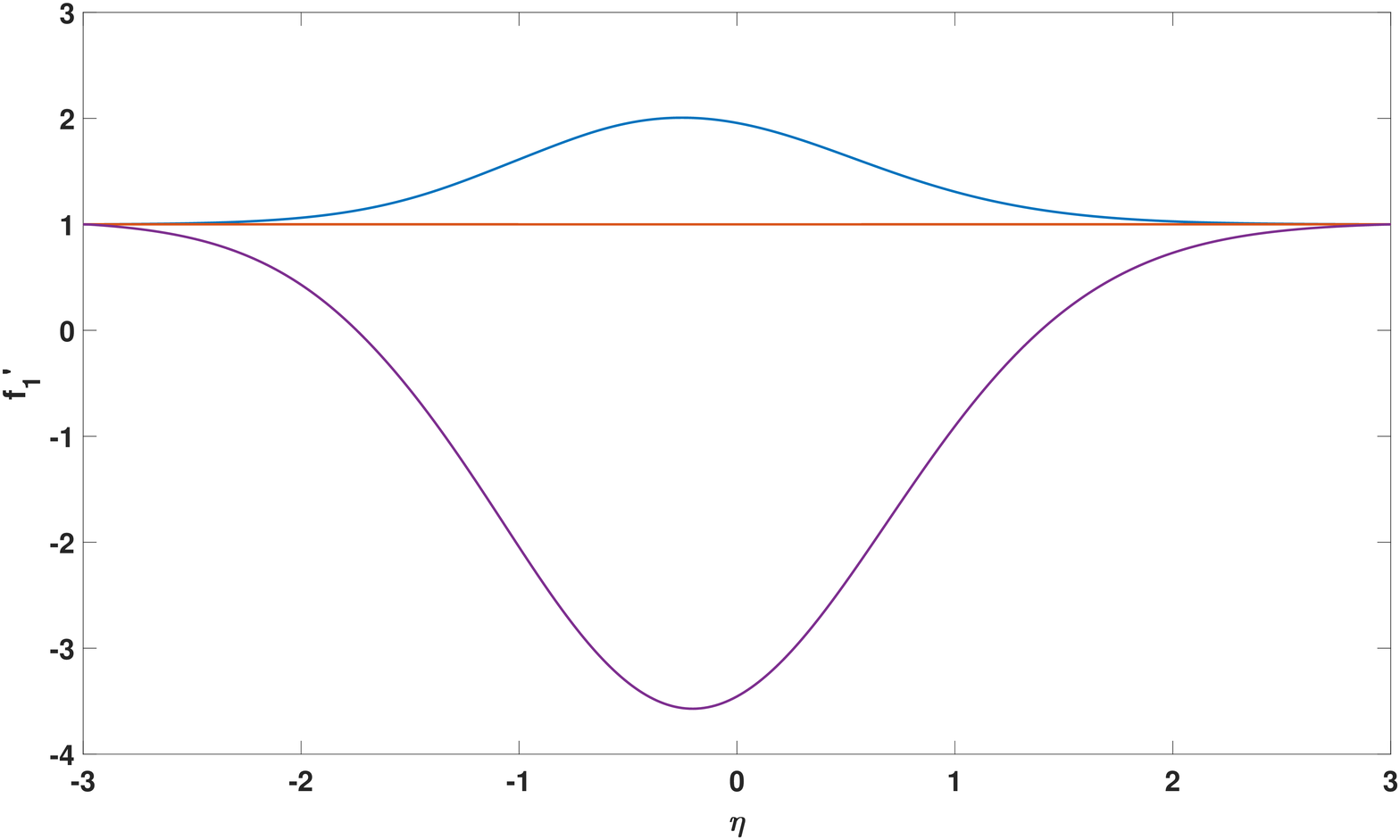}}     \\
  \vspace{0.2cm}
  \subfigure[ velocity component, $f_2' = w/(S_2 z)$]{
  \includegraphics[height = 4.8cm, width=0.45\linewidth]{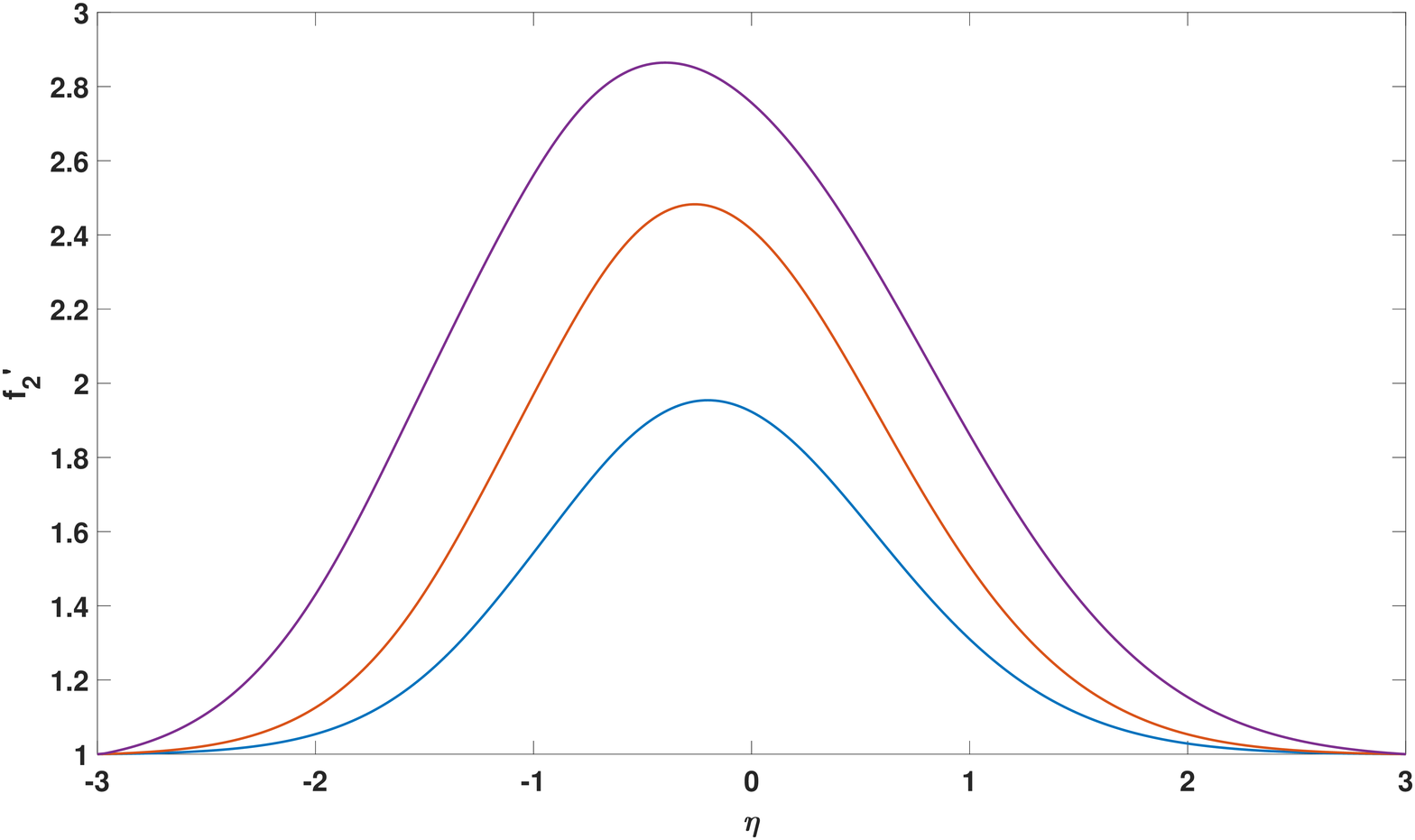}}
  \subfigure[velocity component, $u_{\chi}$]{
  \includegraphics[height = 4.8cm, width=0.45\linewidth]{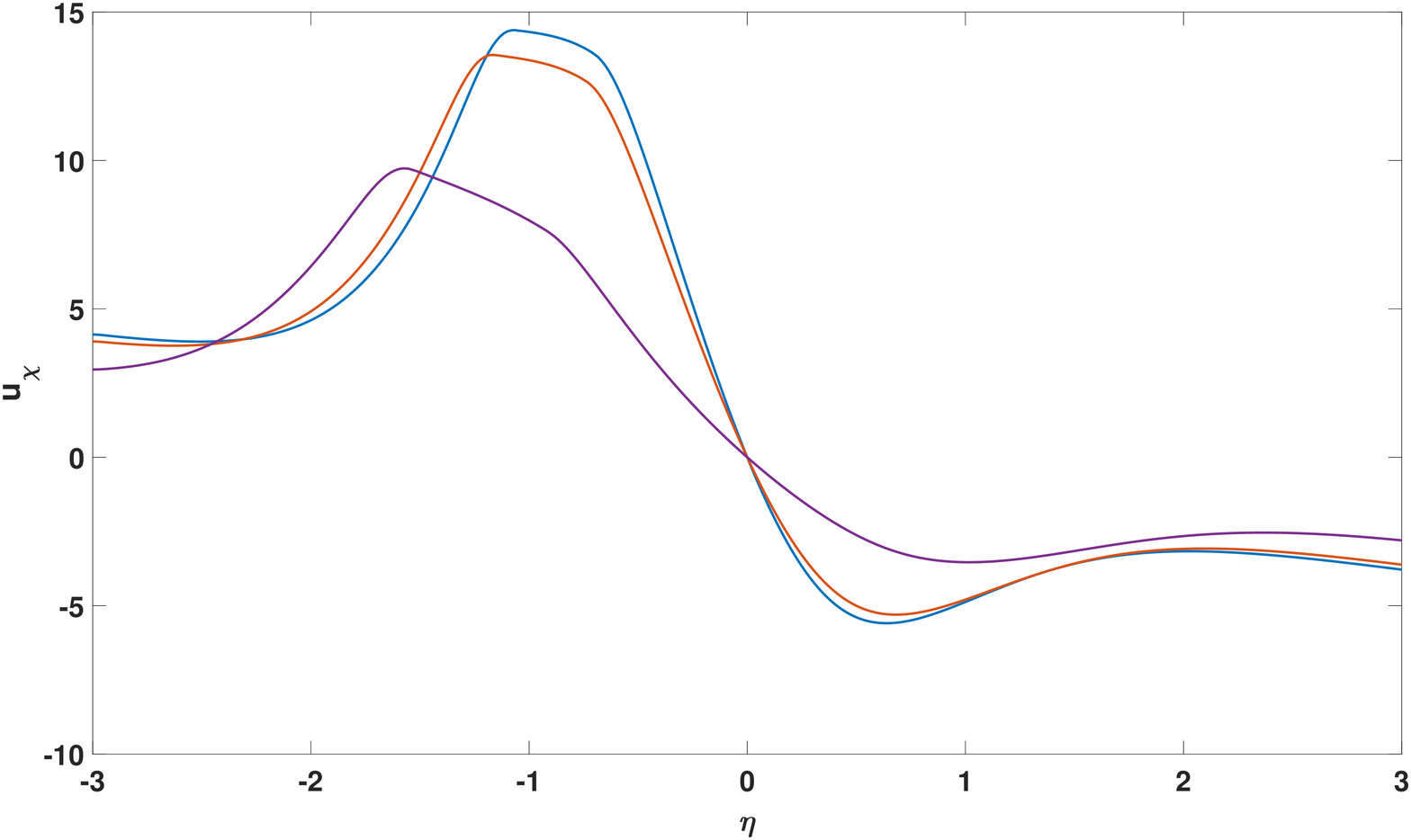}}     \\
  \vspace{0.2cm}
  \caption{Multi-branched flame with varying strain rate.   $K= 0.200,  \omega_{\kappa} = 1.0$. blue $S_1 =0.750, S_2 = 0.250$;   red $S_1 =0.500, S_2 = 0.500$ ;   purple $S_1 =0.333, S_2 = 0.667$  .   }
  \label{S1TripleFlame2}
\end{figure}

\begin{figure}
  \centering
  \includegraphics[height = 11.0cm, width=0.9\linewidth]{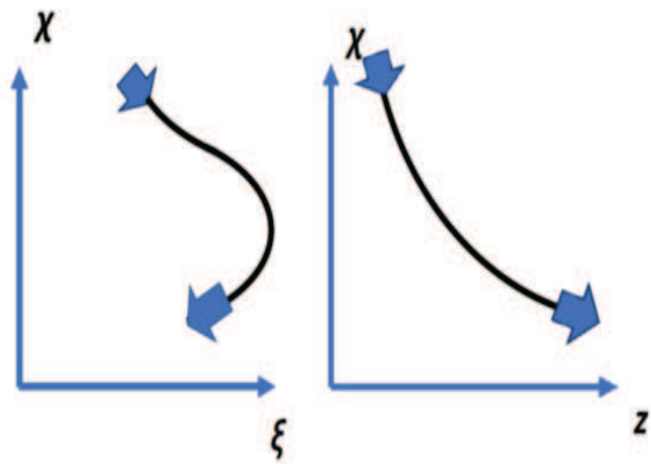}
  \vspace{0.2cm}
  \caption{Multi-branched flamelet with reversed flow.   $K= 0.200,  \omega_{\kappa} = 1.0$.  $S_1 =0.333, S_2 = 0.667$  .   }
  \label{ReversedFlow}
\end{figure}

For both the multi-branched flame and the single diffusion flame,  values of $S_1$ and $S_2$  have  consequence on the behavior. In Figures \ref{S1TripleFlame1} and \ref{S1TripleFlame2}, $S_1$ varies between 0.750 and 0.333. $S_2 = 1 -S_1$ and varies accordingly.  Again, as $S_1 $ decreases and $S_2$ increases, the flame zone moves slightly, the integrated burning rate decreases, and the normalized mass flux $f$ through the counterflow decreases. As $S_1$ decreases, the $u_{\chi}$ velocity component increases while the $u_{\xi}$ velocity component decreases and, below $S_1 = 0.500$, the $u_{\xi}$ velocity reverses direction in the region of highest temperature and lowest density. So, again in that low-density region, there is inflow (compressive strain) in two directions with outflow (tensile strain) only in the $z$-direction. The results show in Figure \ref{ReversedFlow} that for $S_1 =0.333$, an inflowing particle of material enters at first with decreasing magnitude of $\chi$ with Lagrangian time and increasing values of $\xi$ and $z$ in time. However, a reversal for the $\xi$ direction is seen with all the outflow going in the $z$ direction.   As noted earlier, the behavior is consistent with DNS findings concerning alignment of vorticity and principal strain axes for reacting flows by \cite{Nomura1993}.


\newpage
\section{Shear Flow on the Resolved Scale\;\;-\;\;Planar Jet}\label{mixingshearflow}

The ultimate objective is to connect this new flamelet model with RANS, unsteady RANS, and LES computations. Thereby, a limit on dimensionality or time dependence would not be sought. However, in this first application which is aimed towards gathering understanding of the method and not yet towards solving a new problem, a much simpler resolved scale is sought. So, we seek a resolved scale that is time-averaged with minimal dimensionality. This drives the consideration towards a simple mixing-length examination of two-dimensional mixing and reacting flows. In the following discussion, the turbulent planar mixing layer is considered.

We desire a solution for the two-dimensional, time-averaged, turbulent shear layer with variable density. The density can vary due to variations in temperature and/or composition. Pressure gradients in the mixing layer are considered negligible and the boundary-layer approximation is employed. \cite{Pope2000} and \cite{White}  present useful overviews on the topic.  \cite{Gortler1942} treated the non-reacting incompressible case and found a one-dimensional similar solution with $y/x$ as the independent similarity variable.  We have  a long-standing \cite{Illingworth} template with laminar flow to modify the relations for the similar compressible flow. However, the reacting flow will not allow for a similar solution in either a laminar or a turbulent case (except for infinite reaction rate with diffusion flames).  Here, we take a pathway using a time-averaged two-dimensional turbulent planar mixing layer. We will use a mixing-length concept for the eddy viscosity and diffusivities.

\subsection{Reacting Shear-layer Analysis}

We consider the averaged turbulent flow, e.g., steady-state 2D. The density is variable.
The pressure gradient is zero and the boundary-layer approximation is used. The governing equations for the time-averaged velocity components $u, v$ in $x, y$ space are
\begin{eqnarray}
\frac{\partial (\rho u)}{\partial x} + \frac{\partial (\rho v)}{\partial y} &= &0 \nonumber \\
u\frac{\partial u}{\partial x} + v \frac{\partial u}{\partial y} &=&
\frac{1}{\rho}\frac{\partial}{\partial y}\bigg(  \rho \nu_t \frac{\partial  u}{\partial y} \bigg)    \nonumber \\
u\frac{\partial h}{\partial x} + v \frac{\partial h}{\partial y} &=&
\frac{1}{\rho Pr_t}\frac{\partial}{\partial y}\bigg(  \rho \nu_t \frac{\partial  h}{\partial y} \bigg)  - \Sigma_{m=1}^N h_{f, m} \dot{\omega}_{m,rs}   +\frac{1}{\rho}\Phi  \nonumber \\
u\frac{\partial Y_m}{\partial x} + v \frac{\partial Y_m}{\partial y} &=&
\frac{1}{\rho Sc_t}\frac{\partial}{\partial y}\bigg(  \rho \nu_t \frac{\partial  Y_m}{\partial y} \bigg)  +  \dot{\omega}_{m,rs}   \;\;  ;  \;\; m=1, 2, ....., N
\label{shearflow}
\end{eqnarray}

Here, $\nu_t = \mu_t / \rho$ and $\mu_t$ are the kinematic turbulent viscosity and dynamic turbulent viscosity, respectively. It is assumed that the turbulent Prandtl number $Pr_t$ and turbulent Schmidt number $Sc_t$ are uniform through the flow.  The above equations are in a non-dimensional form that uses the resolved scales as reference length and velocity. Thus, quantities such as $ \dot{\omega}_{m,rs}$ and $\Phi$ must be properly scaled before calculations. See Section \ref{scaling}.  Note that we can replace $\nu_t$ in the equations by $\nu +\nu_t$ where $\nu$ is the molecular kinematic viscosity. In most portions of the flow, $\nu_t >> \nu$. For low mean velocities, the dissipation $\Phi$ can be neglected.

  Transform in standard fashion from $x, y$ to $\bar{x} \equiv x, \bar{y}\equiv \int _0^y (\rho/\rho_{\infty}) dy'$.
 $ v$, the transverse  component of velocity, is transformed to the variable $w$ to mimic incompressible flow .
\begin{eqnarray}
w &\equiv& \frac{\rho}{\rho_{\infty}} v +\frac{u}{\rho_{\infty}}\int_{y_{ref}}^y \frac{\partial \rho}{\partial x} dy'\nonumber \\
\frac{\partial u}{\partial \bar{x}} + \frac{\partial w}{\partial \bar{y}} &= &0 \nonumber \\
u\frac{\partial u}{\partial \bar{x}} + w\frac{\partial u}{\partial \bar{y}} &=&
\frac{\partial}{\partial \bar{y}}\bigg(  \frac{ \rho^2}{\rho_{\infty}^2 } \nu_t \frac{\partial  u}{\partial \bar{y}} \bigg)    \nonumber \\
u\frac{\partial h}{\partial \bar{x}} +w \frac{\partial h}{\partial \bar{y}} &=&
\frac{1}{ Pr_t}\frac{\partial}{\partial \bar{y}}\bigg( \frac{ \rho^2}{\rho_{\infty}^2 } \nu_t \frac{\partial  h}{\partial \bar{y}} \bigg)  - \Sigma_{m=1}^N h_{f, m} \dot{\omega}_{m,rs}  +\frac{1}{\rho}\Phi  \nonumber \\
u\frac{\partial Y_m}{\partial \bar{x}} + w \frac{\partial Y_m}{\partial \bar{y}} &=&
\frac{1}{ Sc_t}\frac{\partial}{\partial \bar{y}}\bigg( \frac{ \rho^2}{\rho_{\infty}^2 }\nu_t \frac{\partial  Y_m}{\partial \bar{y}} \bigg)  + \dot{\omega}_{m,rs}  \;\;  ;  \;\; m=1, 2, ....., N
\label{rs}
 \end{eqnarray}

 With the knowledge from Equations (\ref{strain-epsilon}), (\ref{strainscale}), and (\ref{rateintegral}) about the impact of  the Kolmogorov strain rate on the reaction rate and the relation between strain rates at different levels, we may write
\begin{eqnarray}
\rho \dot{\omega}_{m,rs}  =\rho_{dimensional} \dot{\omega}_{m,dimensional} &=& \rho_{\infty, dimensional}S^*\widetilde{\rho \dot{\omega}}_{m} \nonumber  \\
&=&  \rho_{\infty, dimensional} \frac{S^{*3/2}_{rs}\delta}{\nu^{1/2}}\widetilde{\rho \dot{\omega}}_{m}
\label{scalerate}
\end{eqnarray}
Note that the resolved scale reaction rate $\omega_{m,rs}$ is dimensional while the sub-grid averaged term $  \widetilde{\rho \dot{\omega}}_{m}$ is non-dimensional in Equation (\ref{scalerate}).

Here, the layer thickness is described by a constant value of the slope $d\delta / dx$ since a linear growth rate is expected. The constant value can be estimated and adjusted based on the results but is expected to be close to the value for the incompressible, non-reacting mixing layer. Experiments have indicated the range of $0.06$ to $0.11 $ according to \cite{Pope2000}. Here, a constant $\sigma$ is considered as the reciprocal of the slope so that $\delta(x) = x/\sigma.$
 \begin{eqnarray}
  S^*_{rs} &\equiv& \bigg|\frac{du}{dy}\bigg| =     \frac{\rho}{\rho_{\infty}}\bigg|\frac{du}{d\bar{y}}\bigg|
   \nonumber \\
 \nu_t &=& (\delta(x))^2 \bigg| \frac{d u}{dy} \bigg|
 \label{1}
 \end{eqnarray}
Accordingly, the value of $\nu_t$ will vary with both $x$ and $y$.

From Equations (\ref{scalerate}) and (\ref{1}), the dimensional resolved-scale reaction rate $  \dot{\omega}_{m, rs}  $ can be related to the non-dimensional sub-grid-scale reaction rate  $ \widetilde{\rho \dot{\omega}_m}  $.
Equation (\ref{rateintegral}) indicates that the non-dimensional reaction rate is proportional to an integral with $f$ in the integrand. Thus, the dimensional reaction rate will be proportional to the sub-grid-scale strain rate $S^*_1 + S^*_2$. That sub-grid strain rate is larger than the resolved-scale strain rate by a factor $Re^{1/2}$ as shown by Equation (\ref{strain-epsilon}).

The boundary conditions on each of the second-order partial differential equations are given by prescribing the $u$-component of velocity and the scalar properties in the two free streams. In addition, there are the upstream inflow conditions.

The system of equations given as (\ref{rs}) can be  made nondimensional
by using $u_{\infty}$ to normalize velocity components, a downstream length $x_0$ to normalize the $\bar{x}$ and $\bar{y}$ coordinates, and  $\rho_{\infty}$ and $h_{\infty}$ to normalize the corresponding scalar quantities. The turbulent viscosity $\nu_t$ is non-dimensionalized using the kinematic viscosity $\nu$.   The perfect-gas assumption and constant specific heat are assumed so the nondimensional relation $\rho = 1/h$ holds through  this shear layer where the pressure is approximately uniform.  We also define here a Reynolds number $Re \equiv u_{\infty} \delta_0 / \nu   =  u_{\infty} x_0 / (\sigma \nu)$. We will consider cases where the fractional difference  between $\Delta U = u_{\infty} - u_{-\infty}$ and $u_{\infty}$ is not major. The viscous dissipation is considered negligible because the Mach number is low.
\begin{eqnarray}
\frac{\partial u}{\partial \bar{x}} + \frac{\partial w}{\partial \bar{y}} &= &0 \nonumber \\
u\frac{\partial u}{\partial \bar{x}} + w\frac{\partial u}{\partial \bar{y}} &=&
\frac{\partial}{\partial \bar{y}}\bigg(  \frac{ 1}{h^2 } \big(\frac{x}{x_0}\big)^2\frac{1}{\sigma^2}\bigg|\frac{\partial u}{\partial \bar{y}}\bigg| \frac{\partial  u}{\partial \bar{y}}  \bigg)  \nonumber \\
u\frac{\partial h}{\partial \bar{x}} +w \frac{\partial h}{\partial \bar{y}} &=&
\frac{1}{  Pr_t}\frac{\partial}{\partial \bar{y}}\bigg( \frac{ 1}{h^2 }  \big(\frac{x}{x_0}\big)^2\frac{1}{\sigma^2}\bigg|\frac{\partial u}{\partial \bar{y}}\bigg| \frac{\partial  h}{\partial \bar{y}} \bigg)  \nonumber  \\
&&- \Sigma_{m=1}^N h_{f, m} \frac{Re^{1/2}}{(\sigma h)^{1/2}}\frac{x}{x_0} \bigg|\frac{\partial u}{\partial \bar{y}}\bigg|^{3/2}\widetilde{\rho \dot{\omega}}_{m}  \nonumber \\
u\frac{\partial Y_m}{\partial \bar{x}} + w \frac{\partial Y_m}{\partial \bar{y}} &=&
\frac{1}{  Sc_t}\frac{\partial}{\partial \bar{y}}\bigg( \frac{ 1}{h^2 } \big(\frac{x}{x_0}\big)^2\frac{1}{\sigma^2}\bigg|\frac{\partial u}{\partial \bar{y}}\bigg| \frac{\partial  Y_m}{\partial \bar{y}} \bigg)    \nonumber \\
&& + \frac{Re^{1/2}}{(\sigma h)^{1/2}}\frac{x}{x_0} \bigg|\frac{\partial u}{\partial \bar{y}}\bigg|^{3/2} \widetilde{\rho \dot{\omega}}_{m} \;\;  ;  \;\; m=1, 2, ....., N
\label{rs2}
 \end{eqnarray}
 Here, in the  non-dimensional processing, we have taken
\begin{eqnarray}
(x_0/u_{\infty}) \dot{\omega}_{m,rs} & =&  (x_0/u_{\infty})(1/\rho) \frac{S^{*3/2}_{rs}\delta}{\nu^{1/2}}\widetilde{\rho \dot{\omega}}_{m}   \nonumber \\
&=& [ (x_0/u_{\infty})\frac{u_{\infty}^{3/2}}{x_0^{3/2}}\frac{\delta}{\nu^{1/2}}]_{dimensional}
(\rho^{3/2}/\rho) \big|\frac{du}{d\bar{y}}\big|^{3/2}\widetilde{\rho \dot{\omega}}_{m}  \nonumber \\
&=& \big(\frac{u_{\infty}x_o}{\nu}\big)^{1/2}\frac{x}{x_0}\frac{1}{\sigma h^{1/2}}\big|\frac{du}{d\bar{y}}\big|^{3/2}\widetilde{\rho \dot{\omega}}_{m}  \nonumber \\
&=& Re^{1/2}\frac{x}{x_0}\frac{1}{(\sigma h)^{1/2}}\big|\frac{du}{d\bar{y}}\big|^{3/2}\widetilde{\rho \dot{\omega}}_{m}
\label{scalerate2}
\end{eqnarray}
The resolved-scale calculation will provide certain information as inputs to the sub-grid flamelet computation. Specifically, some information is needed to determine through reasonable scaling principles  (i) constraints on the scalar properties (mass fractions of the reactants and the temperature) and (ii) strain rates and vorticity. Then, the sub-grid calculation can give the burning rate and associated energy release rate to the resolved scale.  Since the sub-grid calculations are quasi-steady, they may be performed in advanced with the input-output relation organized in the form of a look-up table or a neural network (NN). Then, the table or NN may be coupled with the resolved-scale computation.

\subsection{Shear-layer Results and Discussion}

Computations are performed for a turbulent diffusion flame in the shear layer. One free stream is pure propane gas while the other is pure oxygen. We assume that the flamelet scale also displays a diffusion-flame character. (Clearly, a need exists for further DNS studies to inform us about the degree of partial mixing in the turbulence cascade.) At the flamelet scale, we take $\int \dot{\omega}_F d\eta =0.4$ which is consistent with results shown in Figures \ref{WeakDiffFlame1} and \ref{S1DiffFlame1}. Furthermore, considering $2b = O(10)$ in Equation (\ref{rateintegral}), we assign $\widetilde{\rho \dot{\omega}}_F = 0.04$.  The burning rate is determined by scaling the flamelet-scale burning rate based upon the large-scale strain rate $S^*_{rs}$; thus, the averaged, large-scale burning rate will vary with $x$ and $y$. Implicitly, the dimensional sub-grid-scale burning rate is thereby varying with $x$ and $y$. The burning rate is forced to zero value where $Y_F = 0$ and / or $Y_O = 0$. In the regions where both mass fractions have positive values, the burning rate is prescribed by Equation (\ref{scalerate2}).

The eddy diffusivity is estimated based on the assumption that the shear-layer density-weighted width grows linearly as 0.11$x$. Solutions start at $x_0 = 10$ and are marched to $x = 20.$  Hyperbolic tangents are used for the $\bar{y}$ profiles at $x_0$, except for the enthalpy where a Gaussian profile is superimposed to serve as an igniter. $Re =1000$ based on layer width at $x_0$.  One free stream, at positive $Y$ values, is five times faster than the other free stream at negative $y$ values.

\begin{figure}
  \centering
 \subfigure[enthalpy, $h/h_{\infty}$ ]{
  \includegraphics[height = 4.6cm, width=0.45\linewidth]{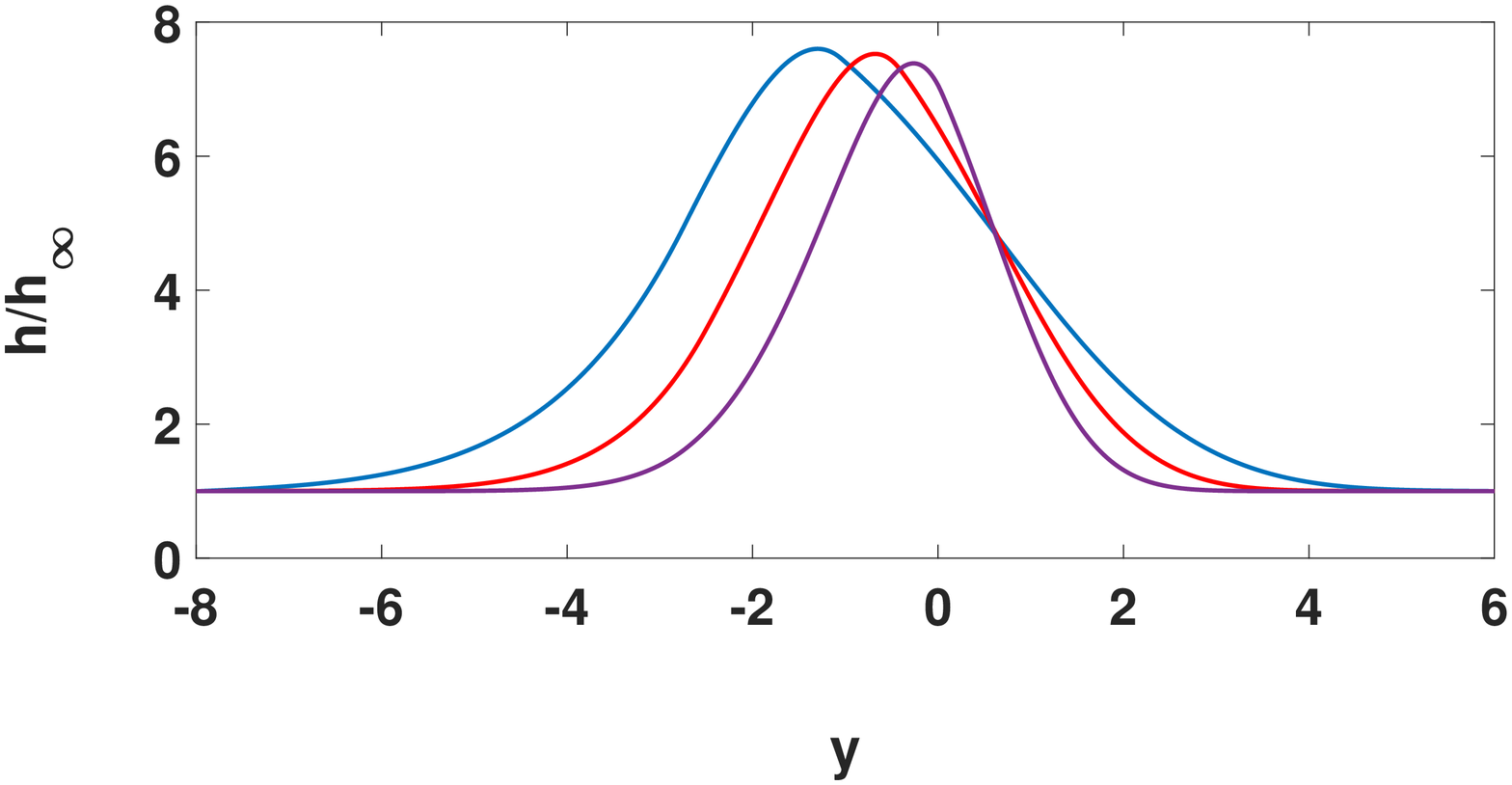}}
  \subfigure[fuel mass fraction, $Y_F$]{
  \includegraphics[height = 4.6cm, width=0.45\linewidth]{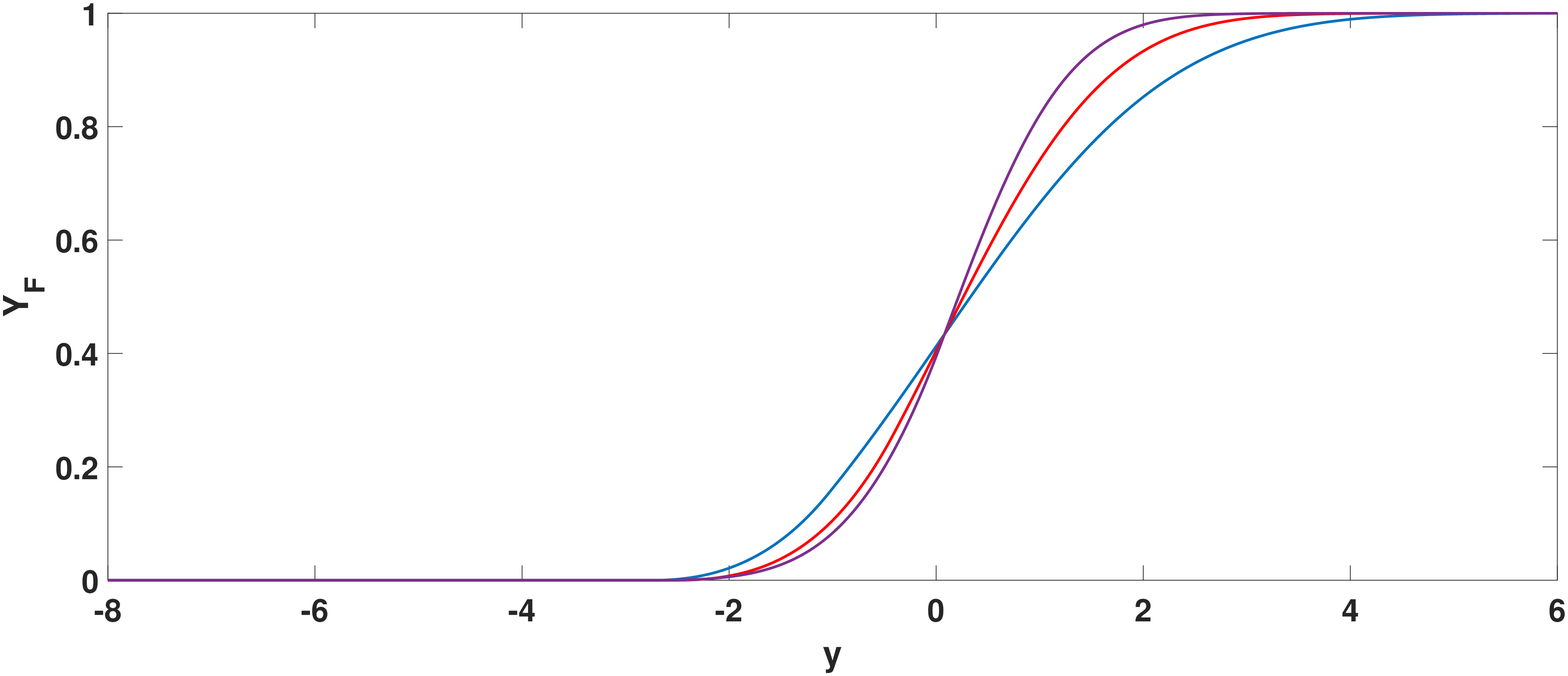}}     \\
  \vspace{0.2cm}
  \subfigure[ mass ratio x oxygen mass fraction, $\nu Y_O$]{
  \includegraphics[height = 4.6cm, width=0.45\linewidth]{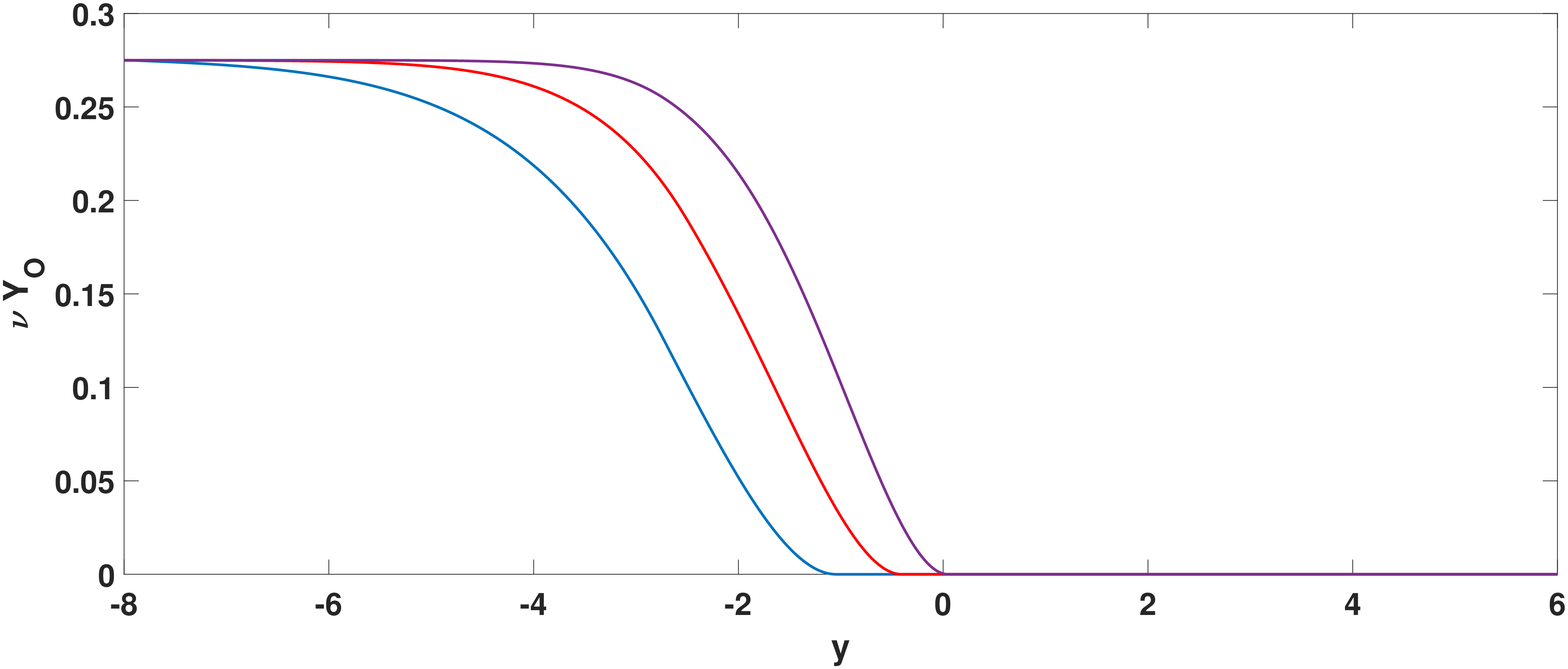}}
  \subfigure[burning rate, $(x_0/u_{\infty})\dot{\omega}_{r,ms}$]{
  \includegraphics[height = 4.6cm, width=0.45\linewidth]{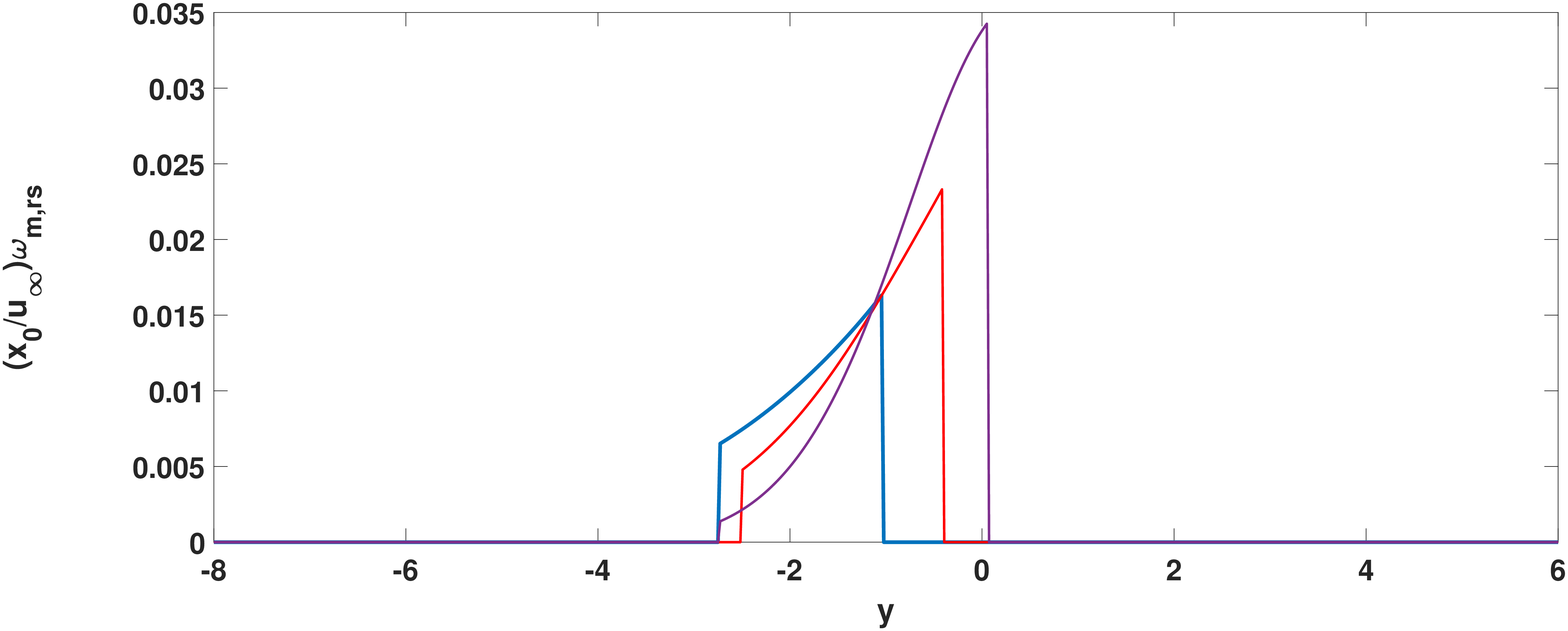}}     \\
  \vspace{-0.1cm}
  \caption{Flow-scale diffusion flame:    blue $x = 20$ , red $x = 15 $ , purple  $x = 12.5$; fuel in higher-speed stream, oxygen in lower-speed stream.                   }
  \label{Flowscale1}
\end{figure}

\begin{figure}
  \centering
 \subfigure[velocity, $u/u_{\infty}$ ]{
  \includegraphics[height = 4.8cm, width=0.45\linewidth]{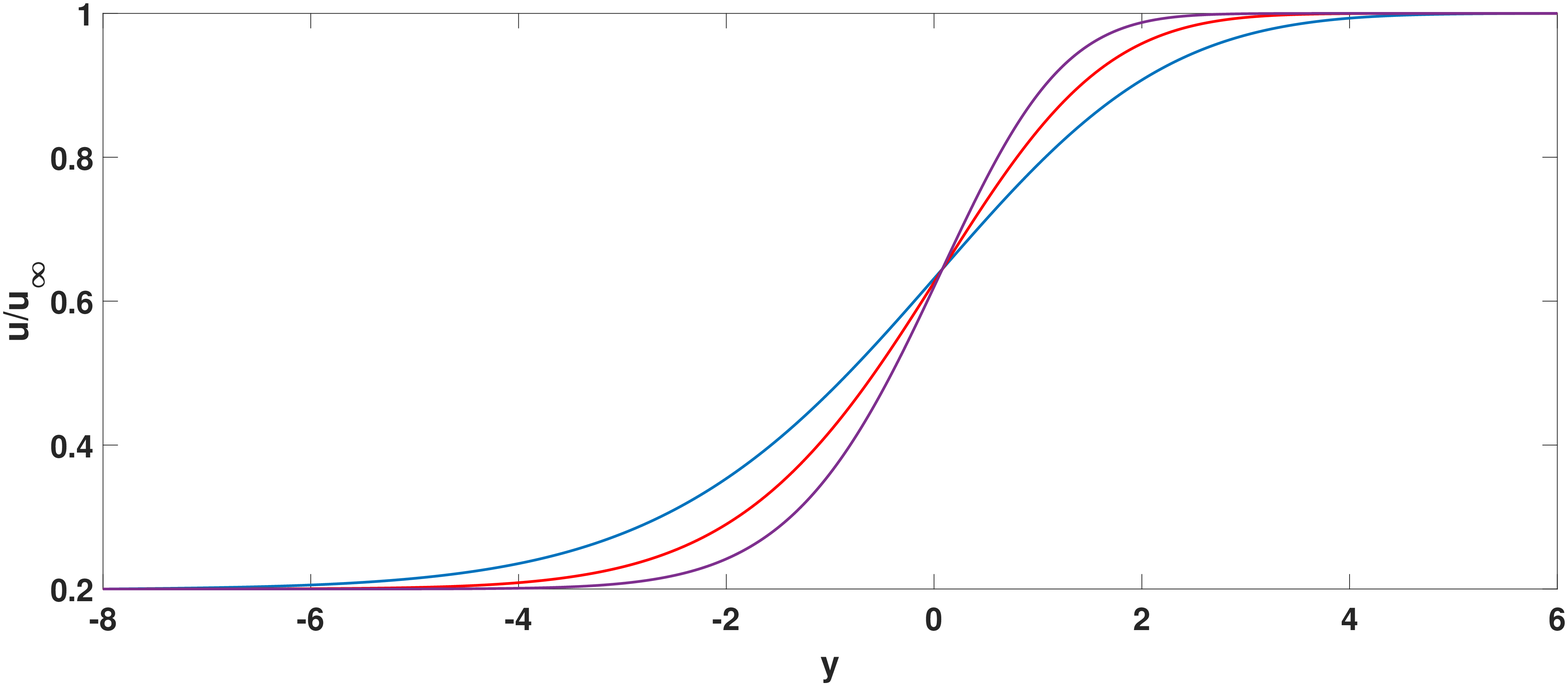}}
  \subfigure[velocity gradient, $|du/dy|$]{
  \includegraphics[height = 4.8cm, width=0.45\linewidth]{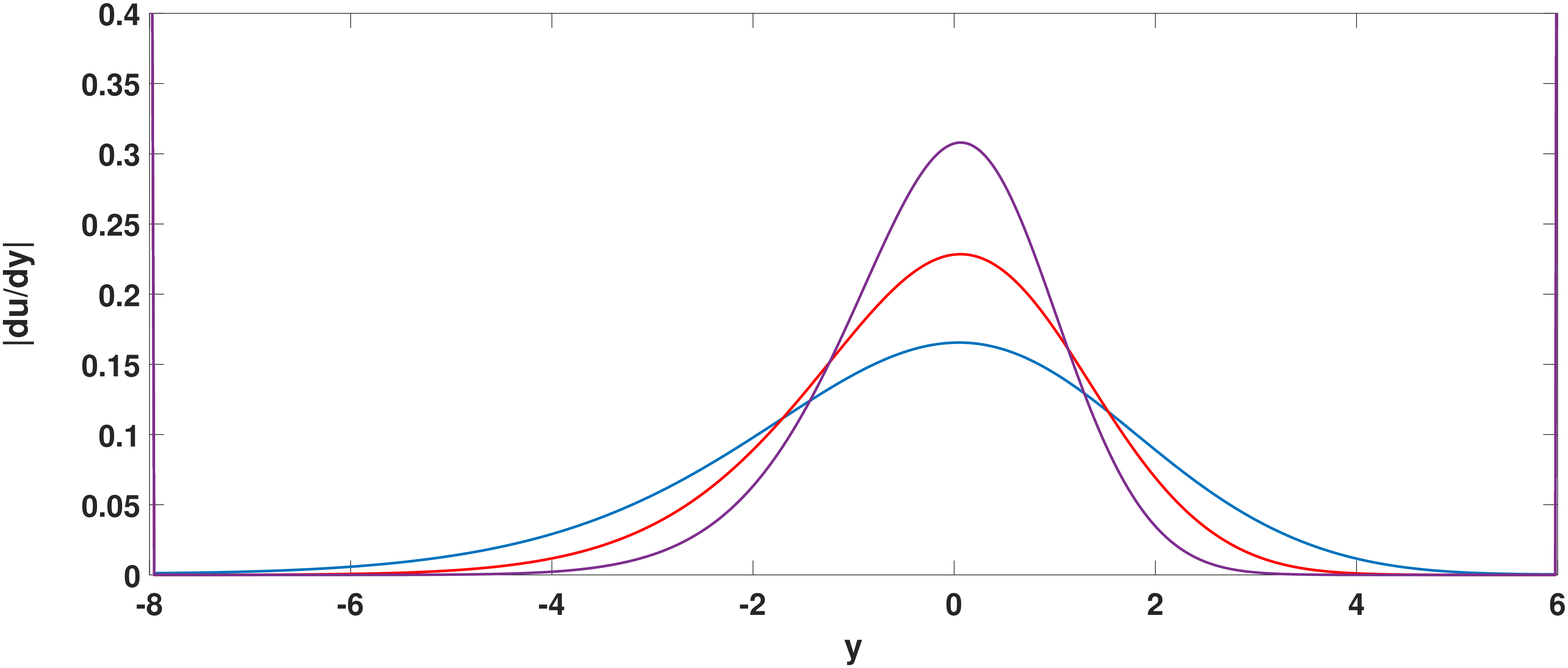}}     \\
  \vspace{0.2cm}
  \caption{Flow-scale diffusion flame:    blue $x = 20$ , red $x = 15 $ , purple  $x = 12.5$; fuel in higher-speed stream, oxygen in lower-speed stream.            }
  \label{Flowscale2}
\end{figure}

Figures \ref{Flowscale1} and \ref{Flowscale2} give results for a case where the fuel stream is the faster free stream while the oxygen stream is the slower stream. The profiles widen with increasing downstream distance for both velocity and scalar properties. The peak enthalpy and temperature values remain approximately unchanged but the locations of the peak value and of the center of the reaction zone shift towards the oxygen-stream side.  The sharp cutoffs in burning rate at certain tranverse positions  occur because the rate is nonzero only where both reactants exist on the large scale but that rate does not depend on the precise mass-fraction values on that scale.

\begin{figure}
  \centering
 \subfigure[enthalpy, $h/h_{\infty}$ ]{
  \includegraphics[height = 4.6cm, width=0.45\linewidth]{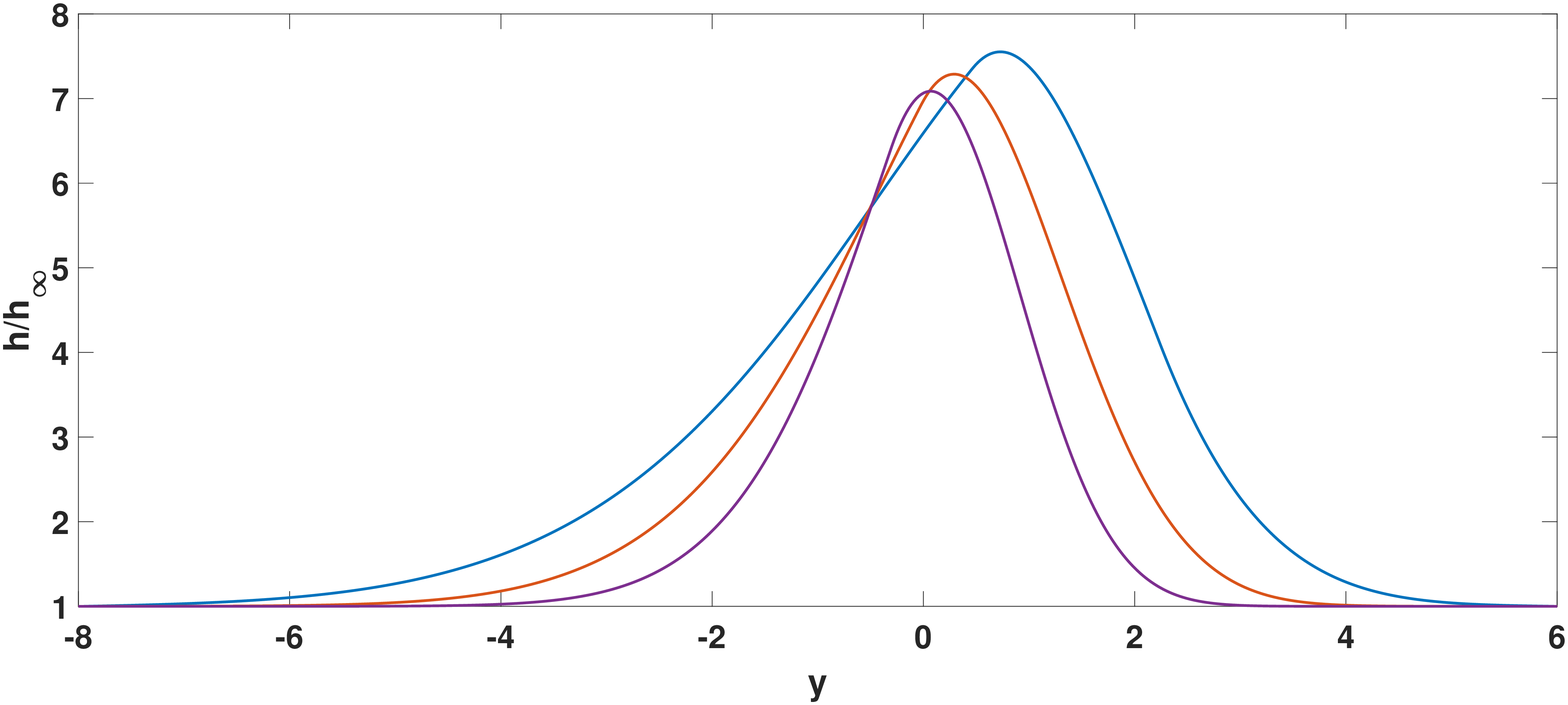}}
  \subfigure[fuel mass fraction, $Y_F$]{
  \includegraphics[height = 4.6cm, width=0.45\linewidth]{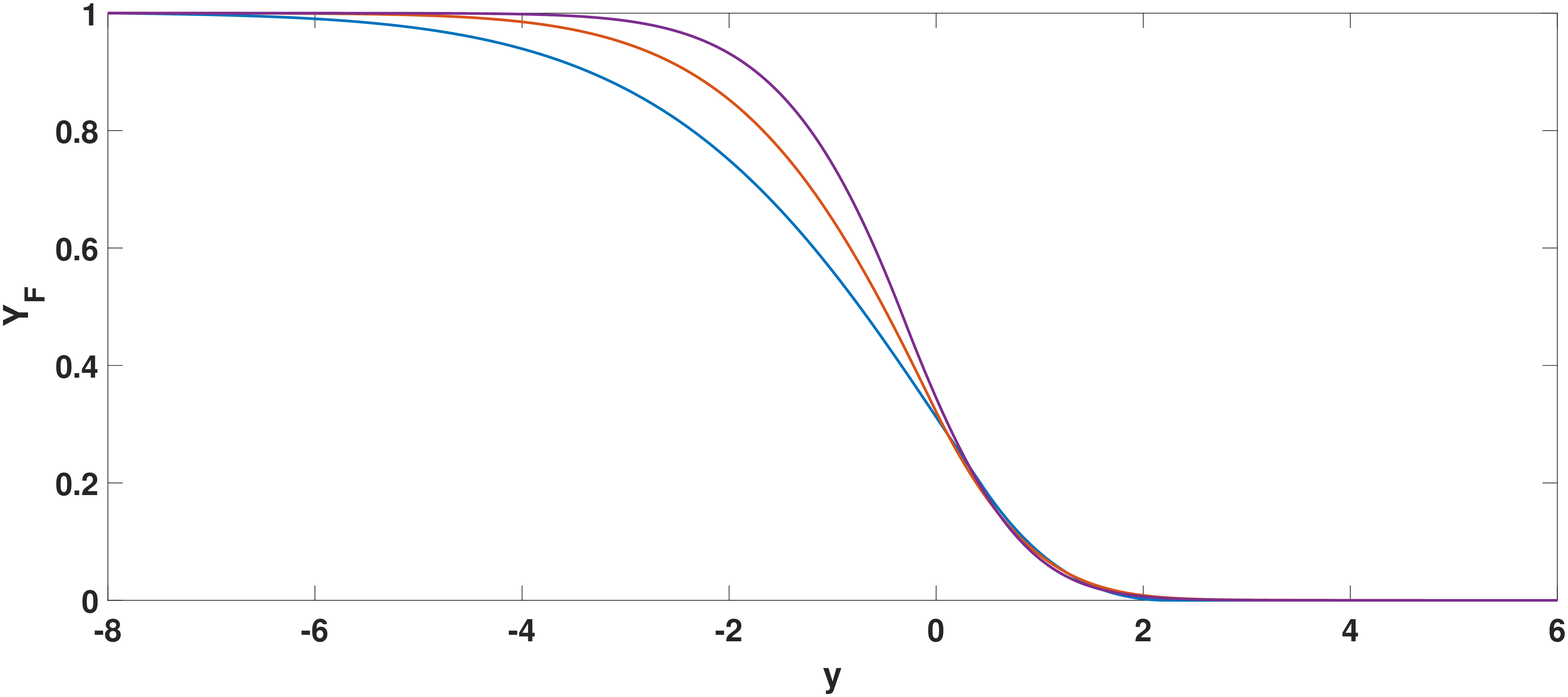}}     \\
  \vspace{0.2cm}
  \subfigure[ mass ratio x oxygen mass fraction, $\nu Y_O$]{
  \includegraphics[height = 4.6cm, width=0.45\linewidth]{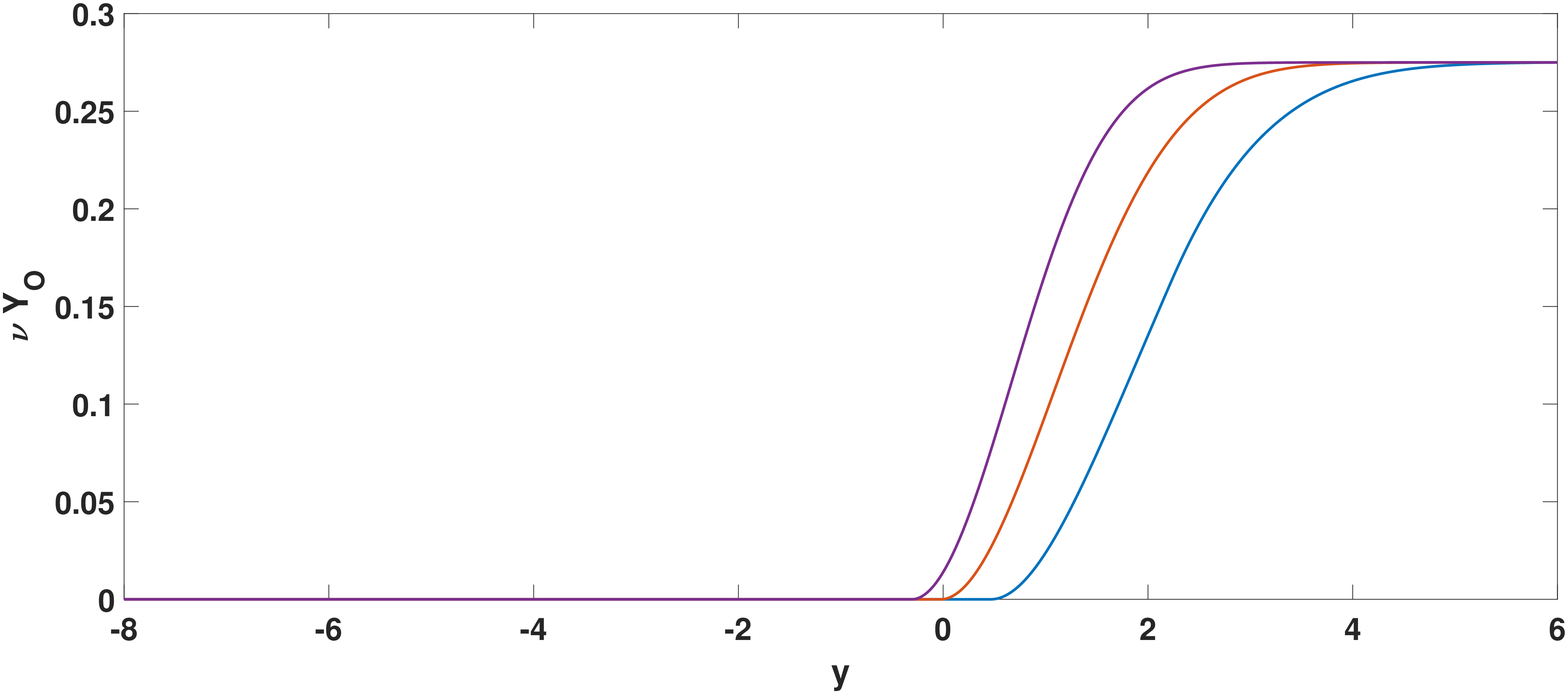}}
  \subfigure[burning rate, $(x_0/u_{\infty})\dot{\omega}_{r,ms}$]{
  \includegraphics[height = 4.6cm, width=0.45\linewidth]{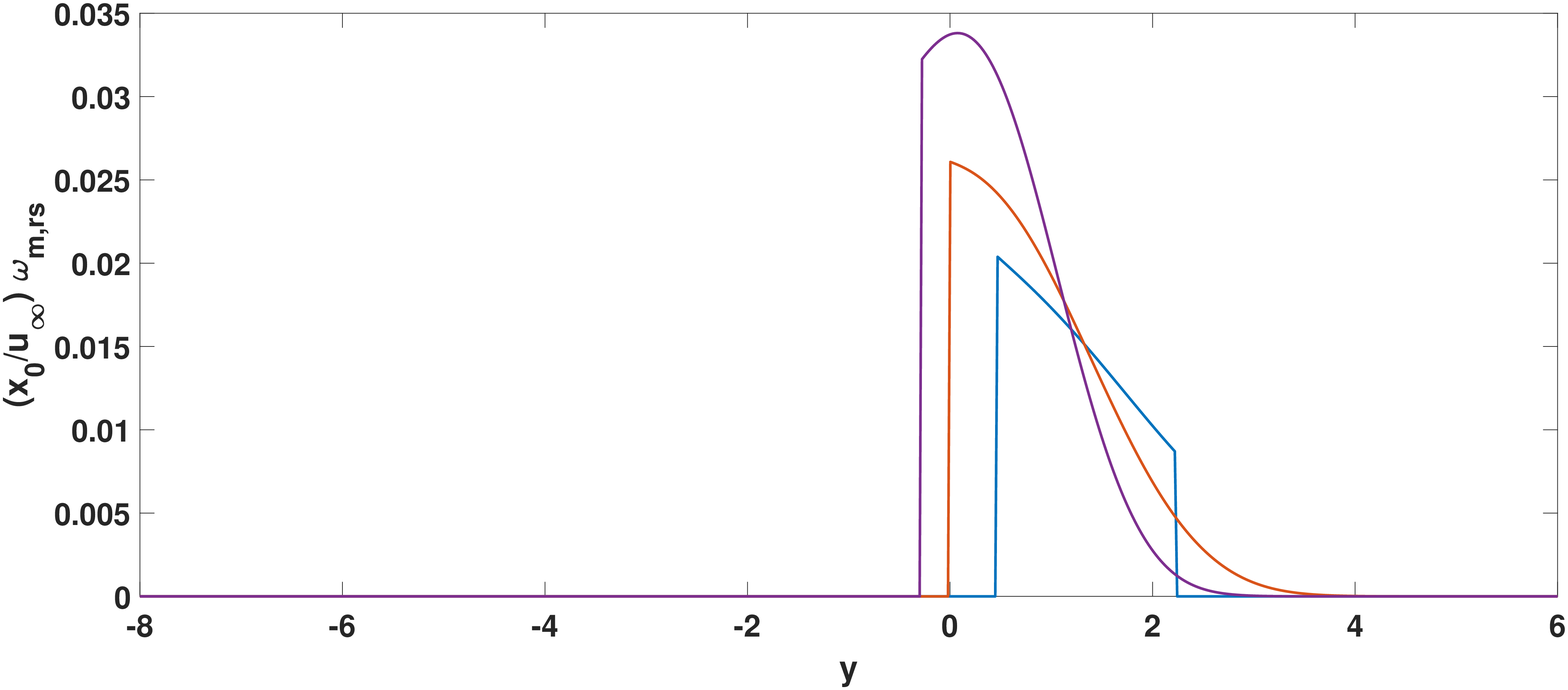}}     \\
  \vspace{-0.1cm}
  \caption{Flow-scale diffusion flame:  blue $x = 20$ , red $x = 15 $ , purple  $x = 12.5$; oxygen in higher-speed stream, fuel in lower-speed stream.          }
  \label{Flowscale3}
\end{figure}

\begin{figure}
  \centering
 \subfigure[velocity, $u/u_{\infty}$ ]{
  \includegraphics[height = 4.8cm, width=0.45\linewidth]{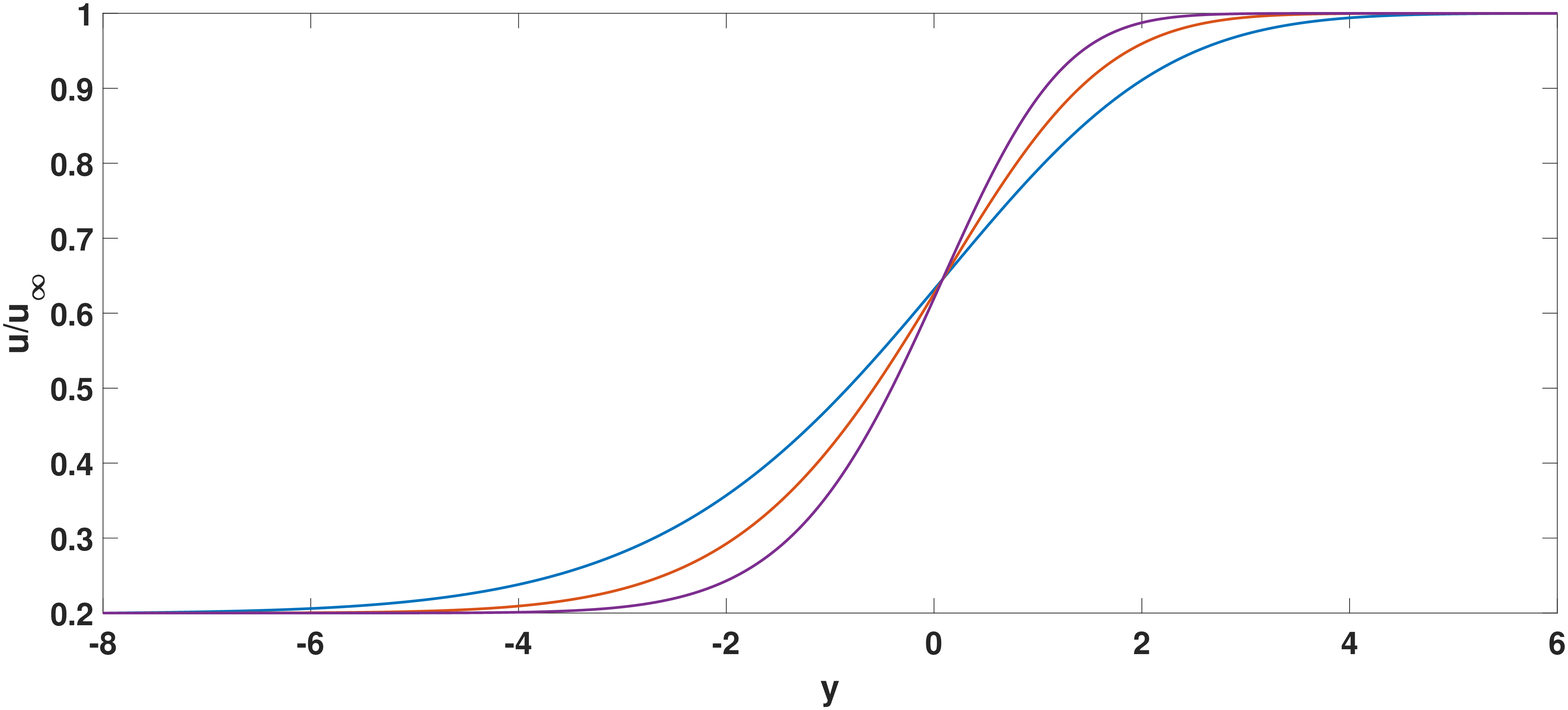}}
  \subfigure[velocity gradient, $|du/dy|$]{
  \includegraphics[height = 4.8cm, width=0.45\linewidth]{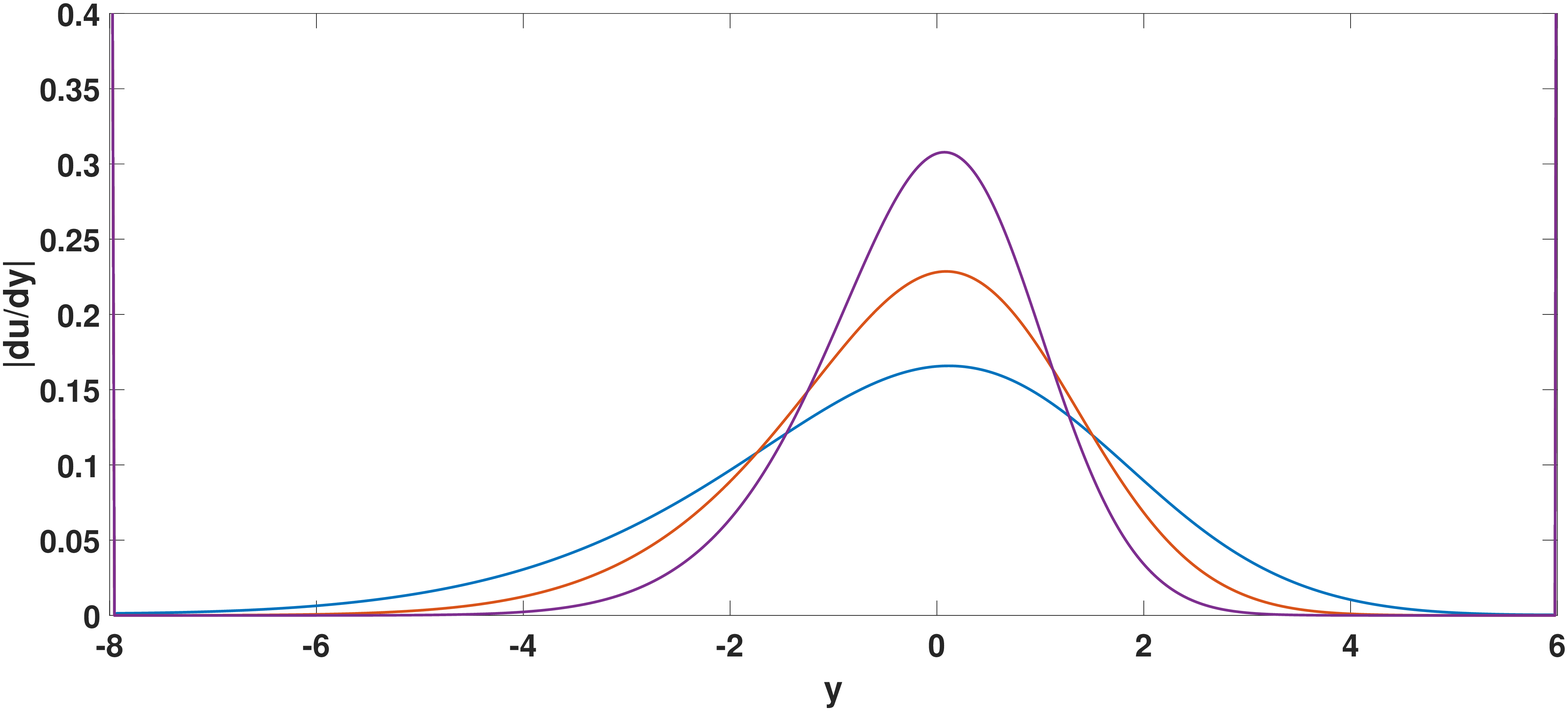}}     \\
  \vspace{0.2cm}
  \caption{Flow-scale diffusion flame:  blue $x = 20$ , red $x = 15 $ , purple  $x = 12.5$; oxygen in higher-speed stream, fuel in lower-speed stream.                }
  \label{Flowscale4}
\end{figure}

Figures \ref{Flowscale3} and \ref{Flowscale4}  give results for a case where the oxygen stream is the faster free stream while the fuel stream is the slower stream. Again, the profiles widen with increasing downstream distance for both velocity and scalar properties. The peak enthalpy and temperature values increase slightly as the locations of the peak value and of the center of the reaction zone shift towards the oxygen-stream side. Note that the burning zone moves towards the oxygen-rich side with increase of the downstream distance whether that stream is the faster or slower stream. This presumably occurs because nearly four times the mass of oxygen (compared to propane mass) is consumed in the reaction.

\newpage

\section{Concluding Remarks}\label{conclusions}

A new flamelet model is developed for use in sub-grid modelling for analysis of turbulent combustion by RANS and LES. This flamelet model presents certain key advances: (i)  non-premixed flames, premixed flames, or multi-branched flame structures are allowed to appear naturally without prescription; (ii) the impacts of shear strain and vorticity (and associated centrifugal effects) on the  flames are determined; (iii) the applied sub-grid strain rates and vorticity are directly related to the resolved-scale strain rates and vorticity without the use of a contrived progress variable; (iv) the flamelet model is three-dimensional without need for assuming axisymmetry or planar geometry, allowing the physically correct counterflow under the vorticity constraint; and (v) variable density is addressed in the flamelet model. The results indicate that each of these five features introduces consequential, vital physics that is missed by current two-dimensional, irrotational, constant-density flamelet models that assume a priori a nonpremixed- or premixed-flame structure and make no direct connection to shear strain or vorticity on the larger turbulence scales.

Information from direct numerical simulations concerning the relative alignments of the vorticity vector, scalar gradients, and principal strain axes provides a basis for a set of assumptions. The analytical framework allows for multi-step, detailed kinetics although the calculations here are limited to one-step propane-oxygen kinetics. The quasi-steady assumption used in previous flamelet models is maintained here.

A similar solution is found for the flamelet model.   Sample computations of the flamelet model without coupling to the resolved flow are presented first to demonstrate the importance of the new features of the model. The rotation due to vorticity creates a centrifugal force that generally decreases the mass-flux through the flamelet counterflow. Thereby, an increase in residence time and a decrease in burning rate occurs. Rotation can thereby allow flames, which  would otherwise extinguish, to survive.

Scaling laws relate sub-grid strain rates and vorticity  to resolved-scale quantities for coupling with large-eddy simulations or Reynolds-averaged flows. Given these relations, the burning rate is determined by the rotational-flamelet physics and the energy release rate and rate of change of species mass fraction are given to the resolved scale by the rotational-flamelet algorithms.  The theory does not introduce any new contrived variable such as a flamelet progress variable. Connection between the two scales is made using long established variables.

For this initial study, a highly challenging turbulent flow is deliberately avoided. Rather, a  simple turbulent flow is resolved with coupling to the rotational-flamelet model in order to explore the interaction across the different scales.   A two-dimensional, multicomponent, time-averaged  planar shear layer with variable density and energy release uses a mixing-length description for the eddy viscosity and is coupled to the new rotational-flamelet model. The eddy diffusivity is  proportional to the local magnitude of the velocity gradient and grows with downstream distance.

The profiles for velocity and scalar properties are seen to broaden with downstream distance in the shear layer, indicating the growing width of that layer. With increasing downstream distance, the zone where burning occurs moves towards the oxygen-rich side whether or not it is the faster stream. That burning zone becomes more narrow with increasing downstream position.

In the future, multi-step kinetics should be utilized with the rotational-flamelet model. The model can be used to produce look-up or neural networks that can be employed with LES or RANS calculations.

There are several important issues that should be addressed in future studies to allow better matching between the closure model for the sub-grid mixing and combustion with the resolved-scale (or time-averaged) flow: (1) the  lags, due to the turbulence cascade, in spatial position and time for the coupling between the large-scale strain and the energy release that it affects; (2) the degree of partial mixing of reactants during the cascade; (3) the proper choice of the $b$ parameter to represent the optimal sub-grid volume-size for averaging; and (4) relative magnitudes of the vorticity and normal principal strain rates on the sub-grid scale. These matters can be examined through carefully designed DNS and perhaps clever experiments. Attention is needed to improve our knowledge of the statistical relations between resolved-scale quantities for vorticity, strain rates, and scalar gradients and those quantities on the flamelet scale.  Of course, flamelets can in principle exist across a range of the smaller scales, not just the smallest scales. Perhaps more attention is needed for the details of the turbulence cascade; for example, strain-rate self-amplification as well as vortex stretching could be relevant \citep{PJohnson}.  Also, better determination is needed of the range of scales where mixing and reaction are prominent.

\section*{Acknowledgements}

An informal review of the manuscript draft by  Professor Said Elghobashi was especially helpful. Suggestions from Professor Perry Johnson  and  Professor Dimitri Papamoschou have been very helpful to the author in writing this article. The effort was supported by AFOSR through Award  FA9550-18-1-0392 managed by Dr. Mitat Birkan.

\section*{Declaration of Interests.} The author reports no conflict of interest.

\bibliographystyle{elsart-harv}
\bibliography {RotationalFlamelet}

\end{document}